\documentclass[fleqn,usenatbib,useAMS]{mnras}

\usepackage{newtxtext,newtxmath}
\usepackage[T1]{fontenc}

\DeclareRobustCommand{\VAN}[3]{#2}
\let\VANthebibliography\thebibliography
\def\thebibliography{\DeclareRobustCommand{\VAN}[3]{##3}\VANthebibliography}

\usepackage[caption=false]{subfig}
\usepackage{graphicx}
\usepackage{subfloat}
\usepackage{csquotes}
\usepackage{grffile}
\usepackage{ulem,color}
\usepackage[dvipsnames,x11names,table]{xcolor}
\usepackage{dcolumn}
\usepackage{bm}
\usepackage{natbib}
\usepackage{xspace}
\usepackage{amsmath,bm}
\usepackage{lipsum}
\usepackage{xfrac}
\usepackage{tikz}
\usetikzlibrary{calc,positioning}
\usepackage{enumitem}
\setlist[itemize]{noitemsep}
\usepackage[switch,mathlines]{lineno}

\usepackage{booktabs}
\usepackage{array}
\usepackage{tabularx}

\definecolor{citeblue}{RGB}{45,85,125}

\hypersetup{
  colorlinks=true,
  citecolor=citeblue,
  linkcolor=black,
  urlcolor=citeblue
}

\usepackage{titlesec}
\titleformat{\section}
  {\normalfont\Large\bfseries\raggedright}
  {\thesection}{1em}{}
\titleformat{\subsection}
  {\normalfont\large\bfseries\raggedright}
  {\thesubsection}{1em}{}
\titleformat{\subsubsection}
  {\normalfont\normalsize\bfseries\raggedright}
  {\thesubsubsection}{1em}{}

\usepackage[cmbtt]{bold-extra}

\def\lesssim{\mathrel{\hbox{\rlap{\hbox{\lower4pt\hbox{$\sim$}}}\hbox{$<$}}}}
\def\gtrsim{\mathrel{\hbox{\rlap{\hbox{\lower4pt\hbox{$\sim$}}}\hbox{$>$}}}}
\newcommand{\bea}{\begin{eqnarray}}
\newcommand{\eea}{\end{eqnarray}}

\newcommand{\dotm}{$\dot M$}

\newcommand{\note}[1]{\textcolor{cyan}{\it #1}}

\newcommand{\outline}[1]{}

\definecolor{darkred}{RGB}{140,20,20}

\usepackage[expansion=false]{microtype}

\newcommand{\Sig}{\Sigma}
\newcommand{\vct}[1]{\boldsymbol{#1}}

\newcommand{\proxy}[1]{\underline{#1}}

\newcommand{\Bp}{\proxy{B_{\rm p}}}
\newcommand{\Pp}{\proxy{P_{\rm p}}}
\newcommand{\Up}{\proxy{U_{\rm p}}}

\newcommand{\phimad}{\phi_{\text{\scshape mad}}}

\usepackage[cmbtt]{bold-extra}

\newcommand{\simfmt}[1]{%
  \mbox{%
    \fontencoding{OT1}%
    \fontfamily{cmtt}%
    \fontseries{m}%
    \selectfont
    #1%
  }%
}

\newcommand{\simchange}[1]{%
  {\bfseries #1}%
}

\newcommand{\DeclareSim}[2]{%
  \expandafter\def\csname sim@#1\endcsname{\simfmt{#2}}%
}

\newcommand{\SimName}[1]{%
  \ifcsname sim@#1\endcsname
    \csname sim@#1\endcsname
  \else
    \PackageError{simnames}{Unknown simulation `#1'}%
      {Add the simulation using \string\DeclareSim.}%
  \fi
}

\newcommand{\PROD}[1]{\SimName{PROD#1}}

\DeclareSim{PROD0}{PNP20Lam}

\DeclareSim{PROD1}{\simchange{NWT}20Lam}

\DeclareSim{PROD2}{%
  PNP20Lam\simchange{BoundCool}%
}

\DeclareSim{PROD4}{%
  NWT20Lam\simchange{CoolInst}%
}

\DeclareSim{PROD5}{%
  NWT20Lam\simchange{CoolInstTgt}%
}

\DeclareSim{PROD6}{%
  PNP20\simchange{Stoch}%
}

\DeclareSim{PROD3}{%
  PNP\simchange{50}Lam%
}

\DeclareSim{PROD7}{%
  PNP50\simchange{Stoch}%
}

\DeclareSim{PROD7.longeq}{%
  PNP50Stoch\simchange{Long}%
}

\DeclareSim{PROD7.nwt}{%
  \simchange{NWT}50Stoch\simchange{Long}%
}

\DeclareSim{PROD16}{%
  PNP20Stoch\simchange{B0.3}
}

\DeclareSim{PROD17}{%
  PNP20Stoch\simchange{B1.5}
}

\DeclareSim{PROD20}{%
  PNP20Stoch\simchange{B3.2}
}

\DeclareSim{PROD19}{%
  PNP20Stoch\simchange{B4.8}
}

\DeclareSim{PROD19hr}{%
  PNP20Stoch\simchange{B4.8HR}
}

\DeclareSim{PROD18}{%
  PNP20Stoch\simchange{B16}
}

\DeclareSim{PROD23.1}{%
  PNP20Stoch\simchange{B1.1}%
}
\DeclareSim{PROD23.2}{%
  PNP20Stoch\simchange{B1.1Low$\nu$}%
}

\DeclareSim{PROD25}{%
  PNP\simchange{40}Stoch\simchange{B0.3}%
}

\DeclareSim{PROD26}{%
  PNP\simchange{40}Stoch\simchange{B3.2}%
}

\newcommand{\minorheading}[1]{%
  \par\addvspace{0.5\baselineskip}%
  \noindent\textbf{#1}\par\nobreak
  \vspace{0.15\baselineskip}%
}

\usepackage{scalefnt}
\usepackage{etoolbox}

\newlength{\shortstackgap}
\makeatletter
\patchcmd{\@shortstack}
  {\lineskip 3\p@}
  {\lineskip\shortstackgap}
  {}{}
\makeatother

\title{A Magnetic Unification of Merging Supermassive Black Hole Binaries}

\author[M. J. Avara et al.]{
Mark J. Avara,$^{1}$\thanks{E-mail: mjavara@gmail.com}
David O'Neill,$^{1}$
and Zolt\'an Haiman$^{1,2,3}$
\\
$^{1}$Institute of Science and Technology Austria (ISTA),
Klosterneuburg, Austria\\
$^{2}$Department of Astronomy, Columbia University,
New York, NY 10027, USA\\
$^{3}$Department of Physics, Columbia University,
New York, NY 10027, USA
}

\date{Accepted XXX. Received YYY; in original form ZZZ}
\pubyear{\the\year{}}

\begin{document}
\label{firstpage}
\pagerange{\pageref{firstpage}--\pageref{lastpage}}
\maketitle 

\begin{abstract}

The detection of an electromagnetic (EM) transient from a supermassive black hole binary (SMBHB) merging within a gaseous environment will constitute a multi-messenger milestone. Numerical simulations of these systems have mostly been performed in 2D with a constant viscosity parameter representing turbulent and magnetic stresses. These simulations predict a steep drop in accretion rate and X-ray luminosity at merger, followed by a slow recovery. However, 3D-GRMHD and 3D-MHD simulations have shown only a modest reduction in accretion rate and a fast post-merger recovery. We present a new 2D pseudo-Newtonian framework with physically motivated prescriptions for stochastic viscosity, relativistic apsidal precession, and large-scale magnetic fields. When suitably calibrated, this 2D framework accurately captures key 3D-GRMHD behavior, demonstrating that it can be used to explore binary accretion without the high computational cost of 3D-GRMHD simulations.
We find that the merger signatures are most sensitive to the strength of the magnetic field, with a drop in the accretion rate at merger progressively less pronounced for stronger fields. Post-merger evolution is determined by both magnetization and relativistic precession of orbiting gas, yielding rapid recovery of accretion. For low-magnetization disks, recovery is driven by self-intersection shocks, and the dip at merger is followed by a distinct second dimming. Large-scale magnetic fields of increasing magnitude lead to magnetically arrested disk (MAD) behavior which disturbs the merging system progressively earlier before merger, and longer after. Our results unify the EM signatures of the merger of embedded coplanar SMBHBs, and provide novel observational constraints on their intrinsic magnetization. 

\end{abstract}


\fixfootnotes

\section{Introduction}
\label{sec:introduction}

The observed cosmological process of hierarchical galaxy mergers is expected to result in the ultimate coalescence of the supermassive black holes (SMBH) originating from the central gravitational potential well of each galaxy \citep{BegelmanBlandford1980, MilosavljevicMerritt2003}. The high, or even near total occupancy fraction of SMBHs in low-redshift galaxies, inferred from kinematics \citep{KormendyHo2013, KormendyRichstone1995} suggests that the black hole merger rate may be intimately tied to the galactic merger history, and provides one of the two dominant mechanisms of cosmological black hole growth \citep{KauffmannHaehnelt2000,MenouHaiman2001,VolonteriHaardt2002}, the other being accretion. 

However, in gas rich galaxy mergers, the role of gas dynamics in delaying or accelerating merger \citep{MacFadyenMilosavljevic2008,ShiKrolik2012,NobleMundim2012,TangMacFadyen2017,MirandaMunoz2017,MunozMiranda2019,MoodyShi2019,TiedeZrake2020,DittmannRyan2021,DittmannRyan2022}, contributing to black hole growth, and affecting the observational characteristics of AGN \citep[see review by][]{DOrazioCharisi2023} encompasses many outstanding problems. SMBHs that sink far enough into the combined galactic potential can become gravitationally bound as a supermassive black hole binary (SMBHB), and even further, reach the gravitational wave (GW) driven inspiral regime. These binaries could become valuable multi-messenger sources \citep[see review by ][and references therein]{BogdanovicMiller2022}. 

The Laser Interferometer Space Antenna (LISA) is expected to directly detect the chirping GW signal of SMBHBs as they inspiral and merge \citep{KleinBarausse2016,LISAColpi2024}, so long as they are not too massive ($\lesssim10^7\rm M_\odot$). The joint detection of mergers with LISA and their EM counterparts provides multiple opportunities that a single method cannot \citep{Baker2019Decadal2019,MangiagliCaprini2022}. For example, the GW detection can provide strong constraints on the binary evolution independent of EM signatures, making a joint detection a strong probe into the physics of the EM emission processes and gas dynamics of the accretion disk \citep{KocsisHaiman2008}. On the other hand, the detection of the EM counterpart to a candidate binary can provide a strong prior for the GW signal analysis \citep{Haiman2017,HaimanXin2023,XinIsi2026}, and help localize the source, which reduces error in binary parameter posteriors. 

The EM signatures of SMBHBs can be understood in relation to their accretion geometry through inspiral and merger. A SMBHB of near equal mass, at wide-separation, with a quasi-circular orbit that is co-planar and prograde with respect to the circumbinary disk (CBD), clears out a cavity in the CBD by destabilizing orbits between about twice the binary semi-major axis (denoted by `$a$'), and the Roche lobes of each black hole (BH) \citep{LinPapaloizou1979,ArtymowiczLubow1994}. Accretion into the cavity occurs despite the disk truncation \citep{ArtymowiczLubow1996,MacFadyenMilosavljevic2008} and a `mini-disk' forms inside each Roche lobe \citep{FarrisDuffell2014,DOrazioHaiman2013} where orbits are approximately stable between the innermost stable circular orbit (ISCO) of each BH and the outer tidal truncation radius \citep{FarrisGold2012,BowenCampanelli2017} ($r_\mathrm{BHi}\sim0.4a$, radius measured from each BH, BH1 or BH2). 

The inner edge of the CBD exists on the edge of orbital stability. Interaction of this gas with the asymmetric gravitational potential leads to a convergence of decelerated and accelerated gas, forming a `stream' feature along which most gas enters the Roche lobes. Gas that falls into the stream but does not make it to the mini-disks is accelerated outward and impacts the CBD.

\begin{figure}
  \centering
  \includegraphics[width=\columnwidth]{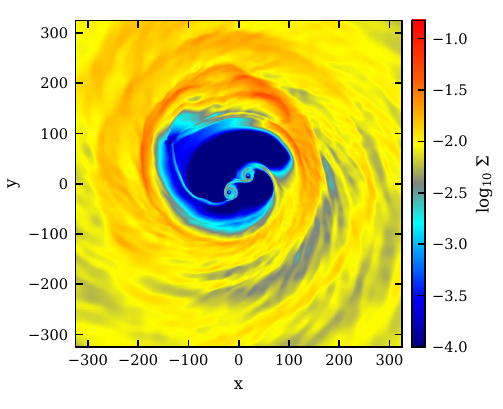} 
  \caption{Density snapshot of our fiducial wide-separation simulation, \PROD{7.longeq} at $\sim 902$ binary orbits plotted in the plane of the disk and binary. The x and y axes are in units of [M], the binary mass. Prominent features include the low-density cavity carved by the binary, the over-dense `lump' located along the middle-right edge of the cavity, eccentric elongation of the cavity from top-left to bottom-right, and mini-disks around each BH. Gas that falls into the cavity collides with gas that has been accelerated outwards by the binary to form `streams' along which most accretion onto the mini-disks occurs.}
  \label{fig:prod7.snapshot1}
\end{figure}

The repeated impact of gas and resonances between the binary and disk drive the inner portion of the CBD into an eccentric configuration that features an azimuthal over-dense region commonly referred to as the `lump', which orbits at $\sim$5-10 times the orbital period and is part of the inner edge of the CBD \citep{NobleMundim2012,Gold2019, DuffellDittmannetal2024}. Figure~\ref{fig:prod7.snapshot1} shows a snapshot of the surface density for one of our viscous hydrodynamic runs where each of these features is apparent\footnote{The lump is generally much more prominent in 3D-MHD simulations than in 2D.}. 

If interaction with the disk removes enough angular momentum from the binary, gravitational wave (GW) losses can start to dominate and shrink the time-dependent binary separation $a(t)$ so rapidly that viscous processes in the disk cannot replenish gas at the CBD inner edge fast enough to keep up.
Equating the GW-driven binary inspiral speed $|\dot a|$ with the radial accretion speed $|v_r|\simeq3\nu/(2r)$ at the CBD inner edge, for kinematic viscosity $\nu$, provides an estimate of the binary separation at which this `decoupling' should occur \citep{DittmannRyan2023}. For an equal-mass quasi-circular binary with the viscosity profile adopted in this work, this decoupling separation is
\begin{equation}
\frac{a_{\rm dec}}{r_g}
\simeq
25
\left(\frac{\alpha}{0.1}\right)^{-2/5}
\left(\frac{H/r}{0.1}\right)^{-4/5}
\left(\frac{\xi}{2}\right)^{1/5},
\end{equation}
with parameterized viscosity $\alpha$, scale height $H/r$ of the disk, and gravitational radius $r_g\equiv GM/c^2$, where $M$ is the total binary mass. We find empirically that for a Mach number (ratio of disk orbital velocity to sound speed, $v_\phi/v_s$) $\mathcal{M}=10$ disk the inner edge of the CBD occurs at $r_\mathrm{in}\approx \xi a$, with $\xi\simeq2$.
This velocity-based criterion has been found to predict decoupling more accurately than a balance of accretion and GW merger timescales in viscous hydrodynamic simulations \citep{DittmannRyan2023}. Timescale-based estimates also tend to over-estimate the decoupling radius measured in 3D general relativistic (GR) magnetohydrodynamic (MHD) simulations \citep[see discussion in][hereafter referred to as MJA24]{AvaraKrolik2024}, though the process of decoupling in binary accretion flows remains under-explored and the decoupling process in turbulent MHD is highly nonlinear. This decoupling can result in a drop in the accretion rate onto the binary. If an accompanying drop in luminosity occurs, it could provide an observable signature.

Prodigious effort has gone into predicting the EM signatures of these highly nonlinear systems using 2D and 3D simulations in both the gas-driven and GW-inspiral regimes \citep[][just to name a few]{ArtymowiczLubow1994, ArtymowiczLubow1996, HayasakiMineshige2007, MacFadyenMilosavljevic2008, ShiKrolik2012,NobleMundim2012,DOrazioHaiman2013,FarrisDuffell2015a,RagusaLodato2016,LopezArmengolCombi2021,ZrakeTiede2021,DuffellDOrazio2020,RoedigDotti2011,SiwekWeinberger2023,WesternacherSchneiderZrake2022,FranchiniLupiSesana2022,FranchiniLupiSesanaHaiman2023,LaiMunoz2023}, and through the merger in 2D \citep{FarrisDuffell2015b,TangHaiman2018,KrauthDavelaar2023,FranchiniPrato2024,ClyburnZrake2025} and 3D \citep{GiacomazzoBaker2012,FarrisDuffell2015b,KellyEtienne2021,BowenMewes2019,CombiLopezArmengol2022,FranchiniBonetti2024,AvaraKrolik2024,EnnoggiCampanelli2025}. 3D general relativistic (GR) magnetohydrodynamic (MHD) simulations of the merger process inside a pre-equilibrated CBD have only recently been accomplished due to the complexity of the codes required, extreme computational expense, and large scale separation intrinsic to the system (see \citealt{EnnoggiCampanelli2025}; hereafter referred to as LE25, references therein, and the reviews by \citealt{Gold2019review} and \citealt{CattoriniGiacomazzo2024}).

Despite this broad effort, the predictions for the EM counterparts for even the simplest case of quasi-circular equal-mass and coplanar binaries from 2D versus 3D-GRMHD simulations have remained in apparent tension. The inherent complexity of the fully turbulent, magnetized, and relativistic simulations has delayed resolution of this tension. 2D viscous hydrodynamic simulations of inspiral and merger like those in \cite{KrauthDavelaar2023} (hereafter referred to as LK23) and \cite{FranchiniBonetti2024} predict decoupling at wider separation, almost complete cut-off in accretion rate and X-ray luminosity, and very slow recovery post-merger. In 2D, a larger and colder cavity at merger is also seen. On the other hand, 3D-GRMHD simulations in LE25 and \cite{MostWang2024} show at most a factor of 10 reduction in accretion during inspiral. LE25 sees a late-time flattening in energy dissipation powering EM emission (first reported in MJA24), and rapid brightening at merger. The accretion rate remains almost continuous through merger and only slowly rises afterwards. 

In 3D-MHD, magnetic fields are central to the behavior of accretion, and therefore offer a strong candidate to account for the contradictory findings above. When accretion disks have more large-scale magnetic flux threading vertically through the system than can be contained on the BH \citep{1974Ap&SS..28...45B,2003PASJ...55L..69N}, this flux is efficiently advected inward along the surface \citep{AvaraMcKinney2016} and piles up from the horizon outwards into the disk. Magnetic flux tubes erupt quasi-periodically off the horizon leading to large variability, and the large-scale flux fundamentally alters the transport of angular momentum by launching magneto-centrifugal winds and driving turbulence \citep[][and references therein]{BegelmanScepi2022}. 
The pile-up results in a self-similar radial magnetic profile extending out to a radius defined by the total large-scale flux available to the disk. Beyond that radius, standard and normal disk evolution (SANE) consistent with gas or radiation-pressure dominated behavior, and angular momentum transport via the magneto-rotational instability (MRI,  \cite{1991ApJ...376..214B}) is expected. So long as this pileup exists to a finite transition radius outside the ISCO, these systems are referred to as Magnetically Arrested Disks (MAD) \citep{1974Ap&SS..28...45B,2003PASJ...55L..69N,2008ApJ...677..317I,2012JPhCS.372a2040T,2012MNRAS.423.3083M}.  Several recent studies in 3D-MHD have begun to explore the MAD accretion regime in binaries \citep{MostWang2024,2025arXiv250816855W,2025PhRvD.112j3050M} . Just as in single BHs (sBHs), this state of accretion can radically alter the behavior of binary disks, blurring the distinction between CBD, mini-disk, and cavity. Binaries in the MAD state will potentially have uniquely different EM signatures than those undergoing SANE accretion, where the MRI dominates. 

In order to make significant progress in understanding the EM emission from merging binaries we need to be able to efficiently explore the binary and disk parameter space in these systems, and resolve the apparent tension between 2D and 3D predictions. 
In this study, we introduce an efficient 2D framework with a simple prescription for key MHD features that allows us to replicate MAD behavior for sBH and SMBHB accretion. Additionally, we account for relativistic precession effects by introducing a novel pseudo-Newtonian potential, and add stochastic fluctuations to our viscosity prescription, motivated by observed AGN variability and the MRI, which imparts important stochasticity to the lightcurves. Finally, we implement, for the first time in 2D simulations of accretion, target-entropy cooling, which limits over-cooling of adiabatic fluctuations of the quasi-turbulent disk and locations of rapidly compressed and shocked gas.

With this new framework we are able to show how prior 2D and 3D predictions for EM merger counterparts are not in tension after all, but exist in a unified framework. We discover a simple relation between large-scale magnetic fields and EM signatures. We also find that precession-induced shocks drive a rapid post-merger recovery, which includes a brief secondary dimming just after merger. Finally, we explore decoupling with magnetic effects for the first time in 2D or 3D when the system starts at pre-decoupling separations for thin disks. Because we can reproduce in 2D simulations many of the key features of magnetized disks only seen before in 3D-GRMHD, we open the opportunity for efficient 2D exploration of the parameter space of binary accretion in this context.

The most significant observational implication of our findings is that the dimming associated with merger has a monotonic correspondence with total net vertical magnetic field present in the binary disk. This suggests that a confirmed EM counterpart to a merging binary, e.g. associated with a LISA detection, could be used to constrain the intrinsic magnetization of AGN disks. 

To present our methodology and results, we organize this paper as follows: in \S\ref{sec:methods} we describe the numerical prescriptions and the physical problem setup.  \S\ref{sec:hydro_results} contains a presentation of our fiducial stochastic 2D viscous hydrodynamic model, results of an unmagnetized suite of simulations, and their comparison to prior work. We compare our prescriptions and parameter choices with those traditionally used. In \S\ref{sec:hydro_bproxy} we describe our models with proxy magnetization that have the same initial binary separation as LE25. In \S\ref{sec:wider_mergers} we describe the evolution of systems with wider initial separation, which ensures we evolve through the complete decoupling process. These are run with and without magnetization. In \S\ref{sec:spectral_evolution} we analyze the spectra and light-curves resulting from our simulations. We discuss implications for EM counterpart modeling and for diagnosing disk magnetization in \S\ref{sec:discussion}, and finally summarize our conclusions in \S\ref{sec:conclusions}. 

\section{Numerical methodology and initial conditions}
\label{sec:methods}

Before introducing the initial conditions of the binary and disk in our simulation suite, we summarize the equations of viscous hydrodynamics that we solve. By augmenting these equations with physically motivated source terms and an additional proxy field for the large-scale vertical magnetic flux, we are able to emulate 3D-GRMHD simulations of single and binary BH accretion disks in 2D. sBH simulations are used to calibrate the magnetic proxy and our novel pseudo-Newtonian potential, as well as empirical normalization of synthetic spectra. We then use binary simulations with this framework to test the relevance of the behavior encoded by each added prescription to the binary accretion problem.

\subsection{Equations solved}
\label{subsec:2hydro}

\begin{figure}
  \centering
  \includegraphics[trim={0 0 0 0}, clip,width=\columnwidth]{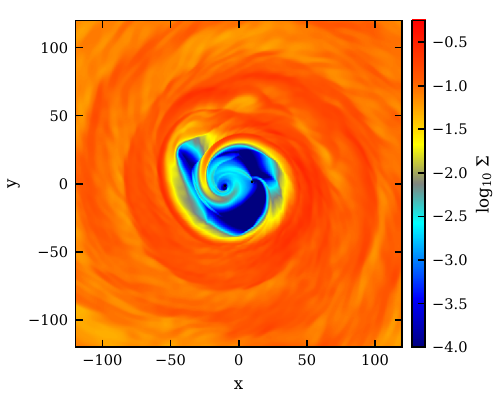}
  \caption{Surface density of \PROD{6} at $t=41600\rm M$, shown on a logarithmic 
scale in the equatorial plane. }
  \label{fig:fiducialsimSig}
\end{figure}
\addcontentsline{toc}{subsubsection}{\note{Figure~\ref{fig:fiducialsimSig}: stochastic disk condition, beta or viscosity and rho.}}

\begin{figure}
  \centering
  \includegraphics[trim={0 0 0 0}, clip,width=\columnwidth]{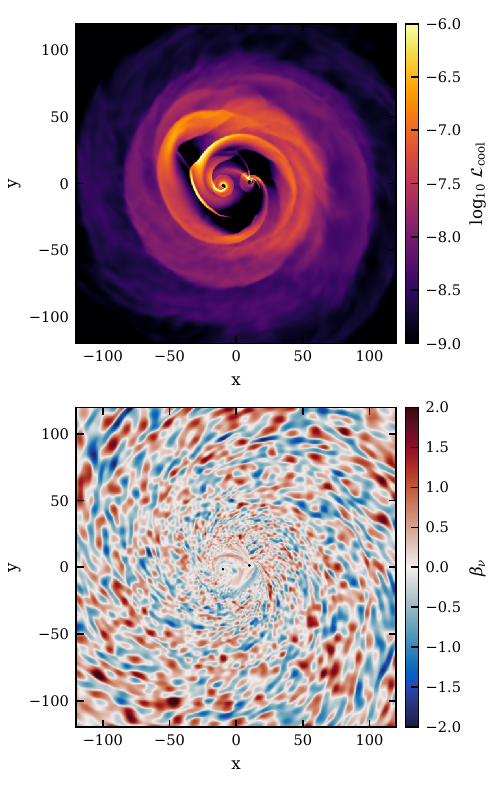}
  \caption{Snapshots showing \PROD{6} at $t=41600M$, matching Fig.~\ref{fig:fiducialsimSig}. The top panel shows the luminosity obtained from the cooling function, described in \S\ref{subsec:cooling}. The bottom panel shows the advected scalar $\beta_\nu$, which controls the fluctuating component of the viscosity, as described in \S\ref{subsec:stoch}. Mesoscale correlations in $\beta_\nu$ introduce correlated $r-\phi$ stress, leading to the propagating surface-density fluctuations visible in Fig.~\ref{fig:fiducialsimSig} as feathering of the more prominent features in $\Sig$.}
  \label{fig:fiducialsimCoolBetanu}
\end{figure}
\addcontentsline{toc}{subsubsection}{\note{Figure~\ref{fig:fiducialsimCoolBetanu}: stochastic disk condition, beta or viscosity and rho.}}

Our time-dependent accretion solution is obtained using the vertically integrated Navier-Stokes equations on an adaptive-resolution Cartesian grid with the code \texttt{Athena++} \citep{Stone2020}. \texttt{Athena++} is a flux-conservative, shock-capturing code, highly scalable and publically available. We select the native Piecewise-Linear-Method (PLM) for reconstruction of fluxes at cell faces, then used by the HLLC Riemann solver, with second-order van Leer time integration. Figure~\ref{fig:fiducialsimSig} shows the density structure of the evolved state of our fiducial unmagnetized simulation.  

We evolve the surface density $\Sigma$, linear momentum terms
$\Sig\vct{v}$, and gas total energy density $E_g$, 
\begin{equation}\label{eqn:Eg}
E_{\rm g}
=
U_{\rm g}
+
\frac{1}{2}\Sig |\vct{v}|^2,
\qquad
U_{\rm g}
=
\frac{P}{\gamma-1},
\end{equation}
where $U_g$ is the internal energy density, $P$ the vertically integrated gas pressure, and $\gamma=5/3$ is the adiabatic constant. 
We adopt geometrized natural units where $G={\rm M}=c=M_b=1$, so mass, length, and time are all in units of the initial binary total mass, $M_b$, which we set to unity. 

For a sub-set of the simulations, we also mimic the large-scale vertical magnetic field with an additional conserved quantity, a scalar magnetic field proxy $\proxy{B_p}$, such that $\Bp=\Sig \underline{b}$,
where $\underline{b}$ is the corresponding primitive scalar density defined in \texttt{Athena++}. We refer to this implementation of magnetization as the `Bproxy' prescription. Bproxy terms are underlined for clarity. This new field is meant to account only for the net vertical flux evolution in accretion disks, so vertical symmetry is implicitly assumed and consistent with the vertical integration of the system. Our full set of coupled equations is essentially that of symmetry-restricted ideal-MHD for a purely vertical field of one dominant sign, i.e. only $B_z$ in the induction equation. We add, however, several effects related to 3D behavior as source terms for the magnetic evolution. Since this prescription is not true 2D-MHD, we refer to it as a proxy field. 

Just as in ideal MHD, the proxy field has associated pressure and energy densities defined by,
\begin{equation}
\Pp
=
\Up
=
\frac{1}{2}\Bp^2
.
\label{eqn:bpress}
\end{equation}
The gradient of the magnetic-pressure is introduced as an explicit cell-centered source term at second order in time rather than through the interface fluxes and Riemann solver. It therefore influences the fluid state update but is not included in \texttt{Athena++}’s characteristic reconstruction. The characteristics of the hydrodynamic state are used in the calculation of the timestep using the native Courant-Friedrichs-Lewy (CFL) condition with a further reduction $\Delta t=0.45 \Delta t_{\rm CFL}$ for stability and accuracy. 

Including the Bproxy prescription, we arrive at a set of evolution equations,

\begin{equation}
\frac{\partial \Sig}{\partial t}
+
\nabla\cdot(\Sig\mathbf v)
=\mathcal{S}^\Sig,
\end{equation}

\begin{equation}\label{eqn:momentum}
\frac{\partial(\Sig\mathbf v)}{\partial t}
+
\nabla\cdot
\left(
\Sig\mathbf v\mathbf v
+
P\mathbf I
-
\boldsymbol{\Pi}
\right)
=
-\nabla \Pp 
-\Sig\nabla \Phi_\mathrm{bin}
+\mathcal{S}^{\Sig\nu}
+\underline{\boldsymbol{F}},
\end{equation}

\begin{multline}
\label{eqn:energy}
\frac{\partial E_g}{\partial t}
+
\nabla\cdot
\left[
(E_g+P)\mathbf v
-
\boldsymbol{\Pi}\cdot\mathbf v
\right]
=
-\mathbf v\cdot\nabla \Pp \\
-\Sig \mathbf v\cdot\nabla \Phi_\mathrm{bin}
+
\mathcal{L}_{\rm cool} 
+\mathcal{S}^{E_g}
+\mathbf v\cdot\underline{\boldsymbol{F}},
\end{multline}

with closures defined by Equations~\ref{eqn:Eg} and \ref{eqn:bpress}, and the stress tensor 
\begin{equation}
\Pi_{ij}
=
\Sig\nu
\left(
\partial_i v_j
+
\partial_j v_i
-
\frac{2}{3}\delta_{ij}\nabla\cdot\mathbf v
\right).
\end{equation}
$\mathbf I$ is the identity tensor, so that $P\mathbf I$ represents an isotropic gas-pressure tensor. We dedicate subsections below to the magnetic source term $\underline{\boldsymbol{F}}$ with its induction equation analogue (\S\ref{subsec:method:bproxy}), the testing and calibration of the Bproxy prescription (\S\ref{subsec:testingproxy}), the kinematic viscosity $\nu$ with stochastic fluctuations (\S\ref{subsec:stoch}), the binary potential $\Phi_\mathrm{bin}$ (\S\ref{subsec:potential}), and the cooling source $\mathcal{L}_\mathrm{cool}$ (\S\ref{subsec:cooling}).

A sink, $\mathcal{S}^j$, for each hydrodynamic conserved quantity $C^j$ is applied to the right hand side of Equations (4-6) at second order in time. These remove mass, momentum, and energy, respectively, in the moving frame of the BHs according to the method described in \cite{DempseyMunoz2020}, to prevent accumulation in active domain cells at the BH centers. The sink has the form,
\begin{equation}
\mathcal{S}^j=-\frac{M_i}{M_b}\frac{C^j}{t_\mathrm{sink}} e^{-(r_i/r_\mathrm{sink})^8},
\end{equation}
with sharp activation at $r_\mathrm{sink}$, and sink rate $t_\mathrm{sink}^{-1}$. The radial position with respect to each BH is $\mathbf r_i$, for i$\in$(BH1, BH2, b) representing each BH and the center of mass of the binary, respectively. The BHs have mass $M_i$. 

We choose the sink radius, $r_\mathrm{sink}=4M_i$, to be inside the ISCO for the initially non-spinning BHs, which justifies removal of angular momentum at the same rate as radial momentum. This corresponds to a `torque-free' parameter of 0, in the terminology of \cite{DempseyMunoz2020}. We choose a very fast sink timescale, $t_\mathrm{sink}=2$M. Our sink differs from \cite{DempseyMunoz2020} in that, for a timestep $\Delta t$, we solve analytically for $C^j$ as a function of time as $\mathcal{S}^j$ and $C^j$ vary together, to get the total reduction. For mass and internal energy we cap the removal so that the conserved variables do not drop below 100 times the floor value. The total accretion rate, \dotm, is calculated by summing the removed mass, so this avoids the injection from floors from biasing the accretion rate, $\dot M_i$, onto each BH. \dotm\ has units of $\Sigma_0r_g^2c$, where $\Sigma_0=1$ is the initial maximum disk density. Because our models exclude density-dependent processes, such as radiation and gas-driven binary evolution, the density normalization remains arbitrary, and in geometrized units, mass and time cancel so it is written in this text without units.

\subsection[Magnetic flux transport equations]
{Magnetic flux transport equations
($\protect\underline{\protect\boldsymbol{F}}$)}
\label{subsec:method:bproxy}

Motivated by sBH 3D-GRMHD simulations of accretion disks with large-scale net vertical flux, specifically the MAD disk state, we introduce three key features. The first, a radial drift with velocity $\mathbf v_{\rm d}$ with respect to the gas, directly enters the magnetic proxy evolution equation,

\begin{equation}
\frac{\partial \Bp}{\partial t}
+
\nabla\cdot
\left[
\Bp
\left(
\mathbf v+\mathbf v_{\rm d}
\right)
\right]
=
0.
\end{equation}
The Bproxy field is evolved conservatively at second order in time, with upwind fluxes used in calculating the advection term. The two additional magnetic source terms encapsulated by $\underline{\boldsymbol{F}}$ are related to the large-scale poloidal magnetic field structure of winds and jets in MAD systems, and is introduced below.

\minorheading{Radial advection}
The radial advection, or drift, of the proxy flux is designed to account for the advection of large-scale field seen in 3D-GRMHD disk simulations. Magnetic flux foot-points can be transported radially along the surfaces of the disk \citep{AvaraMcKinney2016,BegelmanScepi2022}. We define a radial drift velocity,
\begin{equation}
\label{eqn:vd}
\proxy{\mathbf v_{\rm d}}
=
-\frac{3}{2}\proxy{\alpha_{\rm p}}
\left(
\frac{H}{r}
\right)^2
v_K(r)\,
W(r)\,
\mathbf e_r,
\end{equation}
where $v_K(r)$ is the local Keplerian orbital velocity, and $\mathbf e_r$ is the unit radial vector from the binary center of mass defining the direction to be symmetrically inward across the CBD. The rate of this advection is chosen to be faster than the time-averaged radial velocity of gas (governed by our choice for disk viscosity, $\alpha=0.1$), by setting $\proxy{\alpha_{\rm p}}=0.5$. Because the radial advection of the large-scale field is not well understood for the complex binary cavity, we introduce an activation function limiting the advection to the CBD, 
\begin{equation}
W(r;r_0,\epsilon)
\equiv
\frac{1}{2}
\left[
1+
\tanh\left(
\frac{r-r_0}
{\epsilon\,r_0}
\right)
\right],
\end{equation}
where $r_0 = 1.2a_{\rm eff}$ and $\epsilon = 0.1$, and 
\begin{equation}
a_{\rm eff}
\equiv
\left|\mathbf r_1-\mathbf r_2\right|
+
r_{\rm sink},
\end{equation}
is the instantaneous separation of the BHs, floored by the sink radius. After merger, advection is active everywhere outside the sink radius $r_{\rm sink}$, directly mimicking sBH MAD disk behavior.

Qualitatively, we find that the MAD portion of sBH disks under evolution with the B-field proxy are not significantly different with or without the radial advection of flux, but it does make a quantitative difference in capturing the flux eruption cycle.

\minorheading{Wind-induced $r-\phi$ torque}
The second feature of large-scale magnetic field evolution in 3D-MHD which we choose to parameterize is the $r-\phi$ torque associated with loss of angular momentum through a magneto-centrifugal wind \citep{1982MNRAS.199..883B} from the MAD inner portion of highly magnetized disks. 

First, let us consider a self-similar disk model that includes angular momentum and mass loss due to a generic wind. Let the inward accretion rate $\dot M_a$ vary with radius as
\begin{equation}
\dot M_a(r) \propto r^{\zeta},
\end{equation}
so that
\begin{equation}
\zeta \equiv \frac{d\ln \dot M_a}{d\ln r}.
\end{equation}
The wind mass-loss rate $\dot M_w$ from an annulus satisfies $d\dot M_w = \,d\dot M_a$, and the mass-loss rate per unit surface area is
\begin{equation}
\dot\Sig_w(r) = \frac{\zeta(r)\,\dot M_a(r)}{2\pi r^2}. 
\end{equation}

It is straightforward to show that $j_w$, the specific angular momentum carried in the wind, is related to the specific angular momentum in the disk, $j_K$ (which we approximate to be Keplerian), by
\begin{equation}
j_w = \left(1+\frac{1}{2\zeta}\right) j_K.
\end{equation}
This determines the torque per unit surface area from the wind, $\tau_w$. Using the relation $\tau_w \equiv r F_\phi$ to write this loss in terms of an azimuthal force $F_\phi$, we can apply a source term in the equations of evolution, 
\begin{equation}
F_\phi
=
-
\frac{\dot M_a v_\phi}
{2\pi R^2}
\left(
\zeta+\frac{1}{2}
\right).
\end{equation}

We adopt $\zeta=0.1$, the typical value seen in the self-similar MAD part of a scale height $H/r\sim0.1$ disk in 3D-GRMHD simulations \citep{ScepiBegelmanDexter2024}. $\dot M_a= 2\times10^{-2}$ is used for all simulations and is approximately the measured accretion rate at large radii in all single and binary black hole unmagnetized simulations in this study, since they have identical initial conditions for large radii. Finally, $v_\phi$ is chosen to be the local measured azimuthal velocity so that the torque associated with the wind respects the magnitude and sign of the flow in the disk and always removes momentum even if the flow reverses, though we acknowledge the self-similar wind model may not be as well justified for very sub-Keplerian regions. We make an ansatz that mass loss of this magnitude will be driven by the highly magnetized portions of the flow in our 2D simulations, and apply this azimuthal force directly to those regions identified to be highly magnetized with the activation function,
\begin{equation}
{\cal M}(P,\Pp)
=
A_\mathrm{p}\exp
\left(
-\frac{P}{\Pp}
\right),
\end{equation}
where $A_\mathrm{p}$ accounts for the covering fraction of the disk that is magnetically dominated, i.e. where $\exp \left(-P /\Pp \right) \sim 1$. Thus, SANE portions of the flow do not experience wind-related losses.

We add to the right hand side of the momentum (5) and energy (6) equations the azimuthal Bproxy contribution,
\begin{equation}
\underline{\boldsymbol{F}}\cdot\hat\phi=\proxy{F_{\phi}} =F_\phi\ {\cal M}(P,\Pp),
\end{equation}
and $\textbf{v}\cdot (\underline{\boldsymbol{F}} \cdot\hat\phi)$, respectively, projected into the cartesian coordinates of the code. This force is weighted according to its proximity to each BH Roche lobe, in the same way as the final Bproxy feature, a radial force $\underline{\boldsymbol{F}} \cdot\hat r$. The weighting procedure is described for these together, below. 
 
We find in sBH testing simulations that choosing $A_\mathrm{p}=5$ results in a good match to MAD 3D-GRMHD behavior both in terms of variability of flux eruption cycles between the BH and disk, and in terms of qualitative spiral disk features resulting from the interaction of highly magnetized regions with the dense part of the flow. Qualitatively, the production of MAD behavior and its impact on angular momentum transport in the binary is not particularly sensitive to the value for $A_\mathrm{p}$ we select, so long as our magnetic field proxy prescription contributes a torque from a magneto-centrifugal wind at a magnitude on par with the viscous transport from the kinematic shear viscosity.

\minorheading{$r-\theta$ stress near the BHs}

GR-MHD simulations of MAD disks find that large-scale magnetic flux vertically threading the equatorial plane spends little time between the horizon and the ISCO. This is caused by several factors. The gas inside the ISCO experiences plunging trajectories unless it is very magnetically dominated. Also, relativistic reconnection reorients the approximately dipole configuration of magnetic field threading the horizon into a vertical field threading the disk and displacing dense gas, a process generally referred to as flux eruption. These magnetic flux tubes propagate quickly into the disk and almost all flux quickly reaches radii beyond the ISCO. Time and azimuthal averages of the magnetic field in MAD disks \citep{2009ApJ...699.1789T} show that the jet funnel has an approximately parabolic shape (reflected across the equatorial plane) and the field threading the disk out to about $r\sim10M$ is essentially `choked' by the jet funnel. I.e. the accreting gas is squeezed as it moves inwards by the base of each side of the relativistic jet, itself powered by the Blandford-Znajek process \citep{1977MNRAS.179..433B}. Even for non-spinning BHs, this choking occurs because the magnetic field still retains an approximately split-monopole topology. 

The choking process, first described in detail in \cite{2012MNRAS.423.3083M}, results in a bend in any vertical magnetic field threading the near-horizon region, which can induce significant $r-\theta$ stress resulting in an outwardly directed force. Between this direct bending of the field, its coupling to gas on rapidly plunging orbits, and reconnection accelerating material primarily in the radial direction, vertical field is quickly forced out of this region.

During testing we found that the Bproxy field in MAD 2D simulations piles up with roughly the same power-law shape seen in 3D-GRMHD, but the power-law extends very close to the sink, inside the ISCO. To better reproduce the 3D-GRMHD behavior, we introduce the third and final Bproxy feature, an $r-\theta$ stress mimicking the processes described above in the near-horizon region. 

We use 3D-GRMHD simulations of MAD disks (Avara et al. in prep) to measure the constants in a parameterized fit to the jet funnel geometry of the shape,
\begin{equation}
z_{\rm jet}(R)=a_{\rm jet}R^{b_\mathrm{jet}},
\end{equation}
where R is the cylindrical radius. We find that  $a_{\rm jet}=0.1$ and $b_{\rm jet}=2$ match the outline to the jet funnel fairly well in a prototypical simulation. We again found in testing that the behavior of the field in this near-horizon region is somewhat insensitive to the specific values at the order of magnitude level.

Starting with this empirical jet funnel shape, we are then interested in estimating the radial force that would result from the deflection of the vertical field as it runs into the funnel wall. We can assume that at a radius $R_0$ along the disk midplane, the most conservative estimate of the field deflection angle would corresponding to the tangential intersection with the jet funnel wall. A tangent drawn from \((R_0,0)\) in cylindrical \((R,z)\) coordinates grazes the curve tracing the funnel wall, $z_\mathrm{jet}(R)$, at \(R_t=2R_0\), and therefore has a slope determined by,
\begin{equation}
\left.\frac{dz}{dR}\right|_{R_0} =2a_{\rm jet}R_t=4a_{\rm jet}R_0.
\end{equation}
Since the poloidal field-line slope satisfies the approximate relation,
\begin{equation}
\frac{dz}{dR}\approx\frac{B_z}{B_R},
\end{equation}
we obtain
\begin{equation}
B_R=\frac{B_z}{4a_{\rm jet}R}.
\end{equation}
The radial force in the midplane associated with this bending magnetic stress is taken to be
\begin{equation}
\proxy{F_r}\approx B_RB_z =\frac{\Bp^2}{4a_{\rm jet}r}.
\end{equation}
To avoid singular values when r approaches 0 at the BHs, we soften the denominator, and to restrict the choking to the inner region where 3D-GRMHD simulations show the disk thickness is choked by the horizon-threading field, we again make use of our activation function from above to softly taper this force to apply only within the very inner disk, near and inside the ISCO,
\begin{equation}
\proxy{F_r}=\frac{M \proxy{B_\mathrm{p}}^2}
{4a_{\rm jet}(r_i+r_{\rm sink})}
\left[1-W(r_i;6,0.3)\right].
\end{equation}
$R_0$ has been replaced by the radial distance $r_{i}$ from either of the local BHs or the post-merger BH, and the force is scaled to the corresponding BH mass. 

\vspace{4ex}
We assume the radial and azimuthal wind forces on the disk are dominated in the mini-disks by their rapid motion with respect to the closest BH, and in the CBD with respect to the center of mass. 

For a local center denoted by $i$=\{BH1,BH2,b\}, for BH1, BH2, or the center of mass (COM) which transitions to the post-merger BH, respectively, the imposed source terms related to azimuthal wind-induced torque and radial magnetic field bending are decomposed as,
\begin{equation}
\proxy{\vct{F}_{i}}
=
\proxy{F_{r,i}}\hat r
+
\proxy{F_{\phi,i}}\hat\phi
.
\end{equation}
We then apply a single source function during evolution that weighs these total forces according to which source of gravity dominates orbits at that local position, 
\begin{equation}
\proxy{\vct{F}}
=
w_\mathrm{BH1}\vct{F}_{\mathrm{BH1}}
+
w_\mathrm{BH2}\vct{F}_{\mathrm{BH2}}
+
w_\mathrm{b}\vct{F}_{\mathrm{b}}
,
\end{equation}
where,
\begin{align}
w_{\mathrm{b}}
  &= W(r_{\mathrm{b}};M_\mathrm{b} r_{\mathrm{sink}},0.01) 
  \notag\\
  &\quad \times W(r_{\mathrm{b}};1.4a(t),0.1),
  \label{eq:w_com}\\[0.5ex]
w_{\mathrm{BH1}}
  &= W(r_{\mathrm{BH1}};M_\mathrm{BH1} r_{\mathrm{sink}},0.01)
  \notag\\
  &\quad \times
  \left[1-W(r_{\mathrm{BH1}};0.4a(t),0.1)\right],
  \label{eq:w_bh1}\\[0.5ex]
w_{\mathrm{BH2}}
  &= W(r_{\mathrm{BH2}};M_\mathrm{BH2} r_{\mathrm{sink}},0.01)
  \notag\\
  &\quad \times
  \left[1-W(r_{\mathrm{BH2}};0.4a(t),0.1)\right],
  \label{eq:w_bh2}
\end{align}
and $0.4a(t)$ is the approximate outer tidal truncation radius of each mini-disk. The first activation function ensures the radial and azimuthal forces are not applied to the magnetic field inside the sink, which we count as `on the horizon', and the second restricts the activation to be in the mini-disks or the CBD. We account for this source in both the momentum and energy equations of the gas, updating at second order in time.  

We do not account for mass loss associated with the wind. We believe this is a reasonable approximation since the magnitude of $\zeta$ is small and we only consider simulations with relatively small radial extents of the MAD portion of the disk, at most extending into the CBD by a few times the binary semi-major axis.

\subsection{B-proxy testing with single BH disks}
\label{subsec:testingproxy}
Given the admittedly \textit{ad hoc} nature of our Bproxy prescription, we extensively test its efficacy by comparing sBH MAD disk simulations performed with our 2D setup including the Bproxy to published 3D-GRMHD simulations of thin MAD disks. We also compare directly to a new 3D-MHD simulation run with \texttt{Athena++} of a MAD disk around a Newtonian point mass. Many studies of MAD disks using 3D-GRMHD exist, typically using the Kerr metric, but those around Newtonian or post-Newtonian point masses have received little attention, despite this inner boundary condition being used in many binary simulations of MAD circumbinary disks and binaries embedded in AGN disks. 
For our validation, we quantify the 2D MAD state using the following characteristics: qualitative reproduction of the MAD time-dependent flux eruption behavior, and radial cumulative vertical magnetic flux integral. Other characteristics considered but found to be less crucial to the calibration procedure were the variability of horizon-threading flux, and the radial wind torque profile compared to advective and viscous torques. 
\begin{figure}
\centering
\subfloat[]{%
  \includegraphics[width=0.48\textwidth]{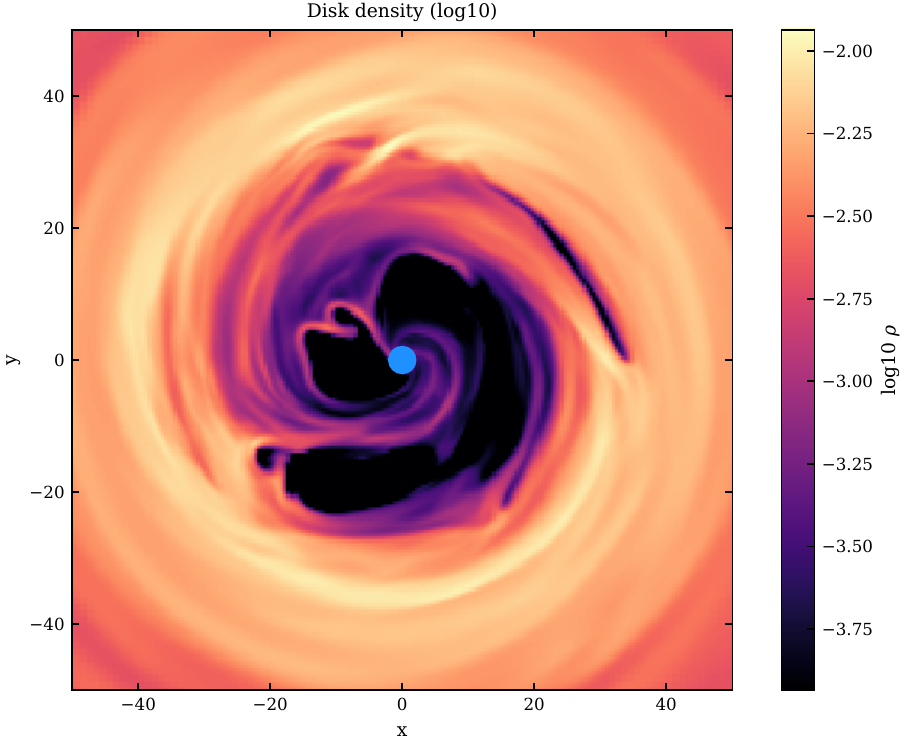}%
  \label{fig:nomerge}%
}
\hfill
\subfloat[]{%
  \includegraphics[width=0.48\textwidth]{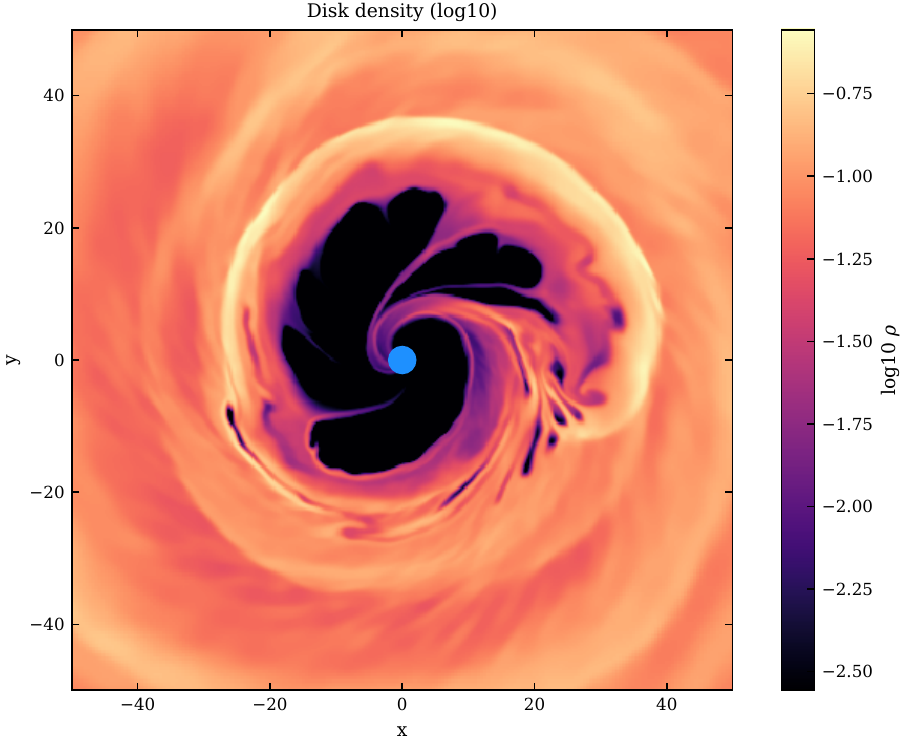}%
  \label{fig:merge}%
}
\caption{(a) Shell-integrated density snapshot of a 3D-MHD MAD disk simulation around a Newtonian point mass. (b) 2D-Bproxy simulation snapshot from a calibration simulation of comparable total magnetic flux. The maximum value of each color-bar is set to the maximum $\Sigma$ across each respective snapshot, and two orders of magnitude in range below each maximum is shown for fair comparison. The absolute scale of $\Sig$ in the simulations is arbitrary. }
\label{fig:magcompare}
\end{figure}

Figure~\ref{fig:magcompare} shows snapshots of $\Sig$ selected from a 2D-Bproxy test run and the 3D-MHD Newtonian run. These snapshots show similar stages in the evolution of a magnetic flux eruption from the sBH. To aid in comparison, we shell-integrate the 3D run and choose comparable density ranges as described in the figure caption. 
To find such similarly matching states between the two simulations took a search by-eye through less than 20 snapshots of the 2D MAD simulation to find a flux eruption at a similar stage as the one selected from the 3D-MHD run. Qualitatively, the flux eruption process results in very similar structural evolution of the disk.

In MAD disks, the magnetic flux is typically characterized using the $\phi$ parameter, a dimensionless measure of magnetization \citep{1999ApJ...522L..57G,2010MNRAS.408..752P,2009ApJ...699.1789T}. To arrive at this quantity, we first calculate the hemispheric integral of absolute flux as a function of spherical radius $r$,
\begin{equation}
\Phi(r)
=
\frac{1}{2}
\int_{0}^{2\pi}\int_{0}^{\pi}
\left|B^r(r,\theta,\varphi)\right|
\sqrt{-g}\,d\theta\,d\varphi.
\end{equation}
For our flat space-time, the term containing the metric determinant, $\sqrt{-g}=1$, and the $1/2$ accounts for the integral being over complete spherical shells -- the vertical flux threading the disk and horizon is approximately half this quantity. 
Then, 
\begin{equation}\label{eqn:phimad}
\phi(r)
\equiv
\frac{\Phi(r)}
{\sqrt{\left\langle\dot{M}(r)\right\rangle r_g^2c}},
\end{equation}
and $\phi_\mathrm{BH}$ is this quantity either measured at the horizon ($R=2$M for zero BH spin) or otherwise inside the plunging region, i.e. inside the ISCO where gas plunges towards the horizon\footnote{In studies where the horizon is covered by few enough cells that the shell integral has significant interpolation error, $\phi_\mathrm{BH}$ is sometimes measured inside the ISCO, but outside the horizon. Also note that there are several version of the unit-less horizon-threading magnetic flux, and $\phi$ is sometimes calculated with the time-averaged $\dot M$, using the signed flux integral, and other variations.}. $\left\langle\dot{M}(r)\right\rangle$ is the time-average of the accretion rate at radius $r$. Though $r_g=c=1$ in our units, we include them here to make explicit the dependence of $\phi$ on the mass of each BH at each stage of evolution in the binary. 

Once a disk advects enough magnetic flux onto the horizon that the ram pressure of matter in the plunging region inside the ISCO cannot trap more, i.e. it reaches force balance with the magnetic pressure, the remaining flux starts to pile up into the disk. This creates an approximately self-similar radial profile out to the MAD transition radius, beyond which the disk remains SANE. 
The limiting value of flux on the horizon is the empirically determined MAD parameter $\phimad$, which is average $\phi_\mathrm{BH}$ measured in the MAD state. This value depends on the disk thermodynamics \citep{AvaraMcKinney2016}, but varies between about 5 to 20 in the units we've chosen\footnote{Gaussian units are also commonly chosen. To convert from our geometrized natural units to Gaussian units, a multiplication of $4\pi$ is necessary. Gaussian units are those chosen in \cite{2009ApJ...699.1789T}.}.

By virtue of the divergence-free constraint of Maxwell's equations, the signed equivalent of $\Phi(r)$,
\begin{equation}
\Phi_s(r)
=
\int_{0}^{2\pi}\int_{0}^{\pi/2}
B^r(r,\theta,\varphi)
\sqrt{-g}\,d\theta\,d\varphi.
\end{equation}
is equivalent to the surface integral of $B_z$ over the entire equatorial plane out to radius r. This provides effectively the same diagnostic for normalized vertical flux in the disk and on the BH (so long as high-order multipole moments don't dominate the structure of the large-scale field, \citep{2012MNRAS.423.3083M}) and from now on will be the form used in our discussion of $\phi$. 

\begin{figure}[t!]
  \centering
  \includegraphics[width=.5\textwidth]{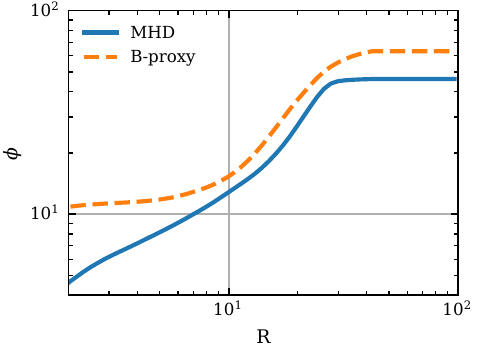}
  \caption{$\phi(R)=\Phi_s(R)/\langle\dot M\rangle$, the dimensionless cumulative integral of magnetic flux inside radius R. The blue solid line represents our 3D-MHD simulation with a softened Newtonian potential. The orange dashed line represents the 2D simulation with approximately the same radial MAD disk extent.}
  \label{fig:radialmad}
\end{figure}

Figure~\ref{fig:radialmad} demonstrates the consistency in $\phi(R)=\Phi_s(R)/\langle\dot M\rangle$ for the MAD portion of the disks in the 3D-MHD and 2D-Bproxy runs featured in Fig.~\ref{fig:magcompare}, where the characteristic MAD radial profile of $\phi(R)\propto R$ is approximately reproduced. This agreement implies the Bproxy prescription captures the fundamental balance of forces that dominate the MAD disk state.

In non-relativistic simulations where the point masses accumulating flux are not represented by the Kerr metric, the $\phimad$ parameter will not be self-consistently determined by the balance of magnetic fields subject to relativistic gravitational forces and gas plunging orbits. Therefore, we have some freedom in choosing what to do with magnetic pressure inside the sink of our pseudo-Newtonian potential (PNP). This enables us to calibrate the Bproxy prescription to match the $\phimad$ value in our 2D simulations to the 3D-GRMHD benchmarks. During testing, we found a linear multiplier for magnetic pressure inside $r_\mathrm{sink}$ which enabled us to reproduce a choice for the $\phimad$ empirical limit. 
Because of the physics of the MAD state, the balance of pressure seems to be more important than the absolute total flux collected in the reservoir on the horizon. Also, as we will see in Section \S\ref{sec:hydro_bproxy}, the relative values $\phi/\phimad$ are key in describing decoupling behavior and so should be somewhat independent of this systematic offset of $\phimad$.  

For the two runs in Fig.~\ref{fig:magcompare}, the total magnetic flux across the disk in the 2D run was chosen to be roughly the same as the 3D-MHD run. However, with different flux reservoir characteristics on the horizon, the same total vertical flux will not necessarily result in the same radial range of the MAD portion of the disk. Since $\phi(r)$ represents the cumulative flux to radius $r$, the extent of the MAD disk is identified by where the radial profile flattens. This is $r\sim30$M and $r\sim40$M, for the 3D-MHD and 2D-Bproxy runs respectively, as seen in Fig.~\ref{fig:radialmad}. 

We save the full accounting of the variability characteristics of 3D-MHD and 2D-Bproxy MAD disks for future work, but note that the magnetic flux eruptions in the 2D simulations result in order unity changes in total flux contained within the ISCO radius. 

\addcontentsline{toc}{subsubsection}{\note{Figure~\ref{fig:radialmad}: radial B-field comparison, 3D-MHD mad vs 2D mad.}}

\addcontentsline{toc}{subsubsection}{\note{Figure~\ref{fig:magcompare}: side by side mhd and bproxy mad disks, snapshots+spectralaz}}

\subsection{Stochastic viscosity ($\nu$)}
\label{subsec:stoch}

Rather than constant $\nu$, we choose a constant Shakura-Sunyaev (SS) \citep{ShakuraSunyaev1973} viscosity parameter $\alpha$. These are related by
\begin{equation}
 \nu
 =
 \alpha\left(\frac{H}{r}\right)^2
 r^{1/2}.
 \end{equation}
We stochastically vary $\alpha$ according to the following prescription. 

The viscous angular momentum transport in real BH accretion disks is expected, if they are in the SANE regime, to be the result of turbulence driven by the MRI. 
This magnetohydrodynamic instability maintains turbulent magnetic field strength in the disk such that the plasma $\beta\equiv 2P_\mathrm{gas}/B^2$ parameter, the ratio of gas to magnetic pressures, is on the order of 10s to 100 \citep{2012ApJ...749..189S}. There is, especially for disks with scale height $H/r\lesssim 0.2$, a cyclical dynamo behavior that creates fields organized on the disk scale height that buoyantly rises to the surfaces of the disk (commonly plotted in `butterfly' diagrams). This dynamo has been shown \citep{HoggReynolds2016} to create correlated Maxwell stress on the same scale. 

It is then natural to question how this correlated stress may translate into the EM flux variability seen in accreting BH systems. The propagating fluctuations model \citep{Lyubarskii1997,UttleyMcHardy2001}, where radially independent fluctuations around a mean viscosity create correlated and compounding fluctuations in surface density and thus accretion rate, matches well to the idea of stochasticity in viscosity being tied to these local correlations in Maxwell stress driven by the MRI. 

Motivated by the behavior of the MRI in geometrically thin disks, and prior 1D models of the propagating fluctuations in time-dependent models of stochastic accretion disks, \cite{TurnerReynolds2023} expanded these models to 2D for a SS type disk around an accreting Newtonian point mass. They used the scale of correlated Maxwell stress fluctuations seen in 3D-MHD simulations to set a scale for stochastic perturbations of the viscosity and were able to reproduce the characteristic red-noise and log-normal bolometric flux distribution observed in many astrophysical black hole systems. 

Here we introduce the first implementation of a stochastic variability prescription in 2D simulations of binary accretion. A full presentation of this powerful tool is in preparation in a companion paper (Avara et al. in prep), but we use it here to incorporate a realistic variability into our binary merger simulations. This allows us to assess how accretion stochasticity might lead to differences in 3D-MHD and 2D viscous hydrodynamic models. 

As in \cite{TurnerReynolds2023}, stochasticity in the viscosity is introduced by multiplying the normal constant SS viscosity $\alpha$ by
\begin{equation}
\alpha = \alpha_0 e^{\beta_\nu},
\end{equation}
where $\beta(R,\phi,t)$ is a scalar field advected with the gas and driven locally under an Ornstein–Uhlenbeck (OU) process with spatial correlation. The evolution equation is,
\begin{equation}\label{eqn:betanu}
{\rm d}\beta_\nu
=
-\omega_0\left(\beta_\nu-\mu_\beta\right){\rm d}t
+
\xi\,{\rm d}W,
\end{equation}
where $\omega_0$ is the fundamental OU frequency, $\mu_\beta$ is the mean to which $\beta_\nu$ is damped at that rate, and stochastic driving occurs with the last term where $\xi$ is a driving amplitude and $\rm dW$ is the discrete derivative of a Wiener process, $N(0,dt)$.

Our binary implementation differs from \cite{TurnerReynolds2023} in that the radial dependence of $\omega_0$ is set in relation to the orbital frequency either around the binary COM or at a given radius from each BH, depending on which dominates the local potential. For gas in the CBD, outside a transition radius defined by $r_{\mathrm{CBD\nu}}(t)\equiv\max\!\left[ 0.8a(t), 2M r_{\mathrm{sink}}\right]$, stochasticity parameters are determined by the radius measured from the COM. For the minidisks around each BH$_i$, a range spanning $r_\mathrm{Roche\nu,i}(t)\equiv(2M_i r_{\mathrm{sink}},0.4a(t))$ is used, such that
\begin{align}
\omega_0(r,r_i,t)
  &=
  \alpha_0
  \begin{cases}
  \left(r_i+M_i r_{\mathrm{sink}}\right)^{-3/2},
  &r_i \in r_\mathrm{Roche\nu,i}(t)
  \\[1ex]
  \left(r+10\right)^{-3/2},
  &
  r>r_{\mathrm{CBD\nu}}(t).
  \end{cases}
  \label{eq:omega_stoch}
\end{align}

Also, to implement the spatial correlations, the driving term $\rm dW$ is a truncated discrete Fourier transform defined in $r-\phi$ coordinates centered in each of these three radial domains.  

Following \cite{TurnerReynolds2023}, we also utilize the convenient relation between the driving magnitude, 
\begin{equation}
\xi
=
\left(
2\omega_0
\left\langle\beta_\nu^2\right\rangle
\right)^{1/2},
\end{equation}
and the variance of the process, $<\beta_\nu^2>$, so that $\xi$ inherits the same radial dependence as $\omega_0$, and so that $\beta_\nu$ has unity variance.

Figure~\ref{fig:fiducialsimCoolBetanu} shows the cooling rate, or emissivity of the disk, $\mathcal{L_\mathrm{cool}}$, and $\beta_\nu$ for a typical snapshot of our fiducial run, \PROD{6}, in the equilibrated state. This snapshot is the same used to plot surface density in Fig.~\ref{fig:fiducialsimSig}. Fluctuations in the density mimicking turbulence are evident, and modulate the accretion rate into the cavity. However, because of the symmetric fluctuations in $\beta_\nu$ the mean viscosity $\alpha$ remains unchanged and we find the time averaged accretion rate matches 
\begin{equation}
\left\langle \dot M\right\rangle = 3\pi\left\langle \alpha \right\rangle \Sig H^2 \Omega_\mathrm{K}
\end{equation}
whether stochastic fluctuations are incorporated or not.

\subsection{The gravitational potential ($\Phi_\mathrm{bin}$)}
\label{subsec:potential}

Prior to merger, two BHs of equal mass, $M_1=M_2=M_{\rm b}/2$, contribute gravitational forces on the gas. After merger the gas experiences a single, stationary gravitational force from the merger remnant of mass $M_m=0.97(M_1+M_2)$, which accounts for the loss of total mass-energy to GWs. We find that not accounting for mass loss at merger only results in a difference at the $\sim10\%$ level in accretion rate and luminosity.

The trajectories of the BHs are that of a circular orbit before inspiral is activated, $t_\mathrm{insp}$ provided in Table~\ref{tab:sims}, after which the orbital phase and separation evolve according to the approximate post-Newtonian treatment of \cite{Peters1964}. Once the BHs are within $0.1M$ of each other, they are combined into a single central point mass. 

Because the source term is implemented as a force, we introduce the gravitational potential in force form. For the selection of our runs designed to be compatible with prior 2D work, each BHs gravitational field is a softened Newtonian, or Plummer potential \citep{Plummer1911}, with radial force,
\begin{equation}
    F^r(r) = \Sig\frac{GMr}{(r^2+s^2)^{3/2}}.
\end{equation}
$s$ is a softening radius set to $s=0.1\rm M$, for local BH mass $M$, unless otherwise stated. 

Any choice of gravitational potential that only approximates the full relativistic space-time must make some sacrifice regarding which relativistic affects to capture and which to ignore. Newtonian potentials offer simplicity but cannot capture relativistic precession, the transition from stable orbits to plunge at the ISCO, or the event horizon as a one-way causal boundary.

The Paczy\'{n}ski-Wiita potential \citep{PaczynskyWiita1980} (PW), in its very simple form almost magically reproduces the exact Schwarzschild ISCO, marginally bound orbits, and circular-orbit specific angular momentum \citep{Abramowicz2009}. However, true precession rates are not accounted for, and precession and the ISCO location have no dependence on BH spin. Because the evolution of the disk in merging binary systems has significant eccentricity, and the spin of the BH is practically guaranteed to be fairly high in the case of a near-equal mass merger \citep[see][and references therein]{FishbachHolzFarr2017}, spin-dependent apsidal precession rates could be important.

We derive a new type of pseudo-Newtonian potential referred to as the PNP in this work. Rather than taking the approach of expansions like that of Post-Newtonian or Post-Einsteinian formulations, we focus on trying to match as well as possible to the precession rates for equatorial orbits in the Kerr metric and thereby include realistic spin dependence.

We start by postulating a hybrid potential resulting in radial force of the form, 
\begin{equation}
    F^r(r) = \frac{GM}{r^2}g(u)
\end{equation}
where $u\equiv 1/r$. We solve for spin-dependent precession and match the form for differential apsidal precession as a function of g(u) to that in the Kerr metric to find an exact solution in the limit of small eccentricity. 

The full derivation can be found in Appendix~\ref{app:gravpot}, but the equation defining the gravitational force per unit surface area around a point mass with normalized spin magnitude $\chi=J/M^2$ is found to be,

\begin{multline}
F^r(r)
=
\Sig\frac{GM}{r^2}
\exp\!\Bigg[
\frac{6GM}{r}
-\frac{16}{3}\chi\left(\frac{GM}{r}\right)^{3/2} \\
+\frac{3}{2}\chi^2\left(\frac{GM}{r}\right)^2\Bigg].
\end{multline}

For numerical stability, and without sacrificing orbital stability or precession frequency anywhere outside the sink radius, this force function is further modified with smooth transition functions near and inside the sink. These transitions are also described in the appendix. 

This potential exactly matches Kerr for small eccentricity, but its function here is to preserve the precessional behavior of moderately eccentric geodesics. Testing shows that our function performs at least as well as some other recent post-Newtonian approaches \citep{AradSari2026} for spin $\chi=0$, and accounts for spin dependence, which that work does not directly explore. Additionally, this potential reproduces the spin-dependent ISCO of the Kerr metric, with the correct spin-dependent radius, a feature important for gas evolution during merger. 

A test comparing these approaches for moderate to highly eccentricity precession is included in Appendix~\ref{app:gravpot}.

\subsection{Cooling ($\mathcal{L}_\mathrm{cool}$)}
\label{subsec:cooling}
We apply a cooling prescription via the source term $\mathcal{L}_\mathrm{cool}$ that balances the viscous heating and other dissipation in the disk to keep it near the target scale height (interchangeable with a disk Mach number, $\mathcal{M}\equiv v_\phi/c_s$). Specifically, we adapt the target-entropy cooling function introduced in \cite{2009ApJ...692..411N}, extended to binaries in 3D with an excised cavity by \cite{NobleMundim2012}, and further extended to cool the mini-disks in the cavity by \cite{BowenCampanelli2017}. This work, and those inheriting the same approach \citep[e.g.][]{BowenMewes2018,LopezArmengolCombi2021,AvaraKrolik2024, EnnoggiCampanelli2025}, start with an isentropic torus of gas and then cool at all radii to that constant target entropy. To our knowledge, this is the first study performed using target-entropy cooling in the 2D regime. 

Our motivation to apply target-entropy cooling rather than black-body cooling (as typically done in 2D; see \citealp{FarrisDuffell2015a,TangMacFadyen2017}, and \citealp{WesternacherSchneiderZrake2022}), another common choice, is two-fold. First, we can directly compare target-entropy-cooled 2D-hydro simulations to the 3D-MHD simulations in those studies mentioned just above. Second, we can enforce target-entropy cooling to act on a timescale approximately equal to the photon diffusion timescale. While black-body cooling has the advantage of approximating, in a very simple way, radiation transport from an optically thick disk, in the highly dynamic region around a binary, the usual prescription used in 2D simulations can cool shock and compression-heated gas in the disk much faster than the photon diffusion timescale would allow. We discuss this lack of self-consistency at length in Section~\ref{sec:discussion}.   

In the case of a 3D constant scale height disk with constant $\alpha$ and a non-relativistic $\gamma=5/3$ polytropic index, the entropy has no radial dependence. However, solving the vertically integrated equations for 2D evolution, and evolving $P=\Sig c_s^2$, for vertically isothermal sound speed $c_s$, the equilibrium entropy becomes,
\begin{equation}
K_{\rm targ}=\frac{P}{\Sig^{5/3}}=K_0 r^{-2/3}.
\end{equation}
We add a specific emissivity sink term to the right hand side of Eqn.~\ref{eqn:energy}, 
\begin{equation}
\mathcal{L}_{\rm cool}
\equiv
-\frac{U_\mathrm{g}}{t_{\rm cool}}
\left(
\frac{K-K_{\rm targ}}{K_{\rm targ}}
\right)_{+}^{q},
\label{eqn:cool}
\end{equation}
where the $+$ subscript denotes that only positive or null values of the entropy difference are used. We adopt the $q=1/2$ of \cite{NobleMundim2012} and use the local Keplerian orbital timescale for $t_\mathrm{cool}$. 

Most simulations of disks using this cooling function around single point masses have reported equilibrated scale heights consistent with a higher entropy than the target. The excess entropy depends on $t_\mathrm{cool}$, but only in the limit of very rapid cooling will an equilibrated entropy at the target be expected. Thus, we expect the disk to have slightly higher entropy in equilibrium than the target. Given that the viscous heating timescale is well defined, one can solve for the thermal equilibrium expected given the cooling function. If $K_0=0.01$ was chosen as in \cite{NobleMundim2012}, a disk of Mach number $\mathcal{M}\sim6$ would be expected. We therefore choose $K_0=0.003$, resulting in an expected $\mathcal{M}\sim10$, and measure values within 10-20\% of this target in regions of the binary simulations where gas has not been shocked or compressed due to interaction with the binary, and at all radii outside the ISCO in the case of sBH calibration runs. 

In a Newtonian ideal gas, the Bernoulli parameter,
\begin{equation}\label{eqn:Be}
\mathcal{B}e
=
\frac{1}{2}|\mathbf v|^2
+
\frac{\gamma}{\gamma-1}\frac{P_g}{\Sig}
+
\frac{\underline{B}^2}{\Sig}
+
\Phi_\mathrm{b},
\qquad
\begin{cases}
\mathcal{B}e<0, & \text{bound},\\
\mathcal{B}e>0, & \text{unbound}
\end{cases}
\end{equation}
can be used to identify which gas is formally gravitationally bound to the instantaneous potential $\Phi_\mathrm{b}$. 
In 3D-GRMHD simulations only activating the cooling source term for bound gas is often used to avoid cooling the atmosphere above and below the disk, which would then rain onto the inner system. This restriction is further motivated by the fact that significantly shock-heated or magnetically dominated regions of the disk would not be expected to cool at the same rate or through the same processes. 
In this study, by default, we also apply this restriction to provide consistency between our magnetized runs and those in 3D-MHD. We report a test of cooling all versus only bound gas in Section~\ref{sec:hydro_results}.

For numerical stability we never remove more than 45\% of a cell's thermal content in a single timestep. Additionally, we do not cool with a simple explicit form given this rate and the timestep, but solved analytically for a cell's reduction in internal energy assuming a continuously varying cooling rate, responding to the evolving entropy, according to Eqn.~\ref{eqn:cool}, over the duration of the full timestep. We explicitly remove that amount from the total energy.

\subsection{Initial conditions}
\label{subsec:physicalsetup}

We initialize a constant scale height $H/r=0.1$ disk, representing a moderately sub-Eddington accretion rate, around binaries of circular orbits (quasi-circular during inspiral), and with semi-major axis between 20M and 50M, depending on the model. 

For all simulations a constant SS-type $\alpha=0.1$ viscosity is chosen, and this remains the mean viscosity for those with stochastic variation of $\alpha$. The surface density profile, consistent with constant scale height and this viscosity is,
\begin{equation}
\Sig=f(r)\Sig_0 r^{-1/2},
\end{equation}
where $\Sig_0$ is simply chosen to be unity, and $f(r)$ is a shaping function used to truncate the disk before reaching our outer boundary condition, and to initially clear a cavity region approximately consistent with the equilibrium binary accretion state. This initial cavity prevents long-lived transients in the disk evolution. Specifically, we adopt
\begin{equation}
f(r)=W(r;r_\mathrm{in},\Delta_\mathrm{in})(1-W(r;r_\mathrm{out},\Delta_\mathrm{out}),
\end{equation}
where $r_\mathrm{in}$ and $r_\mathrm{out}$ are the initial inner and outer edges of the finite disk. $r_\mathrm{out}=500$M in all simulations, and $r_\mathrm{in}=(36, 72, 90)$M for simulations starting with $a_0=(20,40,50)$M, respectively. 

We use directed damping to implement an outer boundary condition, or buffer region, that respects the rotational symmetry of the outer disk. We damp towards $10^{-7}$ times the initial tapered disk primitive quantities in this region, defined by $r_\mathrm{B}=900$M using our shaping function, $1-W(r;r_\mathrm{B},\Delta_\mathrm{B}=0.3)$. This means we evolve the disk over almost all our active domain, which extends to $\pm1000$M in x and y. 

The pressure in the disk is initialized so that the scale height is $H/r=0.1$ assuming vertical hydrostatic equilibrium,
\begin{equation}
P_i=\Sig \left( \frac{H}{r} \right)^2\frac{GM}{r},
\end{equation}
using the Keplerian orbital frequency.

Mimicking the MRI seed perturbations used in 3D-MHD/3D-GRMHD, we perturb the pressure with a Gaussian perturbation at the 10\% level. Though not strictly necessary in 2D, we employ this perturbation for comparative consistency and to encourage the local dissipation of transient shocks and sound waves during the initial equilibration stage. 

Because the disk is moderately cool, we do not find the need to self-consistently balance radial pressure gradients in the initial disk profile with centrifugal support in order to achieve equilibrated evolution in a reasonable number of orbits. We simply start the disk with purely azimuthal velocity with a Keplerian profile about the center of total mass. 

We specify floor values of $\Sig_\mathrm{floor}=10^{-10}$ and $P_\mathrm{floor}=10^{-20}$. When a cell density or pressure goes below these values the quantity is reset to the floor for numerical stability. At $t=0$, once the density is initialized, we reset any value that is less than 100$\Sig_\mathrm{floor}$ to this `atmosphere' value before using density to set the pressure using the target disk scale height $H/r=0.1$. This buffer above the floor prevents the gas from reaching the floor as frequently and improves early numerical stability. Similarly, we do not cool gas when it is below $10\Sig_\mathrm{floor}$. The density and pressure floors are rarely needed. 

In cases where the magnetic proxy is used, the initial proxy magnetic field is introduced as a constant in the inner region, truncated near the binary inner edge so that,
\begin{equation}\label{eqn:Binit}
B_i=B_0[1-W(r;r_\mathrm{B,i},\Delta_\mathrm{B,i})],
\end{equation}
where $B_0$ is provided in Table~\ref{tab:sims}, $r_\mathrm{B,i}=25$M, and $\Delta_\mathrm{B,i}=0.1$ for all magnetized simulations.

We assume a small disk mass and do not account for back-reaction of the gas on the binary evolution. For a moderately sub-Eddington accretion rate the inspiral time for the separations we consider is negligibly small compared to the mass doubling time from accretion.

\subsection{Suite of runs}
\label{subsec:runs}

\begin{table*}
\centering
\caption{}
\label{tab:sims}

\setlength{\tabcolsep}{2.5pt}
\renewcommand{\arraystretch}{1.15}

\begin{tabularx}{\textwidth}{
  @{}
  l
  *{9}{>{\centering\arraybackslash}X}
  @{}
}
\toprule
\shortstack{Simulation\\name} &
\shortstack{Initial\\separation\\$a_0$\\$[\mathrm{M}]$} &
\shortstack{Grav.\\potential\\type} &
\shortstack{Inspiral and\\final times\\
$t_{\mathrm{insp}},t_f$\\$[\mathrm{T}_{\mathrm{orb},0}]$} &
\shortstack{Cooling uses\\target or\\local $P$} &
\shortstack{Stochastic\\initial time\\$[10^3\mathrm{M}]$} &
\shortstack{Cooling\\rate} &
\shortstack{Gas-cooling\\selection} &
\shortstack{Bproxy\\amplitude\\
$\phi_{\mathrm{tot}}/\phimad$} &
\shortstack{Resolution\\ cell size\\$[\mathrm{M}]$\\
at $t_{\mathrm{merger}}$}
\\
\midrule

\multicolumn{10}{@{}l}{
  \textit{Our fiducial simulation, with physically motivated choices,
  used as a reference}
}\\[1pt]

\PROD{6}
& 20 & PNP & 89, 156 & target & \textbf{1} & $\Omega_b/2\pi$ & all & --- & 0.24
\\

\addlinespace[3pt]
\multicolumn{10}{@{}l}{
  \textit{Purely hydrodynamic runs}
}\\[1pt]

\rowcolor{black!4}
\PROD{0}
& 20 & PNP & 89, 159 & local & --- & $\Omega_b/2\pi$ & all & --- & 0.24
\\

\PROD{1}
& 20 & \textbf{Newt} & 89, 150 & local & --- & $\Omega_b/2\pi$ & all & --- & 0.24
\\

\rowcolor{black!4}
\PROD{2}
& 20 & PNP & 89, 167 & local & --- & $\Omega_b/2\pi$ & \textbf{bound} & --- & 0.24
\\

\PROD{4}
& 20 & Newt & 89, 129 & local & --- & \textbf{instant} & all & --- & 0.24
\\

\rowcolor{black!4}
\PROD{5}
& 20 & Newt & 89, 120 & \textbf{target} & --- & \textbf{instant} & all & --- & 0.24
\\

\PROD{3}
& \textbf{50} & PNP & 23, 1565 & local & --- & $\Omega_b/2\pi$ & all & --- & 0.49
\\

\rowcolor{black!4}
\PROD{7}
& 50 & PNP & 23, 318 & target & \textbf{1} & $\Omega_b/2\pi$ & all & --- & 0.49
\\

\PROD{7.longeq}
& 50 & PNP & \textbf{None, 1400} & target & 1 & $\Omega_b/2\pi$ & all & --- & 0.98
\\

\rowcolor{black!4}
\PROD{7.nwt}
& 50 & \textbf{Newt} & \textbf{None}, 1235 & target & 1 & $\Omega_b/2\pi$ & all & --- & 0.98
\\

\addlinespace[3pt]
\multicolumn{10}{@{}l}{
  \textit{Magnetized runs with Bproxy}
}\\[1pt]

\PROD{16}
& 20 & PNP & 89, 118 & target & 1 & $\Omega_b/2\pi$ & bound & \textbf{0.3} & 0.24
\\

\rowcolor{black!4}
\PROD{17}
& 20 & PNP & 89, 119 & target & 1 & $\Omega_b/2\pi$ & bound & \textbf{1.5} & 0.24
\\

\PROD{20}
& 20 & PNP & 89, 139 & target & 1 & $\Omega_b/2\pi$ & bound & \textbf{3.2} & 0.24
\\

\rowcolor{black!4}
\PROD{19}
& 20 & PNP & 89, 116 & target & 1 & $\Omega_b/2\pi$ & bound & \textbf{4.8} & 0.24
\\

\PROD{19hr}
& 20 & PNP & 89, 127 & target & 1 & $\Omega_b/2\pi$ & bound & \textbf{4.8} & 0.12
\\

\rowcolor{black!4}
\PROD{18}
& 20 & PNP & 107, 132 & target & 1 & $\Omega_b/2\pi$ & bound & \textbf{16} & 0.24
\\

\PROD{23.1}
& 20 & PNP & 240, 271 & target & 1$^{a}$ & $\Omega_b/2\pi$ & bound & \textbf{1.1} & 0.06
\\

\rowcolor{black!4}
\PROD{23.2}
& 20 & PNP & 240, 268 & target & 1$^{a}$ & $\Omega_b/2\pi$ & bound & \textbf{1.1}$^{b}$ & 0.06
\\

\addlinespace[3pt]
\multicolumn{10}{@{}l}{
  \textit{Wider separation runs}
}\\[1pt]

\PROD{25}
& \textbf{40} & PNP & 101, 246 & yes & 1 & $\Omega_b/2\pi$ & bound & \textbf{0.3} & 0.24
\\

\rowcolor{black!4}
\PROD{26}
& \textbf{40} & PNP & 101, 247 & yes & 1 & $\Omega_b/2\pi$ & bound & \textbf{3.2} & 0.12
\\

\bottomrule
\end{tabularx}

\vspace{3pt}

\begin{minipage}{\textwidth}
\raggedright

\textbf{Simulation naming.}
Simulations are named according to their gravitational potential
(``PNP'' or softened Newtonian/Plummer, ``NWT''), initial binary
separation ($20\,\mathrm{M}$, $40\,\mathrm{M}$, or $50\,\mathrm{M}$),
and whether the viscosity prescription was purely laminar (``Lam'')
or stochastically varied (``Stoch''). Additional descriptors are
``BoundCool'' for simulations that cool only bound gas according to
the Bernoulli criterion, ``CoolInst'' for simulations in which the
entropy-cooling timescale is instantaneous, and ``Tgt'' for simulations
in which cooling is prescribed according to the target-pressure
approach rather than the local pressure; see
Section~\ref{subsec:merging_hydro} for details. Unless otherwise
specified, simulations of the ``Stoch'' type use target-pressure
cooling, while those of the ``Lam'' type use local-pressure cooling.
``Long'' indicates a very long evolution at fixed separation.
``Bx-y'' indicates a magnetized run with total initial flux set using
the constant $B_0=x\times10^y$, although ``LE25r1/2'' are two special
runs whose magnetization was adjusted to match the simulation in LE25.
Finally, ``HR'' indicates a duplicated run with higher resolution.

\par\smallskip
\textbf{Columns.}
The columns give the run name; initial binary separation; gravitational
potential; time at which inspiral driven by gravitational-wave losses
begins (`None' indicates fixed separation) and the total duration of the simulation, both in units of the
initial orbital period $\mathrm{T}_{\mathrm{orb},0}$; pressure choice
for the cooling prescription; time at which the stochastic variation
in viscosity begins; cooling rate; whether the cooling function is
applied to unbound gas; total initial vertical magnetic flux normalized
by the horizon MAD threshold, with unity denoting the minimum total
flux required for a MAD single-BH disk in equilibrium; and the
resolution represented by the finest cell dimensions covering the
binary. Boldface identifies the parameter or parameters that each run
was designed to explore.

\par\smallskip
$^{a}$ A shifted mean,
$\langle\beta_\nu\rangle=e^{-3/2}$, was used to suppress viscosity at
the inner edge of the CBD, mimicking the suppressed Maxwell stress
measured there in 3D-MHD simulations
\citep{NobleKrolik2021}.

\par\smallskip
$^{b}$ To test the effects of the Bproxy calibration constants,
\PROD{23.2} uses a smaller sink radius,
$r_{\mathrm{sink}}=2.5\,\mathrm{M}_i$, and the magnetic pressure inside
$r_i=4\,\mathrm{M}_i$ is multiplied by approximately 16 relative to
\PROD{23.1}. Consequently, four times more total flux is required in
the cells inside the sink to produce the same pressure balance.

\end{minipage}
\end{table*}

The suite of primary simulations and key model parameters presented in this work are summarized in Table~\ref{tab:sims}, along with a description of our naming convention. The parameter choice that each simulation is designed to address is in bold. For example, our fiducial unmagnetized simulation, \PROD{6}, for which a snapshot of the evolved surface density is plotted in Figure~\ref{fig:fiducialsimSig}, is run with the Post-Newtonian gravitational potential described in \S\ref{subsec:potential}, starts with $a_0=20$M binary separation, and is run with the stochastic viscosity prescription described in \S\ref{subsec:stoch}. Aside from this fiducial run, the other simulations are grouped by whether they are purely hydrodynamic, magnetized with an initial separation $a_0=20\rm M$, or magnetized with a $a_0=40\rm M$. 

In the next three sections, we describe the simulations of each of these groups in turn. 

\section{Equilibrium, decoupling, and merger in unmagnetized disks}
\label{sec:hydro_results}

In the preceding section, we have developed a toolkit with which we can explore the influence of specific numerical and physical parameters. In this section we establish our fiducial hydrodynamic choices, replicate established equal-mass circular binary behavior in purely hydrodynamic simulations, and describe our findings from a suite of simulations varying the numerical prescriptions. We study inspiral, decoupling, and post-merger evolution, and explore the influence of viscous stochasticity, the effect of different thermodynamic prescriptions, the relativistic precession of gas orbits, and differing viscosity prescriptions. 

To prevent the parameter space becoming too large, we only include simulations performed with target-entropy cooling. Since this is the first study to use target-entropy cooling in 2D, we begin our presentation of results with a setup that closely mimics the choices in prior studies, and reproduces established results. 

\subsection{Equilibrium binary accretion}
\label{subsec:fixed_sep}

Nearly all published 2D viscous hydrodynamic simulations of binary accretion are designed to mimic wide binaries. To avoid having to simultaneously resolve the widely disparate scales of the binary semi-major axis and the ISCOs of the BHs, torque-free sink prescriptions are often used as an inner mini-disk boundary condition \citep{DempseyMunoz2020,DittmannRyan2021}. These stand in for the rest of the mini-disk that would extend from this sink radius down to the ISCO. Additionally, in non-relativistic studies, a simple Newtonian or Plummer potential is usually chosen since the dynamics governing such wide binaries is expected to be approximately Newtonian. Finally, the most common thermodynamic prescriptions are isothermal and, more recently, black-body cooling. These choices have been inherited by 2D studies of the inspiral and merger regimes \citep[LK23,][]{ClyburnZrake2025, DittmannRyan2023}. 

We make several different physically motivated methodological choices. First, unlike isothermal accretion, or evolution with black-body cooling, target-entropy cooling allows shocks and rapid compression/rarefaction to occur approximately adiabatically, which we will show has significant advantages. Second, use of a gravitational potential that accounts for gas orbital precession and has a spin-dependent ISCO is needed to capture important behavior of the 3D-GRMHD models. Third, our stochastic viscosity prescription is motivated by observed BH accretion variability and 3D-MHD behavior of the MRI. Each of these is included in our fiducial hydrodynamic setup, which we use to further explore the role of magnetization. Therefore, we will now show that these prescription choices do not significantly alter the nearly universally established characteristics of equal mass quasi-circular binary accretion. 

The spin-dependent precession terms in PNP breaks the scale-free nature of the binary separation of a constant $H/r$ disk. However, if we evolve with wide separation, the gas dynamics are approximately Newtonian, and the behavior of the Newtonian-potential simulations should be recovered. 

Common choices for scale-free binary simulations involve a sink with radial size $0.1a_0$ and a torque-free parameter near unity \citep{DempseyMunoz2020}, corresponding to a minimal removal of angular momentum in the frame of each BH. Simulations \PROD{7.longeq} and \PROD{7.nwt} approximate these choices with a sink of that size and a torque-free parameter of 0.9, with an initial binary separation of $a_0=50$M. Since we normally have a sink radius of $r_\mathrm{sink}=4M_i$, this is just over twice that for each BH at $10M_i$. This places the sinks outside the ISCO, replacing it as the physical inner boundary condition. 

For both simulations, the target-entropy cooling prescription is employed to relax towards a target Mach$\sim$10 disk state, corresponding to a scale height $H/r\sim0.1$. 

\PROD{7.nwt}, identical to \PROD{7.longeq} except for the gravitational potential, provides a way for us to confirm that, in the wide-separation limit, PNP reproduces the expected Newtonian behavior. We find similar quantitative evolution in both runs and focus on presenting only \PROD{7.longeq} for clarity. This run also provides an effective test of our full code base by reproducing the established equilibrium state for these disk and binary parameters, exhibiting each of the features described in the Introduction. 

The eccentricity of the cavity, the over-dense orbiting lump, and persistent mini-disks are all evident in the snapshot of $\Sig$ from \PROD{7.longeq} shown in Figure~\ref{fig:prod7.snapshot1}. 

\begin{figure}
  \centering
  \includegraphics[width=\columnwidth]{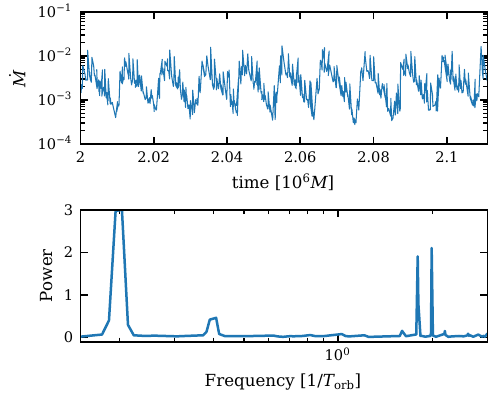}
  \caption{Top panel: Combined accretion rate from the two BH sinks during the equilibrium state of simulation \PROD{7.longeq}. Bottom panel: Power spectrum of the variability of \dotm. \PROD{7.longeq} is evolved for 1400 binary orbits at a fixed separation of $a_0=50$M. Primary modulatory features are the orbit of the lump driving sawtooth-shaped accretion oscillations at $\sim0.2/T_\mathrm{orb}$ (for orbital period $T_\mathrm{orb}=2\pi/\Omega_b$), a harmonic at twice that, the interaction of the binary with the periapsis of gas orbits in the CBD at $\sim 2/T_\mathrm{orb}$, and the modulation of accretion rate through close-passage of the lump and the BHs at $\sim 1.7/T_\mathrm{orb}$, all standard features of circular equal-mass binary accretion.
  }
  \label{fig:prod7.periodogram}
\end{figure}
\addcontentsline{toc}{subsubsection}{\note{Figure~\ref{fig:prod7.periodogram}: power spectra of 50M with PNP}}

Figure~\ref{fig:prod7.periodogram} provides a time-series from the equilibrated state of \PROD{7.longeq} of the total accretion rate onto both BHs, \dotm. The top panel demonstrates the sawtooth modulation from the orbit of the lump on top of the higher frequency modulation from the orbit of the binary interacting with the lump and eccentric cavity. The bottom panel shows the power spectrum of the accretion rate. This power spectrum is calculated using Welch's method with a Hamming window taken over more than 10 non-overlapping segments, each of length $\sim108$ orbits. A power law is fitted to \dotm$(t)$ to measure the long-term reduction of accretion as the reservoir of gas in the finite disk depletes. This fitted function is subtracted from \dotm\ and used to normalize the residual before calculating the power spectrum. We find that a fit of the form $A t^B+C$ is accurate with \{A,B,C\}=$\{3.8\times10^4, -1.1, -5.6\times10^{-4}\}$ when applied to the equilibrated state after t=100,000M. The parameters for \PROD{7.nwt} and its power spectrum are very similar. 

The power spectra reveal modulation by pericenter passages of the eccentric orbit of the lump at frequency $f\approx0.2(\Omega_b/2\pi)$, modulation of accretion as BHs pass close to the pericenter of the CBD at $f\approx2(\Omega_b/2\pi)$ (BH mass symmetry means that over long periods, most power at $f=\Omega_b/2\pi$ is averaged out, leaving only broad-band noise centered around $\Omega_b/2\pi$), and where they pass near the orbiting lump at $f\approx2(1-0.2)\Omega_b/2\pi$. Some prominent harmonic peaks of the fundamentals are due to phase modulation of the primary modes and finite window effects. 

The similarity of the \PROD{7.longeq} and \PROD{7.nwt} power spectra suggests that our PNP gravitational potential, while adding the physical realism of orbital precession, does not significantly alter the accretion flow in the CBD, or fundamentals of episodic modulation of the accretion rate through the BH mini-disks onto each BH.

Though this is the first time that either viscous stochasticity or a target-entropy cooling function has been used in 2D simulations of binaries, we have demonstrated that the primary features of accretion rate modulation seen in prior studies are robustly reproduced in this framework. We conclude that this as a reasonable baseline model. Our fiducial simulation \PROD{6} keeps these choices fixed for the exploration of magnetization in the second set of runs listed in Table~\ref{tab:sims}. 

Next, before introducing magnetic fields, we will use the unmagnetized simulations of the first set in Table~\ref{tab:sims} to explore the effects of hydrodynamic prescriptions on the behavior of accretion as it responds to the binary losing energy and angular momentum due to GW emission, inspiraling, and ultimately merging.

\subsection{Decoupling in 2D hydrodynamics}
\label{subsec:merging_hydro}

\begin{figure*}
  \centering
  \includegraphics[width=\textwidth]{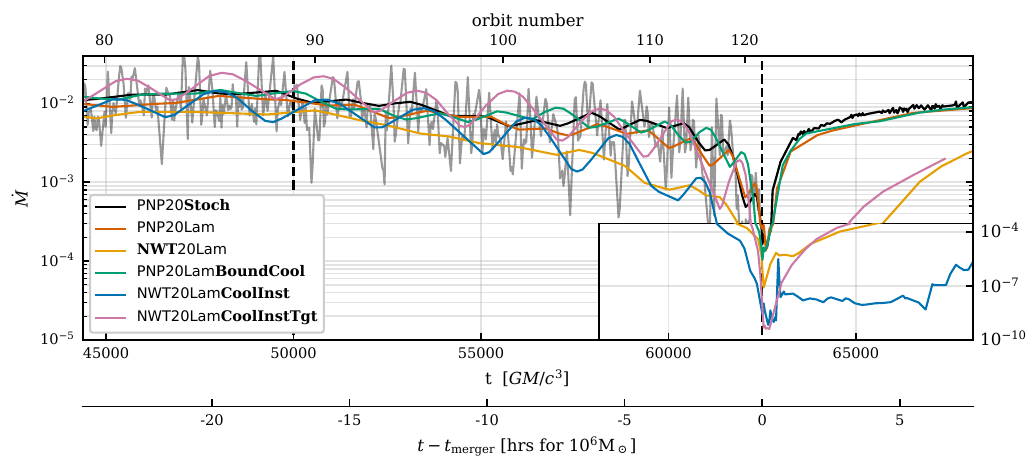}
  \caption{Sum of the mass accretion rates measured at the sinks for both BHs for the unmagnetized simulations. Vertical dashed lines indicate the time of inspiral and merger, $t_\mathrm{insp}$ and $t_\mathrm{m}$, respectively. \PROD{6} is the fiducial run, and \PROD{0} is a matching run with laminar viscosity. \PROD{0} is shown with full time resolution as well as smoothed over a window of 5 binary orbits that evolves with the binary separation until merger. All other simulations only have the smoothed averaging for visual clarity, and to highlight the lump orbital modulation, which is preserved by this window size. Each simulation represents a run with the same configuration as \PROD{6} but with one or two fundamental changes: Newtonian rather than hybrid potential (\PROD{1}), only bound material is cooled (\PROD{2}), Newtonian potential and instant cooling to target entropy with local-pressure viscosity (\PROD{4}) and target-pressure viscosity (\PROD{5}). The drop at merger is so large for some runs that it is shown in an inset with a different vertical scale marked on the right, and matching along its upper boundary to the larger plot. The number of complete binary orbits leading to merger is shown along the top axis. \dotm\ is plotted in code units and shares the same arbitrary scaling as the surface density.}
  \label{fig:20Maccretionrates}
\end{figure*}
\addcontentsline{toc}{subsubsection}{\note{Figure~\ref{fig:20Maccretionrates}: hydro-only accretion rates vs time, 20M seps}}

Our fiducial simulation \PROD{6} and most others listed in Table~\ref{tab:sims} are initially evolved at a fixed binary separation $a_0=20$M until we initiate the inspiral at $t_\mathrm{insp}=50000\rm M\simeq 89 \mathrm{T_{orb,0}}$, where $\mathrm{T_{orb,0}}$ is the initial orbital period. This is a few $\sim10s$ of orbits after noticeable modulation of accretion by the lump begins, just as in MJA24. Other fiducial hydrodynamic choices in \PROD{6} are a sink radius of  $r_\mathrm{sink}=4\rm M_i$, smaller than the ISCO, stochasticity drives fluctuations in the viscosity after an initial laminar period of 1000M, and the viscosity uses the target-pressure prescription where the local pressure calculated from the target entropy is used to calculate the viscosity,~$\nu$. 

Equilibrating with fixed binary separation, then transitioning to inspiral, is the same procedure as in the 3D-GRMHD sims of LE25, where a larger initial torus required a longer equilibration period, closer to 165 binary orbits, and the disk may be slightly more equilibrated compared to MJA24. However, we find that in 2D the CBD eccentricity and lump evolve more quickly than 3D-MHD so equilibration is reached sooner. After equilibration, we evolve through the inspiral, merger, and post-merger phases for all runs with finite $t_\mathrm{insp}$.  

\PROD{6}, has similar prescription and parameter choices as \PROD{7.longeq} except for $a_0$, $t_\mathrm{insp}$, and $r_\mathrm{sink}$. We find that the fixed-separation portion of \PROD{6} exhibits qualitatively similar behavior in accretion disk morphology and time dependence compared to the equilibrated state of \PROD{7.longeq}. However, the closer proximity of the ISCO to the tidal truncation radius for $a_0=20$M compared to $a_0=50$M results in a restriction of stable mini-disk orbits, and \PROD{6} exhibits less sustained and less massive mini-disks as a result. For instance, in Fig.~\ref{fig:fiducialsimSig}, the right BH has no minidisk and gas directly plunges past the ISCO.

The $a_0=20$M purely hydrodynamic simulations of the first grouping listed in Table~\ref{tab:sims}, including \PROD{6}, are compared by their relative accretion rates vs. time in Figure~\ref{fig:20Maccretionrates}. \dotm\ is smoothed over a moving window of five orbital periods prior to merger for visual clarity. This shows the lump orbital modulation but not orbital timescale variability. We do show both the smoothed and full time-series for \PROD{0} for reference. No smoothing occurs after merger.

The drop in \dotm\ due to decoupling appears to begin at or close to $t_\mathrm{insp}$, a shared feature among all the hydrodynamic runs starting at 20M separation. This suggests $a_0=20\rm M$ may already be in the decoupling regime. 
Through inspiral, merger, and post-merger evolution \PROD{6} experiences a drop in accretion rate \dotm\ to about 1\% its initial equilibrium value, most steeply plunging just before merger. After merger the accretion rate recovers quickly to the original value. This is less of a drop in \dotm, and a \textit{much} quicker recovery than reported by LK23. To explore contributions to this difference, \PROD{6} represents our point of comparison as we vary methodological parameters roughly in order of increasing impact on the evolution: stochastic variation of the viscosity, the cooling of unbound gas, gravitational potential, cooling timescale, and scaling of viscosity with pressure. 

\minorheading{Stochasticity in viscosity (\PROD{6})} 
\PROD{0} has constant $\alpha$ viscosity whereas the stochastic modulation is included in \PROD{6}. The accretion rate evolution is nearly unchanged in overall shape and depth of the dip, or decoupling magnitude, whether stochastic or constant viscosity is used. \PROD{0} exhibits slightly greater reduction of the accretion rate before merger, and slightly slower recovery post-merger, but is within the variance of the profiles. This provides further evidence that including stochasticity introduces background variability without obscuring the effects of varying magnetic-evolution parameters. Overall, we find that stochasticity alone does not qualitatively change decoupling and post-merger recovery in these regimes. 

\minorheading{Boundedness of the gas (\PROD{2})}
In LE25 the relativistic version of the Bernoulli parameter, Eqn.~\ref{eqn:Be}, is used to limit cooling to bound gas. On the other hand, LK23 and all 2D-hydro simulations of binaries known to the authors do not apply cooling cuts based on the Bernoulli criterion. In order to measure the impact of this choice, we perform an explicit comparison of cooling all or only bound gas using simulations that are otherwise the same.  

The accretion rate evolution in \PROD{2}, where only bound gas is cooled, does not significantly differ from \PROD{6} and \PROD{0}, where all gas is cooled. Possible differences seen in other characteristic quantities are small compared to the effects from varying the gravitational potential, cooling timescale, magnetization, etc. In light of the relatively small difference cooling all gas makes, and to align with the choices in many 3D-GRMHD simulations invoking target-entropy cooling, our magnetized simulations only cool bound gas. In regions where high magnetization leads to a positive $\mathcal{B}e$, the lack of cooling can cause the gas to become very hot, adding somewhat to the total pressure support in highly magnetized regions. 

\minorheading{PNP vs Newtonian gravity (\PROD{1})}

\PROD{1} is run with a gravitational field defined by the Plummer potential \citep{Plummer1911}. This is a softened form of the Newtonian potential that exhibits minor retrograde apsidal precession for eccentric orbits. It is slightly shallower at a given radius compared to PW and our PNP. \PROD{1} otherwise matches the parameter choices of \PROD{0}, but produces a greater drop in the accretion rate. It has a factor of $\sim100$ lower minimum in accretion rate at merger, and a very slow monotonic post-merger recovery. We conclude that the precession and deeper potential well of PNP strongly affects accretion during decoupling and merger.

\minorheading{Cooling timescale (\PROD{4})} 

The largest difference from our fiducial model is seen when conditions resembling those in LK23 are chosen. Starting with the parameter choices of \PROD{1}, in \PROD{4} we cool to the target entropy almost instantly by setting $t_\mathrm{cool}=\Delta t$. 
Though we do not report either locally isothermal or black-body cooled simulations here, we found in exploratory tests that forcing the disks to be locally isentropic by reducing the entropy-type cooling timescale to be the fastest disk timescale mimics the behavior of both isothermality and black-body cooling. (See the Discussion, \S\ref{sec:discussion}, and Appendix~\ref{app:entvst4} below, for the physical reasoning for this effect.) 

We find a similar drop in \dotm\ in \PROD{4} compared to \PROD{1} during most of the inspiral, but in the last $\sim1000$M before merger, \PROD{4} sees a dramatic final plunge to reach a total of seven orders of magnitude below the pre-merger equilibrium, effectively a complete shut-off. Also, consistent with the highest resolution run of this type in LK23, we see an initially flat, and then extremely slow post-merger recovery. 

Accretion into the cavity and onto the mini-disks occurs through a complex combination of gravitational torques, angular momentum transport through turbulent or parameterized shear stress, or through Reynolds stresses and shocks incurred when the binary gives significant angular momentum to some gas and slingshots it outwards into the CBD \citep{ShiKrolik2012, NobleMundim2012,NobleKrolik2021,TiedeZrake2022}. These processes are sensitive to the local kinematic viscosity in 2D simulations of binaries, and to the thermodynamic conditions in shocks. In the approximately isentropic case (not exact isentropy since we cool but do not heat the gas), local excess heat created in shocks formed during stream impact is instantly removed. Compared to no cooling, this both reduces the ability for compressed material to accelerate gas back into the cavity through pressure gradients, and localizes the deposition of momentum by reducing the formation of pressure waves. The overall result is the efficient deposition of angular momentum from the binary (via the streams) to the very inner edge of the CBD. In 2D, the only route to accrete onto the binary is through this region. 

Additionally, tidal truncation of the CBD results in gas at the edge of the cavity being subject to unbalanced radial pressure gradients. The net flow of gas from the CBD into the cavity results in an adiabatic expansion for which we measure a pressure drop of more than two orders of magnitude in simulations like \PROD{4}. When kinematic viscosity is tied to this local pressure, it also drops and the viscous dissipation leading to accretion 
onto the binary is reduced. The larger cavity placing the gas further from the binary and the lower viscosity both contribute to a faster and more complete reduction in accretion rate in \PROD{4}.

\begin{figure*}
  \centering
  \includegraphics[width=\textwidth]{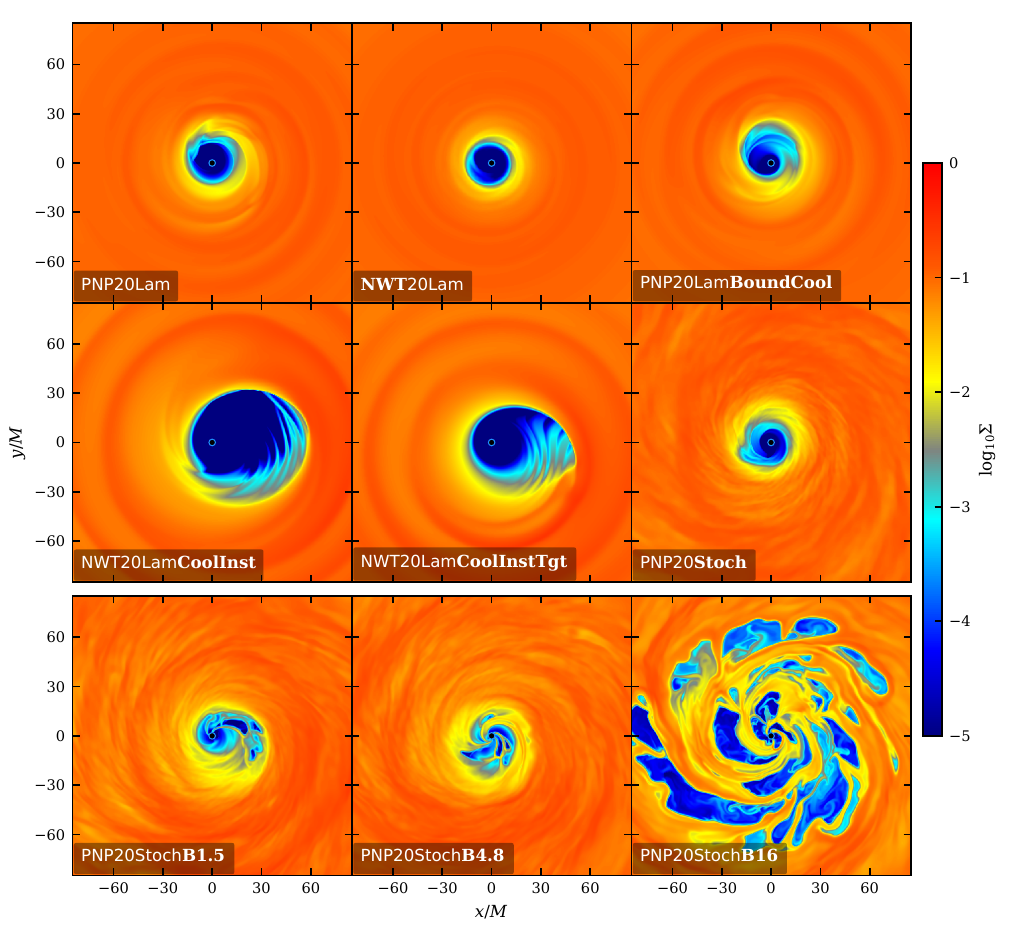}
  \caption{Surface density snapshots plotted equatorially at merger for, from top left, across and down, \PROD{0}, \PROD{1}, \PROD{2}, \PROD{4}, \PROD{5}, \PROD{6}, \PROD{17}, \PROD{19}, and \PROD{18}. The bottom row represents our fiducial configuration, \PROD{6}, but with a magnetic field proxy of increasing magnitude to the right. The proxy field accounts for large-scale magnetic field threading vertically through disks.
  }
  \label{fig:6snapshotsdecoupling}
\end{figure*}

The large relative size of the cold cavity formed in the traditional approach of LK23, matched by the run \PROD{4}, is shown in Figure~\ref{fig:6snapshotsdecoupling}. The surface density distributions of our unmagnetized runs are plotted at the time of merger, $t_\mathrm{merger}$, in the two upper rows. The lower row of Fig.~\ref{fig:6snapshotsdecoupling} contains $\Sig$ at merger for three magnetized runs demonstrating the different regimes of magnetized accretion described in the next section. 

\minorheading{Pressure choice for kinematic viscosity (\PROD{5})}
To recap, \PROD{4} has a Plummer potential, approximately isentropic disk, and a viscosity which is determined using the local cell-measured thermal pressure. We will refer to this as the `local-pressure' viscosity prescription. In \PROD{5}, we apply a single change and calculate $\nu$ using the target pressure,
\begin{equation}
P_\mathrm{targ}\equiv K_\mathrm{0}r^{-2/3}\Sig^{5/3}.
\end{equation}
Under this prescription choice, the adiabatic cooling taking place at the inner edge of the CBD in the ways described above, is not tied to as significant of a drop in the viscosity, and the low-density cavity material can accrete more efficiently. The resulting dip in \dotm\ from decoupling seen for \PROD{5} in Fig.~\ref{fig:20Maccretionrates} is less severe compared to \PROD{4} except right at merger, 
where they reach comparable values. However, the post-merger recovery of \PROD{5} is more similar to \PROD{1}, where the thermodynamics of gentler cooling results in a smaller cavity at merger and a warmer inner CBD edge to feed the cavity viscously. 

In 3D-MHD simulations even if the local thermal pressure is greatly reduced, and even if turbulence is suppressed, the magnetic stress contribution can remain large. Prior studies \citep[e.g.][and MJA24]{ShiKrolik2012,NobleMundim2012,NobleKrolik2021} have found that the Reynolds and Maxwell stresses differ by at most about an order of magnitude in turbulent equilibrium in the CBD outside the inner edge, and the Maxwell stress in the cavity can be quite high. Maxwell stress is generally reported to be lower in the inner portions of CBDs where the binary drives extreme dynamics. We would expect that as the material peels off the CBD and looses thermal support the Maxwell stress will take over and remain roughly comparable to its prior values. For these reasons, we propose that the local-pressure method for calculating parameterized viscous stress underestimates its magnitude in real systems. This reasoning is further supported by \PROD{5} and \PROD{6} (\textbf{Stoch} runs use target-pressure by default) showing much more similar behavior to 3D-GRMHD simulations than \PROD{4} or \PROD{1}.

\vspace{0.7\baselineskip}

\addcontentsline{toc}{subsubsection}{\note{Figure~\ref{fig:6snapshotsdecoupling}: snapshot grid of 20M sep runs at merger.}}

\subsection{Cavity shape at merger}
Comparing the accretion rate dips during decoupling (Fig.~\ref{fig:20Maccretionrates}) to their respective cavity morphologies at merger (Fig.~\ref{fig:6snapshotsdecoupling}), we find a general intuitive trend where larger and more eccentric cavities correspond to greater drops in the accretion rate and slower post-merger recovery. 

Other than sub-structure in the surface density of the CBD in \PROD{6}, the cavity shape and radial density profile appears very similar to \PROD{0} and \PROD{2}. All three evolve with PNP and thus include precession effects, but differ in cooling and stochasticity. The three most similar profiles of \dotm\ share similar cavity structure at merger. 

Evolved with the Plummer potential, \PROD{1} also has a somewhat similar \dotm\ profile and cavity morphology to PNP runs, but a more distinct drop in density at the inner edge of the CBD. We are not able to distinguish if this is primarily due to the orbital precession from the PNP and the resulting ISCO, its deeper potential well changing the conditions under which gas is captured by the BHs, or some other aspect of the differing gravitational forces. It is at least consistent that simulations with a potential approximating relativistic precession are more similar to those evolved with GR-MHD. If decoupling were to happen at much wider separations, as has been suggested for high Mach disks \citep{TiedeZrake2025}, the Newtonian regime may be more applicable. However, the effects of stochasticity, cooling, and stress previously described would still apply and perhaps dominate.

The largest cavities belong to the runs with the most severe cut-off in accretion and slowest post-merger recovery, \PROD{4} and \PROD{5}.

One final interesting feature in Fig.~\ref{fig:6snapshotsdecoupling} -- it is apparent that rings of higher density in the non-stochastic runs are not obviously present in the presence of surface density fluctuations from stochastic viscosity, despite the fact that the mean viscosity in the disks are the same. If the ring features are not physically realistic, which is supported by their absence in the stochastic and 3D-MHD runs, then removing their potential influence on the synthetic spectral signatures is an added benefit of the  physicality of the stochastic viscosity. 

\subsection{Post-merger evolution}
\label{subsec:merger_entropy_stoch}

After merger, GW losses result in a remnant BH of mass $0.97\rm M$. During testing we found the response of the system to this loss was negligible compared to the other parameter choices. This is consistent with LK23. We do not include a kick response from asymmetric GW emission which could make a qualitative difference \citep{CorralesHaiman2010}. Under these conditions, the process of re-brightening and recovery of the accretion rate is determined by disk conditions at merger and by the mass and spin of the remnant BH. We set the post-merger spin parameter of PNP to $\chi=0.8$, approximately the value expected for the merger of two non-spinning BHs of equal mass and minimal eccentricity.

\begin{figure*}
  \centering
  \includegraphics[width=\textwidth]{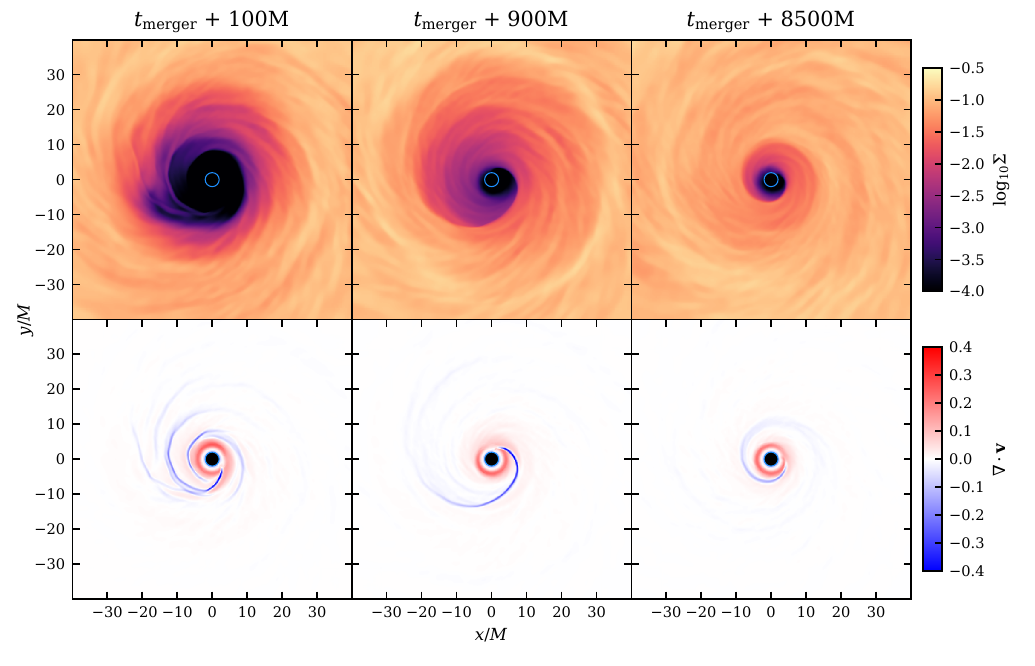}
  \caption{Upper and lower rows show surface density and divergence of the velocity, respectively, for \PROD{6} at three times following merger. In both characteristics the time evolution of the self-intersection shock from eccentric gas in the cavity can be seen.}
  \label{fig:prod6postmergerrhoframes}
\end{figure*}

Figure~\ref{fig:prod6postmergerrhoframes} shows the post-merger surface density evolution in three successive snapshots of \PROD{6}. The density profile increases out to the radius of decoupling, approximately coincident with the radius of maximum $\Sig$. The eccentricity of orbiting gas diminishes for radii beyond that. The inner edge of the CBD and much of the gas in the cavity has significant eccentricity, or order 0.1-0.2. 
Because the gas is eccentric near the BH, apsidal precession leads to self-intersecting flow clearly seen in the divergence of the velocity field in the lower panel of Fig.~\ref{fig:prod6postmergerrhoframes}. 

\addcontentsline{toc}{subsubsection}{\note{Figure~\ref{fig:prod6postmergerrhoframes}: snapshots showing prod6 spiral post-merger}}

The temporal evolution in Fig.~\ref{fig:prod6postmergerrhoframes} shows the rapid formation of a self-intersection shock in the cavity, evidenced by strongly negative $\nabla \cdot \mathbf{v}$. The first frame shows the eccentric structure of the cavity with apoapse towards the upper left. In the corresponding lower panel there are several convergence features in the disorganized flow of the gas near the BH, but these organize quickly into a single dominant shock as gas orbits converge among adjacent radii. The initial dominant shock in the middle frames approximately traces the long edge of the cavity on the more azimuthally advanced side, the result of pro-grade apsidal precession. This structure equilibrates by the last frame into a quasi-steady spiral shock as nonlinear pressure forces and internal stresses establish a radially twisted and eccentric configuration \citep{Ogilvie2001,BarkerOgilvie2016,ZanazziOgilvie2020}.  This tight spiral provides additional angular momentum transport in the flow and has already shrunk significantly by the last frame due to circularization. Accretion returns to normal sBH behavior all the way down to the ISCO as the shocks and the eccentricity of the flow ultimately vanish. 

The equilibrium shock phase is strongest close to the ISCO, where there is a large apsidal advance per orbit, creating a high angle of incidence at intersection and thus a strong shock. This feature is rapid, organizing on the orbital time, and so can quickly enhance angular momentum transport across the entire region of high precession. 
For our $a_0=20$M runs, this includes most of the cavity and the spiral feature can be seen to graze the inner edge of the CBD. 

The rapid shock formation and large extent relative to the cavity size at merger is the key to rapid recovery seen in the unmagnetized simulations with PNP. This dominates the post-merger recovery in \PROD{6}, \PROD{0} and \PROD{2}.

The initial cavity formed through the binary equilibrium has eccentric gas orbits with approximately aligned arguments of periapsis, possibly due to streams of gas pulled off the CBD at periapsis and flung outward, landing approximately at apoapsis \citep{ShiKrolik2012,FarrisDuffell2014}. If the cavity edge is large, this somewhat aligned eccentric flow has very slow precession. If significant precession is absent from the potential, this organized flow remains aligned over many orbits no matter the radius of decoupling, and only decays through Reynolds and viscous stress. This explains the long-lived eccentric cavity in Newtonian simulations \PROD{1}, \PROD{4}, and \PROD{5}, and their slow post-merger recoveries. 

An extreme case occurs when there is no enhancement of accretion by the self-intersection shock from precession, and very low gas pressure in the cavity results in low viscosity. This explains the sustained lack of accretion in \PROD{4}, resembling LK23. This likely also explains why they found recovery to be very sensitive to resolution -- numerical viscosity could dominate in the cavity over all other processes.

The similarity we see in the spiral shock structure evolution, as well as the similar recovery profiles in Fig.~\ref{fig:20Maccretionrates} between the PNP runs, \PROD{0}, \PROD{2}, and \PROD{6}, demonstrates that this behavior is largely independent of viscous stochasticity (at the magnitude we drive) and energy loss (i.e. cooling function). 

In summary, the self-intersection shocks dominate the recovery process in our unmagnetized simulations making proper accounting of precession necessary in the prediction of EM signatures immediately following the merger.

\subsection{Light-curves: inspiral, merger, and post-merger}
\label{subsec:lum_hydro_20M}

\begin{figure*}
  \centering
  \includegraphics[width=\textwidth]{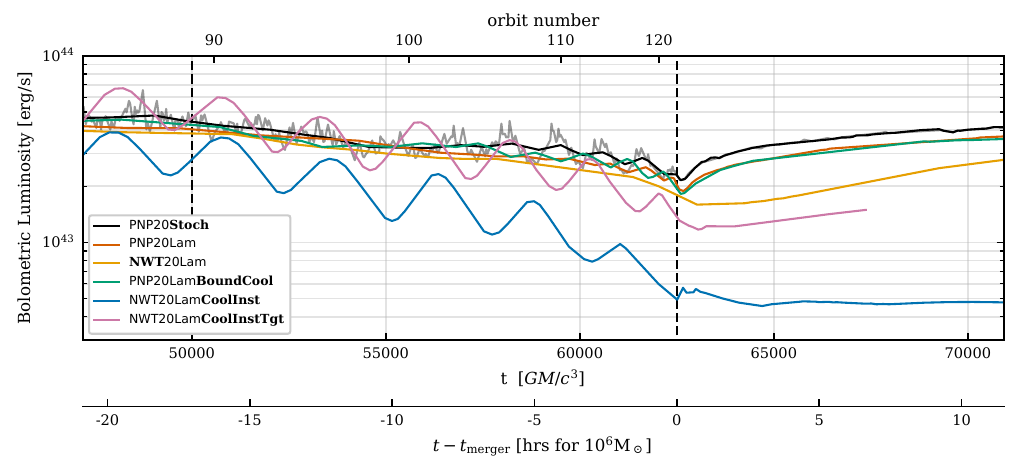}
  \caption{Bolometric luminosity calculated from the surface integral of the cooling function source term over entire system (or the bound portion of the gas only, as relevant), for the runs shown in Fig.~\ref{fig:20Maccretionrates}. The curves prior to merger are smoothed over an evolving window of 5 binary orbits. \PROD{0} is also plotted at full time resolution for comparison. Dashed black lines indicate the times of onset of inspiral and merger. }
  \label{fig:bol_lum_hydro_20M}
\end{figure*}
\addcontentsline{toc}{subsubsection}{\note{Figure~\ref{fig:bol_lum_hydro_20M}: hydro-only bol-lum vs time, 20M seps}}

It is not guaranteed that the bolometric luminosity of the merging binary system, the sum of all energy dissipated by viscous, compressional, and shock heating in the disk, captured by the cooling term $\mathcal{L_\mathrm{cool}}$, will exactly track the mass accretion rates measured at the horizon(s). Figure~\ref{fig:bol_lum_hydro_20M} shows the evolution of the total bolometric luminosity for the unmagnetized runs. We plot light-curves scaled to a $10^6M_\odot$ binary accreting at 0.3\dotm$_\mathrm{Edd}$, i.e. mildly sub-Eddington and shining with a radiative efficiency of $\eta=0.1$. (See Section~\ref{sec:spectral_evolution} for further details)

The overall evolution of the luminosity turns out to follow the qualitative behavior of the accretion rate quite closely, though with a smaller fractional suppression compared to the pre-decoupling equilibrium values. This is not unexpected since the total luminosity contains a significant flux of energy from the CBD in optical and UV bands. However, the finite extent of the disk does mean the high energy emission in the curves shown in Fig.~\ref{fig:bol_lum_hydro_20M} contribute more to the total than they would otherwise. The comparatively large decoupled cavity of \PROD{4} and its very slow refill after merger results in a suppressed luminosity both early in decoupling and over long timescales post-merger.

\section{Evolution in magnetized disks}
\label{sec:hydro_bproxy}

In the previous section, we have shown that stochastic and target-pressure viscosity, target-entropy cooling on roughly the orbital timescale, and a potential accounting for apsidal precession, are physically motivated prescriptions, and result in behavior more similar to that seen in the 3D-GRMHD runs of MJA24 and LE25. These prescriptions make a significant difference in decoupling and post-merger evolution. Even though the decoupling in \PROD{6} is less severe, and the post-merger recovery is much more efficient, compared to our 2D models which more closely mimic LK23, \PROD{6} still differs significantly from published 3D-MHD and 3D-GRMHD simulations.

In the evolution leading up to the merger, the accretion rate in LE25 decreases by only about one order of magnitude (rather than two in \PROD{6}), and it stays nearly constant through the merger itself. The emission inside $r<15$M shows an immediate jump at the time of merger, unlike the slower and gradual increase in bolometric luminosity in Fig.~\ref{fig:bol_lum_hydro_20M}. None of our hydrodynamic models even qualitatively match the 3D-MHD simulations in the MAD state \citep{MostWang2024}. For instance, the MAD simulations exhibit nearly unchanged \dotm\ through merger. 

We hypothesize that large-scale magnetization may explain the remaining differences. Given the factors described above, we progress with \PROD{6} as our default hydrodynamic setup for exploring magnetization with a suite of simulations using the Bproxy prescription. These are listed in the second group in Table~\ref{tab:sims} with names \PROD{16} through \PROD{18}, and are initialized with increasing total vertical magnetic flux according to Eqn.~\ref{eqn:Binit}. 

As a reminder, $\phimad$ is the empirical threshold where additional magnetic flux cannot be accumulated onto the BH and instead piles up in the disk. For our magnetized runs, which approximately share the same equilibrium accretion rate as the hydro runs, $\dot M \approx 10^{-2}$, we measure $\phimad\approx50$. In Table~\ref{tab:sims} we list the total global Bproxy flux in each run, normalized by the empirical MAD limit, $\phi_\mathrm{tot}/\phimad$, and encode this value in the tail portion of the simulation names. 

In this section, we report how our suite of magnetized simulations captures key large-scale magnetic field behavior in a 2D framework for the first time. Our range of magnetization results in binary accretion in the SANE and MAD regimes, and an intermediate regime with complex behavior we call `mini-MAD'.

\begin{figure*}
  \centering
  \includegraphics[width=\textwidth]{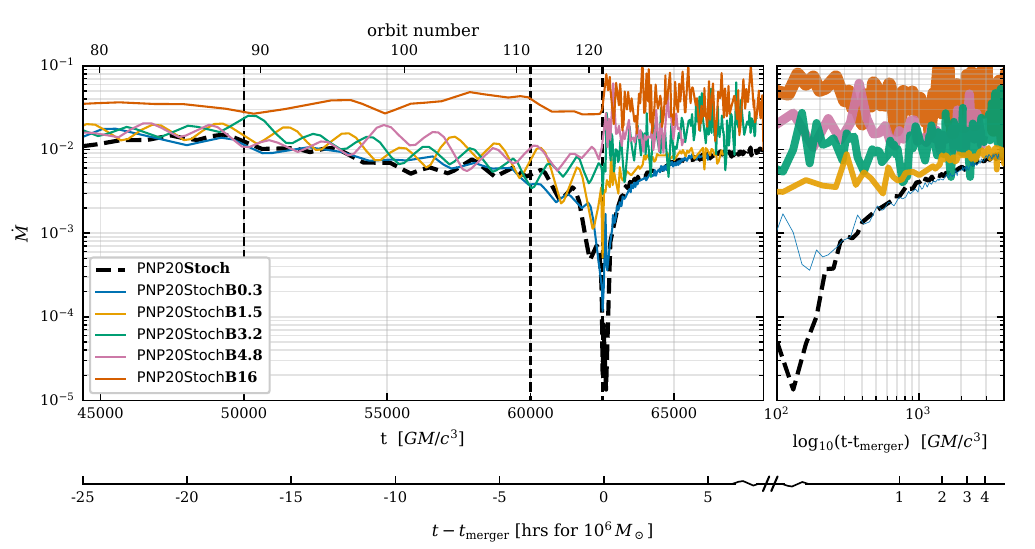}
  \caption{Left panel shows mass accretion rate as a function of time for a series of runs with proxy net vertical magnetic fields of varying magnitude (colors labeled in the legend). The last portion of the magnetized run names indicates the numerical value of the ratio $\phi_\mathrm{tot}/\phimad$, the total magnetic flux across the domain normalized by the empirical MAD limiting value on the horizon. Our fiducial unmagnetized run \PROD{6} is shown for comparison in the black dashed line. Time is plotted in units of $\mathrm{GM/c^3}$, as well as hrs (counting down to and past t$_\mathrm{merger}$, scaled for a total binary mass of $M_b=10^6 \rm M_\odot$. The top x-axis tracks the number of full binary orbits from t=0. The right panel shows the post-merger evolution of the accretion rates with time measured from merger and plotted on a log$_{10}$ scale. The thicker the line, the greater the total integrated Bproxy inside r=15M immediately after merger. Vertical black dashed lines indicate 20M and 10M separation, and merger.}
  \label{fig:dotmbproxy}
\end{figure*}

Figure~\ref{fig:dotmbproxy} shows accretion rate profiles as in Fig.~\ref{fig:20Maccretionrates}, but for the magnetized runs, with \PROD{6} shown for reference. In addition to plotting in geometrized natural units of time and orbits leading up to merger, we scale the binary to a total mass of $10^6 \rm M_\odot$ and show the time in hours before and after merger. To highlight the dependence of post-merger evolution on the magnetic field, we also show the accretion rate for $t>t_\mathrm{merger}$ in log$_{10}$ scale. For the range of $\phi_\mathrm{tot}/\phimad$ values of the runs, we see a corresponding range of behavior, spanning from closely matching our hydrodynamic model, to \dotm\ showing almost no influence from the evolving embedded binary.  

To quantify this relation of global magnetization to accretion, we report the BH-threading magnetic field as a fraction $f_m \equiv \phi_\mathrm{BHi}/\phimad$ (in each normalization made, $\phimad$ is scaled by the appropriate $1/M_\mathrm{BHi}$ factor accounting for the BH mass). $\phi_\mathrm{BHi}$ is calculated using the instantaneous accretion rate and $\Phi$.  
So long as the net magnetic flux threading the BHs is far below the MAD threshold, $f_m\ll 1$, the magnetic field will not significantly affect the accretion dynamics. However, if either variability or decoupling leads to the accretion rate dropping low enough that $\phi_{\rm BHi}$ becomes comparable to $\phimad$, magnetic flux will start to erupt off the BHs, alter minidisk dynamics, and sometimes erupt into the cavity. 

Ignoring variability for the moment, let us assume that a magnetized system attempts to follow the behavior of hydrodynamic-only evolution, prototyped by \PROD{6}, and magnetic fields disturb this behavior according to the same ram-pressure/magnetic pressure balance used to characterize MAD disks. Let us further assume the magnetic flux on the BHs changes slowly compared to decoupling. Given the expected hydrodynamically-driven drop in \dotm, and measuring the pre-decoupling equilibrium on each BH to be $\phi_\mathrm{BHi}$, we can predict whether and when $f_m$ will reach order unity, and the magnetic field becomes dynamically important in the vicinity of each BH. The scalings of $\phi$ in Eqn.~\ref{eqn:phimad} mean that the accretion rate can drop by a factor of $\dot M_\mathrm{t} / \dot M_b = (\phi_\mathrm{BHi}/\phimad)^2 $ before crossing this threshold, where $M_\mathrm{t}$ is the accretion rate at the transition from SANE to MAD accretion near the horizons, and $\dot M_b\approx10^{-2}$ is the initial accretion rate we measure. We assume symmetric behavior since the BHs are of equal mass, and factors of 1/2 for the accretion portion to each BH cancel.

The logic applies in reverse for the post-merger recovery, where the accretion rate will increase as the cavity shrinks and $\phi_\mathrm{BH}$ will drop back below the MAD threshold ($f_m\sim 1$) if there is not too much magnetic flux in the system. 

In an initially SANE system, where MAD behavior is not evident in the CBD or the mini-disks, the more magnetic flux initially stored in the reservoir of the BHs, the earlier in the decoupling regime this threshold is reached, and the greater magnetic disturbance and degree to which angular momentum loss from winds in the minidisks, cavity, and even CBD will effect the merger signature. We will now explore the SANE, mini-MAD, and MAD limits using $f_m$ to tell a unified story of magnetized binary accretion.

\subsection{The SANE limit}
\label{subsec:full_sane}

Our lowest magnetic flux model, \PROD{16} does not show evidence of significant magnetic disturbance of gas in the cavity, mini-disks, or CBD, and contribution of the magnetic flux to angular momentum transport in the disk is negligible throughout evolution. This is expected for $\phi_\mathrm{tot}/\phimad=0.3$, indicating there is not sufficient magnetic flux in the entire system to reach the MAD state in equilibrium. It is not surprising, then, that this case behaves very similarly to the purely hydrodynamic run \PROD{6}. 

$\phi_\mathrm{BHi}\sim5$ is measured for both black holes in early quasi-equilibrium and the first half of inspiral, and then rises to a peak of about $15$ at merger, before dropping as the accretion rate increases again. The multiplicative change in \dotm\ allowed before onset of flux eruptions is then $(\phi_\mathrm{BHi}/\phimad)^2\sim(5/50)^2=0.01$ times the pre-decoupling accretion rate. 
\PROD{16} sees about a factor of 10 less suppression in $\dot M$ compared to \PROD{6} at merger, curtailing the drop almost exactly as it reaches a factor of 0.01 of the pre-decoupling accretion rate.  

By a few 100M after merger, the accretion rate rises past this threshold and the subsequent evolution of \PROD{16} matches almost exactly to \PROD{6}. After merger, any magnetic flux that has managed to erupt off the horizons at the lowest accretion rate is very quickly re-accumulated by the ram pressure of the accreting gas, which has particularly high radial in-fall speed due to the self-intersection shock. In terms of absolute flux, $\Phi_\mathrm{BHi}$ reduces by about half during decoupling, and the post-merger BH quickly rises to contain the combined pre-decoupling contributions from both BHs. A change of only a half is broadly consistent with our assumption of slowly changing $\Phi_\mathrm{BHi}$ during decoupling.  

\subsection{Intermediate magnetization: ``mini-MAD'' binaries}
\label{subsec:mag_middle}

If the BHs are not MAD prior to decoupling but more flux is present in the system than the purely SANE case, the systems are in an intermediate regime where the pre-decoupling equilibrium $\phi_\mathrm{BHi}$ value is high enough to predict a mini-disk SANE to MAD transition before merger. We refer to this transitional state where magnetic eruptions disturb the minidisks over a significant portion of decoupling, or even prior to decoupling, as ``mini-MAD'' binary accretion. 

Before both mini-disks are permanently disturbed, the periodic variability of the accretion rate will cycle the mini-disks from below to above the MAD threshold on approximately the orbital timescale. It is suggestive that in MJA24 the mini-disks were found to transition from magnetically dominated to gas-pressure dominated regimes in a very similar way. 

Magnetic flux eruptions can disturb the mini-disks to a degree that they never fully form beyond some point of the progression of decoupling. However, in the mini-MAD regime there is too little magnetic flux to erupt regularly into the CBD itself and so evolution there remains similar to the unmagnetized case. 

\PROD{17}, \PROD{20}, and \PROD{19} represent mini-MAD behavior where the magnetically disturbed phase exists for differing portions of the decoupling and post-merger evolution. 
Considering these in order of increasing magnetization, \PROD{17} has an initial equilibrium value of $\phi_\mathrm{BHi}\sim20-30$, reaches as low as 10 towards the end of inspiral, becomes more variable before merger, and settles to just below 10 after merger. We thus predict a transition threshold at approximately $(25/50)^2=0.25$ times the equilibrium \dotm, which is again consistent with the minimum accretion rate observed, within 10s of percent. Each $\phi_\mathrm{BHi}$ exhibits relatively smooth behavior except for just prior to merger, when there are large oscillations between $\phi_\mathrm{BHi}\sim5-35$, consistent with the onset of flux eruptions only when the MAD threshold is exceeded surrounding the time of merger. 

With more total magnetic flux in the disk, \PROD{20} exhibits an initial equilibrium value of $\phi_\mathrm{BHi}\sim20-40$, going as low as to 10-20 during inspiral, and rising back to $\sim20-25$ about 2500M before merger, but at times as high as 50 or more. The post-merger BH has $\phi_\mathrm{BHi}\sim20-30$. Thus, estimating an initial quasi-equilibrium value of $\phi_\mathrm{BHi}\sim30$, we predict a drop to 0.4\dotm. The system reaches this value at about $t\sim70000$, corresponding to a binary separation of 10M (note that inspiral for \PROD{20} begins at 60kM, so it is shifted to merge at the same time as the other simulations in Fig.~\ref{fig:dotmbproxy}).

Finally, \PROD{19} exhibits large fluctuations encompassing $\phi_\mathrm{BHi}\sim10-50$, but usually closer to the average of 30 in quasi-equilibrium and early inspiral, dropping to an average of $\phi_\mathrm{BHi}\sim20$ towards later inspiral once transient minidisks no longer form. The average value 30 predicts the same MAD threshold \dotm\ fraction as \PROD{20}, and unsurprisingly, decoupling profiles are very similar for these runs. The disk saturates at a post-merger value of $\phi_\mathrm{BHi}\sim20-30$. The post-merger state has a MAD inner region out to $r\sim25$, reminiscent of the post-merger disk in LE25, which is short-lived once the accretion rate rises significantly. 

Overall, our more highly magnetized mini-MAD simulations exhibit qualitatively similar evolution to LE25, where more often than not $\phi_\mathrm{BHi}$ exhibits a slight decrease from the early equilibrium in the inspiral regime, flattens or even rises slightly just before merger, and exhibits a jump to the MAD threshold at and just following merger. 

Comparing these values to the $\phi_\mathrm{BHi}$ calculated in LE25, however, there are a few important differences to be aware of. First, in their units, $\phi_\mathrm{BHi}$ was calculated with an extraneous factor of $1/(4\pi)^{1/2}$, increasing an equilibrium $\phimad$ value at merger of $\sim2.5$ in their calculation to $\sim8.5$, in line with the thin-disk MAD limit first reported in \cite{AvaraMcKinney2016} and confirmed in subsequent work\footnote{Confirmed through private communication with the authors of LE25.}. Additionally, they integrated the signed flux instead of the absolute magnetic flux over the horizon, which can diminish the value of $\phi_\mathrm{BHi}$ measured greatly if there is significant structure higher than that of the dipole \citep{2012MNRAS.423.3083M}. Finally, they normalize by the gravitational radius $r_g$ rather than the true spin-dependent horizon area which results in a factor of 2 smaller $\phi_\mathrm{BHi}$ prior to merger compared to the way we make the calculation.  

While our reported values of $\phi_\mathrm{BHi}$ are about a factor of 4 higher in magnitude than those from LE25 (after applying the correction), or the literature studying MAD thin disks, our empirical $\phimad$ is higher by this same factor. Since we report evolution of the system with respect to the empirical MAD threshold, this factor does not make a significant difference in the behavior we observe. In $f_m$, for instance, this factor cancels out. 

For simulations starting with 20M binary separation and the PNP gravitational prescription, the Roche lobes of each BH are only transiently filled with circularized gas even in fully hydrodynamic simulations. Distinguishing between this transient mini-disk formation and the disruption of mini-disks by magnetic flux can be difficult. Starting at wider separation, the larger radial range of stable orbits results in sustained mini-disks where the influence of magnetic disturbances and the transition to MAD mini-disks during decoupling can be seen more easily. In \S\ref{sec:wider_mergers} below, we will explore the influence large-scale magnetic flux on binaries with larger $a_0$.

\subsection{MAD binaries}
\label{subsec:mad}

Finally, two other regimes should exist in which even during pre-decoupling equilibrium, $\phi_\mathrm{BHi,eq}>\phimad$ for the individual black holes: a) the minidisks and cavity are magnetically dominated but not enough flux is available to erupt into the CBD and lead to MAD CBD dynamics, and b) the globally MAD case where the MAD to SANE transition radius in pre-decoupling equilibrium resides inside the CBD. The former is the extreme limit of the mini-MAD regime and the latter is the case studied in \cite{MostWang2024}. We do not present a simulation in this study that represents the limit of the mini-MAD regime. We leave the exploration of that regime for future work and focus now on the globally MAD case.

Our most magnetized run, \PROD{18} falls in this regime, and has a pre-decoupling average of $\phi_\mathrm{BHi}\sim$35, with fluctuations ranging from 20-50. $\phi_\mathrm{BHi}$ drops to $\sim$20 during the early inspiral, rises back to about 30 prior to merger, and is about 35-45 after merger. 

\textit{How sensitive are the values above to the bounds chosen for the $\Phi$ integral, and method of calculating the instantaneous accretion rate?} The post-merger values in \PROD{18} are calculated using a flux integral out to $r=1.4M$ (the horizon radius for post-merger spin of $\chi=0.8$) and the accretion rate is calculated from the mass flux at $r=6M$ (near the transition to plunging orbits, but still large enough that the azimuthal sampling of the grid is very high around the BH, and not subject to any potential delay that the sink rate might give). To estimate the sensitivity of our measurements to these choices, we also consider two other radii for the $\Phi_\mathrm{BHi}$ integral. If we integrate inside $r=M$ ($r=4M$), then $\phi_\mathrm{BHi}\sim25$ ($\phi_\mathrm{BHi}\sim45$ ). Given the non-relativistic nature of the calculation in our 2D framework, the specific threshold magnitudes or empirical $\phimad$ values in our simulations should be considered as estimates. Because the MAD state is sensitive to reconnection effects, in principle, even full 3D-GRMHD is limited from calculating these values fully self-consistently. 

In \PROD{18}, the low-density, high-magnetization bubbles, supported by magnetic pressure, are distinctly visible in Fig.~\ref{fig:6snapshotsdecoupling} (lower right panel). Throughout decoupling and merger, bubbles of magnetic flux and intermittent eruptions from the BHs transport enough angular momentum away from the innermost portion of the CBD that the cavity is hardly evident at the time of merger. 

\PROD{18} shows almost no change in the accretion rate throughout the equivalent SANE decoupling period in Fig.~\ref{fig:dotmbproxy}. The only notable feature is a temporary increase in \dotm\ and luminosity lasting $\sim 500\rm M$ after $t_\mathrm{merger}$.

\subsection{Matched conditions to 3D-GRMHD Simulation of LE25}
\label{subsec:20M_intermediate}

\begin{figure*}
  \centering
  \includegraphics[width=\textwidth]{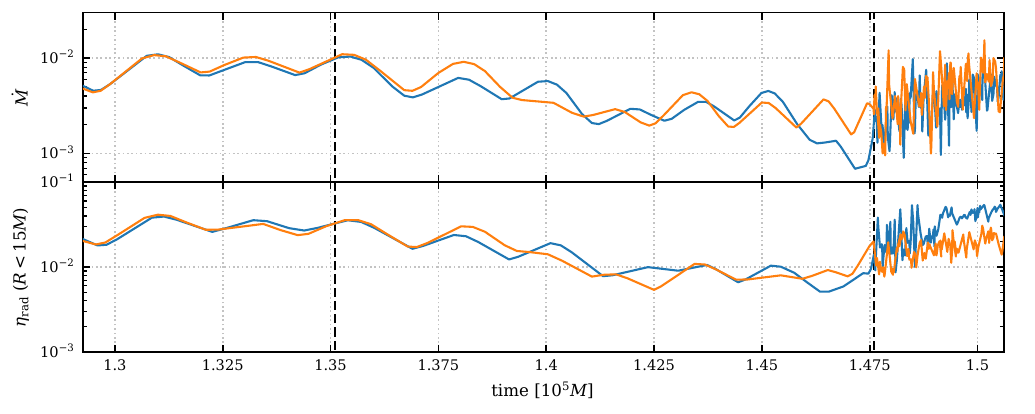}    \caption{Mass accretion rate in code units (top) and contribution to radiative efficiency from $r<15\rm M$ (bottom) for \PROD{23.1} (blue) and \PROD{23.2} (orange). \PROD{23.2} is a restart of \PROD{23.1} from just before inspiral. It tests the effect of different magnetic BH calibration parameters, leading to a different empirical MAD limit, $\phi_\mathrm{MAD}$. The efficiency is calculated by normalizing the total cooling rate within this region by the equilibrium accretion rate at large radii.
  In both panels, prior to merger, a time-average over a moving window of 5 binary orbits smooths the curves for visual clarity. The vertical dashed lines indicate the time of inspiral onset and of merger. In both simulations, the accretion rate is approximately constant through merger with a slow post-merger rise. However, the radiative flux from the $r<15\rm M$ region jumps right around merger to nearly the pre-decoupling value, the same behavior observed in LE25, Figure 14. 
  }
  \label{fig:ennoggicompare}
\end{figure*}
\addcontentsline{toc}{subsubsection}{\note{Figure~\ref{fig:ennoggicompare}: Accretion rate and emission inside 15M radius in Ennoggi comparison run.}}

While several of our intermediate magnetization runs come close to matching the accretion rate and other characteristics in LE25, there are still a few remaining differences. LE25 demonstrates a nearly unchanged continuation of accretion rate through and following merger, whereas our magnetized simulations all show a rather prompt recovery. We do, however, match the post-merger variability in \dotm\ well, suggesting we correctly capture the magnetic evolution in the decoupled cavity, and that it remains MAD for some period of time. 

In order to provide an even closer match to the behavior in LE25, and to account for this prolonged suppression of the accretion rate after merger, we first consider two aspects of the 3D-GRMHD runs in MJA24. The behavior leading up to merger is nearly equivalent in that work and LE25. There is a drop in accretion rate resulting from the evolution intrinsic to the torus that is independent of the effects of inspiral, and is instead due to the depletion of the reservoir of gas and equilibration of the radial density profile, and is estimated to account for a third of the overall drop in \dotm\ through the GW inspiral period (see MJA24 for how this is measured). While the inspiral in LE25 is turned on after a longer equilibration period, and so this contribution may be less, the effect should still be present (see their Figure 3). Additionally, MJA24 note that in their simulations, as in all SANE CBD simulations of similar type of which the authors are aware, the MRI becomes under-resolved in the high density region of the inner portion of the CBD, especially in the lump. The continuing accumulation of mass at those radii over the period evolved in both that work and in LE25 could be considered a result of diminished turbulence and thus lower effective viscosity. 

While we do not have these resolution issues associated with the MRI, we can account for a reduced viscosity near the inner CBD edge. \PROD{23.1} and \PROD{23.2} use the target mean $\mu_\beta$ of the evolution of $\beta_\nu$ in Equation \ref{eqn:betanu} in order to lower the mean viscosity by $e^{-3/2}$ between $(2-4)a_0$. This increases the density during the equilibration of that region (to maintain the same net accretion rate at reduced viscosity), more similar to that seen in the 3D-GRMHD simulations. These runs contain an intermediate magnetic flux magnitude, resulting in mini-MAD behavior. Because our calibration lead to an empirical $\phimad$ value in sBH 2D simulations that is greater than typically measured in GRMHD runs, to test the effect of any potential systematic bias, once \PROD{23.1} has equilibrated, we perform a restart with different horizon-scale calibration of Bproxy that leads to a $\phimad$ reduced by about 50\%. This restart, \PROD{23.2} evolves in parallel with \PROD{23.1}, the latter of which shares the same Bproxy calibration as the other magnetized runs in this work. 

The accretion rate for both runs is plotted in Figure~\ref{fig:ennoggicompare}. We also plot the total radiative efficiency of the region $r<15M$, calculated as 
\begin{equation}
\eta_\mathrm{rad}\equiv\frac{\mathcal{L}_\mathrm{cool}(r<15M)}{<\dot M>_\mathrm{eq}},
\end{equation}
where $<\dot M>_\mathrm{eq}$ is the time-averaged accretion rate onto the BHs in the equilibrated pre-inspiral regime. This plot mimics that of Fig.~14 in LE25. Compared to the mini-MAD simulations, the region of suppressed viscosity makes the decoupling more immediate once inspiral begins, and then shows the type of flattening first characterized in MJA24. There is then a further dip just prior to merger in \PROD{23.1}. In both cases the post-merger accretion is sustained at a value comparable to that from the final inspiral, and the inner disk is seen to exhibit magnetic flux eruptions, signature of a sustained MAD inner disk.

Despite the similarity in the two runs, $\phimad^a \sim 2 \phimad^b$, where $\phimad^a$ and $\phimad^b$ are the empirical MAD limits for \PROD{23.1} and \PROD{23.2}, respectively, and the small difference in \dotm\ near merger likely indicates that the decoupling and merger evolution is somewhat sensitive to the $\phimad$ calibration. However, the differences in evolution are near the magnitude of variability, and if we consider the factor of $\sim$three difference in accretion rate minima, this is very close to the factor of $(\phimad^a / \phimad^b)^2\approx 4$ change in predicted minimum \dotm\ from the onset of magnetic flux eruption. Therefore, even if our calibration of $\phimad$ is somewhat inaccurate, the relation between accretion rate and total magnetic flux remains robust.

The total energy removed by the cooling function inside $r<15M$ also shows very similar behavior to that of LE25 (other than a constant scale offset), with a factor of $\sim4-5$ reduction through inspiral followed by an immediate jump at merger. However, we find that the jump magnitude is sensitive to the inclusion or exclusion of material in the total emissivity calculation at the scale of the horizon. If we exclude the thermal loss very close to the sink and inside the plunging region, the jump is about 50\% less. Because most actual EM emission in this region would likely be beamed towards and captured by the BH, we suggest that, in both LE25 and this Fig.~\ref{fig:ennoggicompare}, the magnitude of the jump and the total post-merger bolometric luminosity from the plunging region be viewed as an upper limit. Full radiation transfer in 3D-GRMHD simulations are needed to more definitively address this. However, a sudden increase in emission from this region does seem to be a robust prediction. This emission appears to come from the stabilization of orbits around the merger remnant, and the associated increase in surface-density. 

\begin{figure}
  \centering
  \includegraphics[width=\columnwidth]{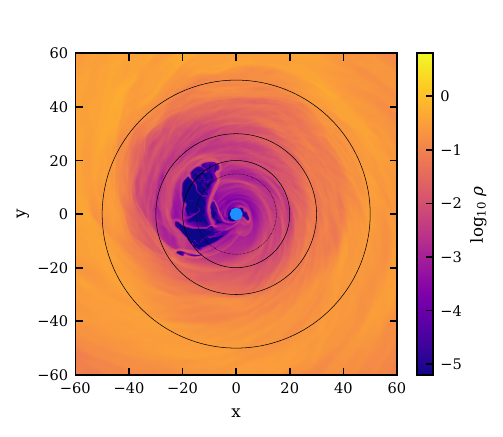}  
  \caption{Surface density snapshot of \PROD{23.2} just after merger. Magnetic flux eruptions from the MAD state of the inner disk are seen as low-density evacuated regions, the 2D equivalent of vertical magnetic flux tubes. }
  \label{fig:prod23snapshot}
\end{figure}
\addcontentsline{toc}{subsubsection}{\note{Figure~\ref{fig:prod23snapshot}: density snapshot of ennoggi match run}}

A snapshot of the post-merger disk of \PROD{23.1} is plotted in Fig.~\ref{fig:prod23snapshot}, showing low density regions from the flux eruptions, and plotted in a similar way to Fig. 8 in LE25. Despite these additional similarities between our \PROD{23.1} and \PROD{23.2} and LE25, the CBD inner edge is still seen in their Figure 8 to be more relatively dense. Considering this difference, a slightly less complete shut-off in accretion rate and a faster post-merger recovery can be expected for our runs. We do, however, qualitatively reproduce the fast decoupling and flattening, and near constancy of the accretion rate through merger. 

We conclude that the suppressed turbulence in the inner edge of the CBD in LE25 results in a larger effective decoupling radius, larger than the starting separation of the runs at 20M, and thus there is a more immediate drop in accretion rate at the onset of inspiral, followed by a reduced decoupling resulting from mini-MAD behavior. In the next section, we will explore the relation of initial binary separation to observed decoupling behavior.


\section{Wide-separation runs: binary - disk decoupling}
\label{sec:wider_mergers}

The simulations of inspiral and merger presented so far all inspiral from an initial $a_0=20 \rm M$ separation. This provides a fair point of comparison to most 3D-GRMHD simulations. However, the analytically predicted separation at which decoupling roughly begins, $a_\mathrm{dec}$, is predicted to be closer to 25-30 for our disk thermodynamic state. Numerical simulations starting from wider separation support this prediction \cite{DittmannRyan2023}. 

In this section we explore the first simulations performed to date that capture the full decoupling and merger process for a binary accreting in the mildly sub-Eddington regime (i.e. thin disk), and which includes magnetic field effects.

We introduce three wide-separation simulations that evolve with stochasticity, and our fiducial thermodynamic and viscosity prescriptions. One is purely hydrodynamic and two include large-scale magnetization via the Bproxy placing them in the mini-MAD regime. By virtue of their wider initial separation, these evolve through many more orbits and confirm our findings in a more equilibrated regime than 3D-GRMHD simulations have been able to reach. 

We have already introduced the equilibrium state of an unmagnetized binary evolving with fixed separation $a_0=50 \rm M$ with run \PROD{7.longeq} in \S\ref{subsec:fixed_sep}. \PROD{7} shares its first 50kM of evolution with \PROD{7.longeq}, but a restart with GW inspiral is initialized at 50kM and we evolve \PROD{7} through merger. The CBD becomes equilibrated after only a small change in binary separation, long before reaching $a_\mathrm{dec}$, so we do not need a pre-equilibration period at fixed separation. 

\begin{figure*}
  \centering
  \includegraphics[width=\textwidth]{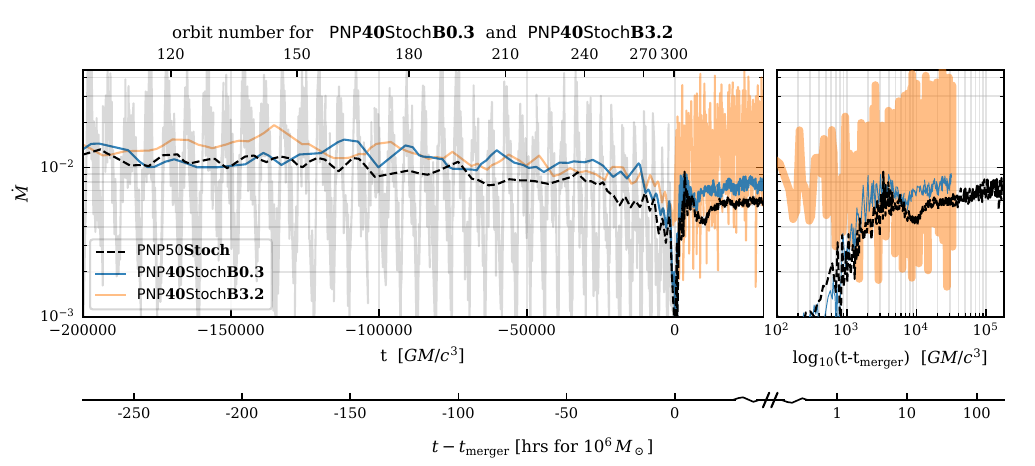}
  \caption{Mass accretion rate as a function of time for \PROD{7}, \PROD{25}, and \PROD{26}, the latter two of which are in the mini-MAD regime. Time is plotted along the lower axes in units of [M]. For the two simulations with matching $t_\mathrm{merger}$, the total number of evolved orbits prior to merger is indicated on the upper axis. The curves are running averages with a width of 10 binary orbits for evolving separation $a(t)$ before merger, and show full time resolution of 10M after merger. To see the variability characteristics, the full resolution \dotm\ of \PROD{7} is also plotted in light gray before merger. The right panel shows the post-merger evolution with a log$_{10}$ time axis. The thicker orange line corresponds to the more magnetized disk in \PROD{26}. \PROD{7}, unmagnetized, is plotted with a dashed black line. A `second dimming' is seen after merger in the unmagnetized and low-magnetization runs at $t\sim 10$kM. \PROD{26} is MAD after merger and magnetic disruption in the cavity disturbs the self-intersection shock and erases this feature. }
  \label{fig:mdotlong}
\end{figure*}

Both magnetized runs, \PROD{25} and \PROD{26}, start with $a_0=40$M separation, and inspiral only after reaching a well-equilibrated state at fixed $a_0$. Figure~\ref{fig:mdotlong} shows the mass accretion rates for these three runs.

\subsection{Unmagnetized evolution}
\label{subsec:wider_50M}

\begin{figure*}
  \centering
  \includegraphics[width=\textwidth]{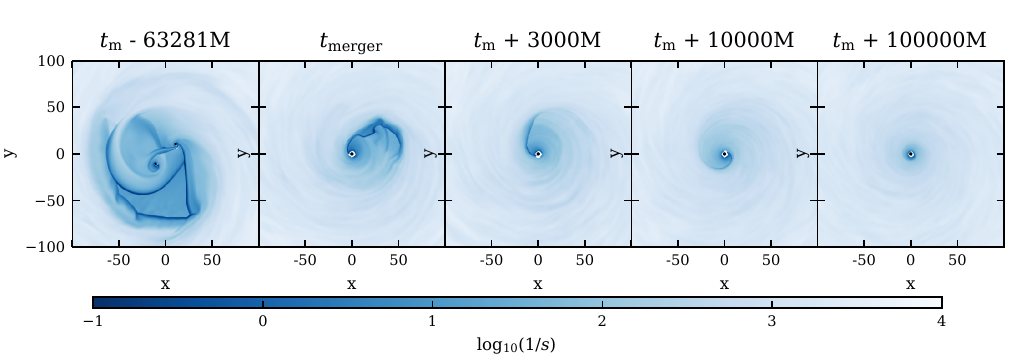}
  \caption{The log$_{10}$ of the inverse of the entropy, $s^{-1}$, is plotted for snapshots spanning from the equilibration separation of 50M, to the fully recovered post-merger state for \PROD{7}. Very low values of the $s^{-1}$ mark the shocks from stream impacts and self-intersections in mini-disks and the post-merger circum-single disk. This quantity complements the velocity divergence plotted in Fig. \ref{fig:prod6postmergerrhoframes} (lower panel) and more directly illustrates shock heating. A well-organized strong shock exists by 3000M after merger, and by $t_m+$10000M transitions into a well-defined and stable spiral structure. By the last snapshot most of the eccentricity has dissipated. }
  \label{fig:prod7framesthroughmerger}
\end{figure*}
\addcontentsline{toc}{subsubsection}{\note{Figure~\ref{fig:prod7framesthroughmerger}: snapshots showing prod7 spiral through merger}}

In the 20M separation runs, the drop in accretion rate due to decoupling is seen to start immediately once inspiral begins. On the other hand, in \PROD{7}, \dotm\ of the equilibrated disk is steady at \dotm$\approx10^{-2}$ for a significant period of time despite the shrinking binary separation. The binary passes the predicted analytic decoupling threshold, $a_\mathrm{dec}=25\rm M$, $\sim31$kM  before merger. In Fig.~\ref{fig:mdotlong} this predicted decoupling threshold roughly coincides with an accelerating drop in accretion rate for all three simulations, though the first noticeable suppression occurs before this, indicating the method of predicting decoupling described in the Introduction should be considered only a rough approximation for the true timescale of the process. 

The remainder of the decoupling process in \PROD{7} looks similar to the 20M separation runs, including a maximum drop in the accretion rate of about an order of magnitude. Post-merger recovery is also fairly rapid, but even by $t_m+$150000M it has not reached the pre-decoupling accretion rate. 
Inspiral from wider than the expected decoupling separation, $a_0>a_\mathrm{dec}$ results in more self-consistent evolution of decoupling, and a significantly larger cavity. The peak surface density of the CBD is between R$\sim$70-100M at merger, as opposed to less than half that in our $a_0=20$M runs. In those runs nearly the entire cavity experienced significant self-intersection shocks. These shocks spanned the entire cavity, reaching the inner edge of the dense CBD gas. \PROD{7} has a large enough cavity that only the inner region experiences significant apsidal precession (which falls off steeply with radius), limiting the influence of the spiral shock structure, and suppressing its ability to bridge the entire low-density gap. Most CBD gas is beyond the reach of this structure, and so the self-intersection shocks can only accelerate accretion of the inner portion of the cavity, and the immediate post-merger rise in accretion pauses before full recovery. In fact, there is a distinct second dip in the prolonged recovery process. This can be understood by taking a more detailed look at the shock evolution. 

Fig.~\ref{fig:prod7framesthroughmerger} shows a sequence of snapshots of $s^{-1}$, the inverse entropy of the gas, highlighting the shock evolution in \PROD{7} from just before merger to a nearly-recovered post-merger state. Here, as in Fig.~\ref{fig:prod6postmergerrhoframes}, we see a shock develop quickly after merger, but it only extends to about R=50M. We split the development of the post-merger shock into two frames at $t_m +3000$M and $t_m+10000$M. These frames demonstrate that the initial shock structure, linearly elongated, is short lived and transitions into a quasi-equilibrium spiral that evolves slowly with less than one orbital precession in phase over the full recovery of the accretion rate. 

\begin{figure*}
  \centering
  \includegraphics[width=\textwidth]{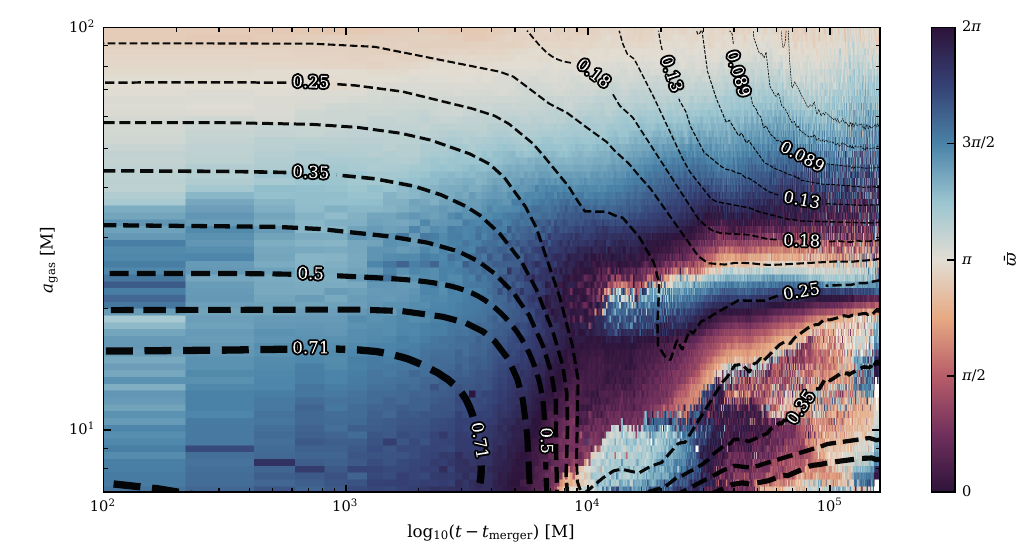}  
  \caption{Argument of periapsis, $\bar{\varpi}_i$ (see Eqn.~\ref{eqn:argper}), as a function of semi-major axis and time after merger. Overlain contours trace the corresponding average eccentricity magnitude, $\left\langle e\right\rangle_{i}$ (see Eqn.~\ref{eqn:eccmag}). Initially aligned and highly eccentric gas in the cavity self-intersects, and, by a few orbital periods for each semi-major axis, reorganizes into a eccentric spiral structure by $\sim10^4$M. Eccentricity then slowly diminishes as the cavity fills in from its high-density edge between radii of about 70-100M.  }
  \label{fig:prod7ecc}
\end{figure*}

The initial linear-shock phase exhibits a steeper self-intersection angle resulting in a stronger shock and larger change in the entropy, compared to the more equilibrated spiral shock. The enhanced angular momentum transport in this region results in a very high effective viscosity. Most gas in this region is accreted before the outer cavity, with much weaker shocks, and the CBD, accreting via normal viscous processes, can replenish it. A drop in the local surface density and thus the accretion rate results, which we refer to as the `second dimming' or second dip. This newly discovered feature can be seen in Fig.~\ref{fig:mdotlong} for runs \PROD{7} and \PROD{25}. 

A final detailed look at the transition of the disk during the second dimming is see in Figure \ref{fig:prod7ecc}, where the eccentric phase of gas orbits is plotted as a function of semi-major axis and time. Overlain contours show the evolving eccentricity magnitude. To produce this plot we calculate, for all cells $n$ belonging to the osculating semi-major axis bin $a_i$, the mass-weighted mean apsidal phase,
\begin{equation}\label{eqn:argper}
\bar{\varpi}_i =
\operatorname{arctan}\left(
       \sum_{n\in i}\frac{\Sig e_{y,n}}{e_n},
    \sum_{n\in i}\frac{\Sig e_{x,n}}{e_n}
\right),
\end{equation}
where $e_{x,n}$ and $e_{y,n}$ are the magnitudes of the x and y components of the eccentricity vector, with total magnitude, $e_n = \left(e_{x,n}^2+e_{y,n}^2\right)^{1/2}$. The full eccentricity vector of a cell $n$ is defined, 
\begin{equation}
\boldsymbol{e}_n
\equiv \frac{
    \boldsymbol{v}_n
    \times
    \left(\boldsymbol{r}_n\times\boldsymbol{v}_n\right)
}{GM}
-
\frac{\boldsymbol{r}_n}{r_n},
\qquad
e_n=\left|\boldsymbol{e}_n\right|.
\end{equation}

The mass-weighted mean eccentricity magnitude in the same bin is, for cells of area A,
\begin{equation}\label{eqn:eccmag}
\left\langle e\right\rangle_{i}
=
\frac{
    \displaystyle\sum_{n\in i}
    \Sig_n\,A\,e_n
}{
    \displaystyle\sum_{n\in i}
    \Sig_n\,A
}.
\end{equation}

The plot of $\bar{\varpi}$ in Fig.~\ref{fig:prod7ecc} shows the radial apsidal phase alignment just after merger, and how, after several orbits of each semi-major axis, reorganizes into a spiral structure of advancing argument of periapsis with increasing semi-major axis, and decreasing eccentricity. This state of well-organized eccentric spiral orbits, reached over a large radial range by $t_m + 10^4\rm M$, is described in detail in \cite{ZanazziOgilvie2020}. The radial eccentricity profile, which we do not show explicitly but can be inferred from the regularity of the vertical spacing the the log-spaced eccentricity contours in Fig.~\ref{fig:prod7ecc}, is very close to the power law predicted for our target disk thermal state. 

The relation of the second dimming to a distinct process of eccentricity evolution suggests it may be a robust feature of binary mergers when the pre-decoupling disk is significantly eccentric. This important behavior was hidden by the small cavity size in the $a_0=20$M runs, and is only captured by simulations that include precession effects. This may explain why it was not seen in prior studies.

\subsection{Mini-MAD evolution}
\label{subsec:wider_mad1}

\PROD{25} and \PROD{26} have the same initial total proxy magnetic flux as \PROD{16} and \PROD{20}, respectively. However, these runs start with $a_0=40$M and equilibrate for much longer, reaching 178 orbits by the time $a=33$M, 266 by $a=20\rm M$, and 302 at merger. Despite these differences, the degree of suppression in \dotm\ at merger and the recovery immediately post-merger are effected by the level of magnetization in the same way seen before.

The very low magnetization of \PROD{25} leads to nearly identical \dotm\ evolution to \PROD{7} through decoupling, a SANE disk after merger, and a second dimming, aside from a general upwards shift in \dotm\ seen in Fig.~\ref{fig:mdotlong} for the magnetized run. Like \PROD{16}, \PROD{25} represents the extreme lower limit of mini-MAD behavior, with the BH minidisks only affected by flux eruptions at the very lowest \dotm\ values. 

On the other hand, \PROD{26} has ten times the total magnetic field in \PROD{25}, and it shows repeated flux eruptions as the $\phi_\mathrm{BHi}$ values fluctuate between the SANE and MAD regime, driven by the accretion rate varying by a factor of nearly 10 from hydrodynamic interaction with the lump and eccentric CBD. The disruption of gas during inspiral results in a minimum \dotm\ that is only about half the equilibrium value. The large field leads to a sustained post-merger MAD inner disk with high variability amplitude, evident in Fig~\ref{fig:mdotlong} by more than an order of magnitude greater variance for \PROD{26} (orange line) than \PROD{25} (blue line) around the second dimming.

We do not include in our suite a wide-separation fully MAD simulation because the concept of decoupling is altered and beyond the scope of this study. In any case, the effective viscosity of order $\alpha\sim1$ or higher in MAD disks brings the analytical estimate for decoupling to much smaller separation. We expect a full 3D-GRMHD treatment is necessary to correctly evolve the thermodynamic feedback of the magnetic field dissipation on the gas. In fact, heating in MAD disks is subject to kinetic plasma effects even beyond the 3D-(ideal)GRMHD treatment and we leave this regime of decoupling for future work.


\section{Temporal and spectral EM signatures}
\label{sec:spectral_evolution}

We now turn our attention to the EM signatures associated with the dynamical behavior in our models, spanning the hydrodynamic (SANE), mini-MAD, and fully MAD binary accretion regimes. These states are represented by \PROD{16}, \PROD{20} and \PROD{18}, respectively.

For all runs, including the three we choose to focus on here, we scale the systems to an accretion rate of 0.3$\dot M_\mathrm{Edd}$, binary mass of $10^6 \rm M_\odot$, and radiative efficiency $\eta=0.1$. The black-body emissivity from a given grid cell is calculated using its gas temperature and the total rate of energy removal by cooling. This frequency-dependent emissivity is logged every timestep into 100 log-spaced frequency intervals, integrated for five regions: each individual Roche lobe of the BHs ($r_i<0.4a$), the CBD ($r_\mathrm{b}>1.7a$), the remaining space in the cavity, and all emission inside $r_\mathrm{b}<15\rm M$. Only gas with surface density greater than 10 times the floor is cooled.

We recover fundamental features expected for optical through soft-X-ray emission. Emission in long-wavelength optical bands may be dominated by gas beyond our finite disk, or at large radii where we do not fully reach inflow equilibrium. High-energy emission in hard X-rays is expected to be dominated by coronal emission in AGN, and possibly their SMBHB counterparts, and is sensitive to the details of particle acceleration, far beyond the scope of 2D simulations, and difficult to capture even in 3D-MHD. Optical and near-UV emission for the binary separations considered in our simulations is essentially constant. Therefore we will focus specifically on the UV and X-ray bands, which are dominated by thermal emission, mostly emitted by the equilibrated gas in the runs. We expect our predictions for these bands to be the most robust to model limitations.

\subsection{Spectra of binary emission}
\label{subsec:spectral_comparison}
\begin{figure*}
  \centering
  \includegraphics[width=\textwidth]{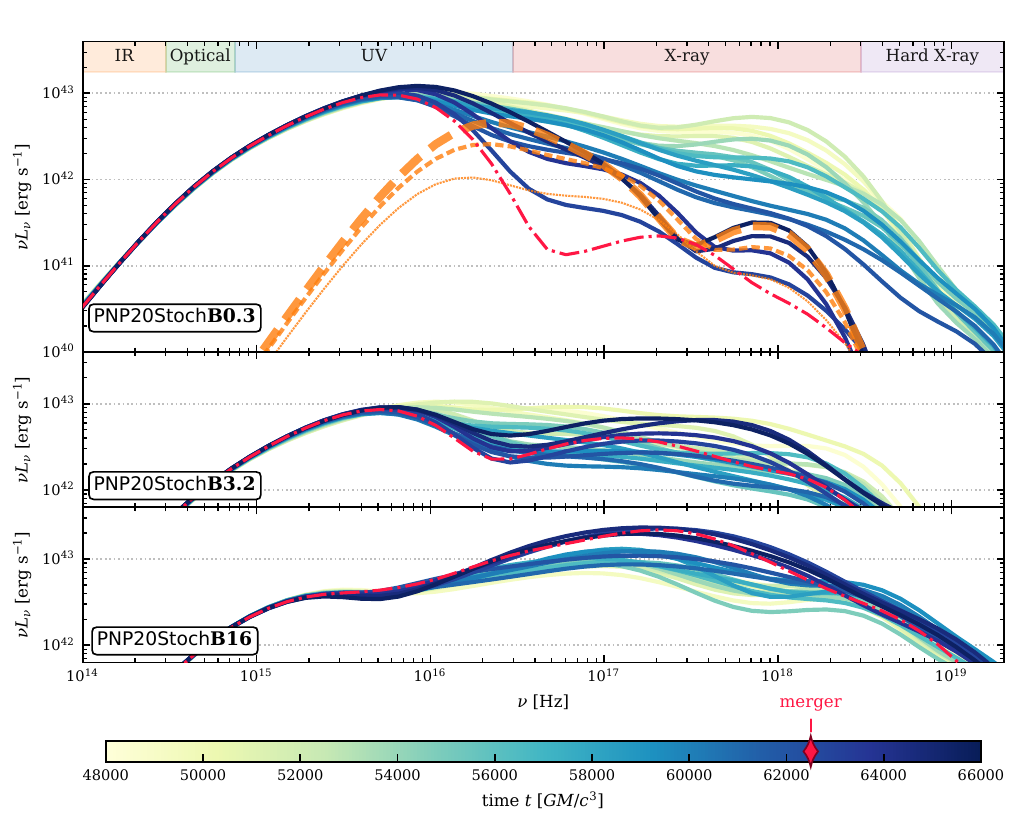} 
  \caption{The spectra for \PROD{16}, \PROD{20}, and \PROD{18}. A remapping of numerical to physical units is made so that the disk accretes at 0.3 times the Eddington accretion rate, for a binary mass $10^6M_\odot$, and radiative efficiency of $\eta=0.1$. The color indicates evolution in time, and the time of merger is marked on the colorbar with a diamond, and for each simulation with a red dot-dashed line (all times for \PROD{20} are offset by 10kM to align with the other two runs at merger). Approximate spectral bands are displayed in blocks for reference. The orange dashed lines of increasing thickness are averages of the spectrum of \PROD{16} taken from radii $R<15M$, and for the windows t=(0,1000)M, t=(1000,2000)M, and t=(2000,3500)M in close succession after merger, respectively. These lines reveal the bolometric contribution from the self-intersection shock, with soft and hard X-ray components.  
  }
  \label{fig:spectraprod16}
\end{figure*}
\addcontentsline{toc}{subsubsection}{\note{Figure~\ref{fig:spectraprod16}: spectra blend for PROD16}}

In Figure~\ref{fig:spectraprod16} we plot the total emission from the four distinct regions for \PROD{16}, \PROD{20}, and \PROD{18}. The equilibrium state and early inspiral of \PROD{16} has peaks in the UV and X-ray associated with the CBD and mini-disks, respectively. As the mini-disks disappear during inspiral the X-ray emission drops more than an order of magnitude, and the post-merger emission in the X-rays has a distinct X-ray bump centered around $\sim10^{18}\rm Hz$ from the self-intersection shock.

The more highly magnetized mini-MAD run, \PROD{20}, features similar pre-decoupling behavior, but the X-ray peak from the mini-disks is much more broadly distributed. The fully MAD simulation, \PROD{18}, has suppressed UV luminosity, consistent with energy and angular momentum transport occurring partially through the parameterized wind. However, the X-ray emission is enhanced compared to the mini-MAD and purely SANE runs.

\subsection{UV and X-ray Light curves}
\label{subsubsec:lightcurves_periodograms}

\begin{figure*}
  \centering

  \label{fig:xraysvstime}{
    \includegraphics[width=\columnwidth]{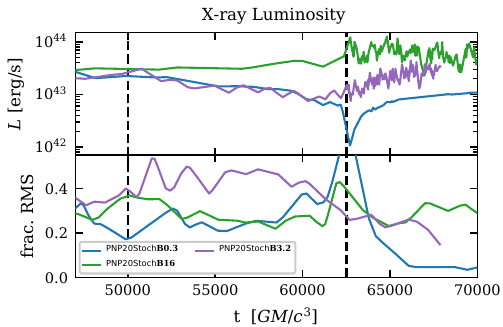}
  }
  \hfill
  \label{fig:UVvstime}{
    \includegraphics[width=\columnwidth]{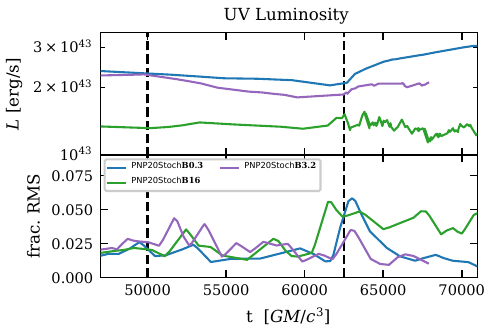}
  }

  \caption{
    X-ray and UV variability for the SANE (\PROD{16}; blue line), mini-MAD (\PROD{20}; purple line), and MAD (\PROD{18}; green line) simulations, respectively. The X-ray band is 0.1-10keV, and UV extends from the edge of optical to 0.1keV. For visual clarity we smooth over a moving 10 orbit window prior to merger, and fully sample after merger. The behavior of the X-ray and UV luminosities at merger, shown in the top panels, follow approximately the same relation to magnetization as the accretion rates seen in Fig. \ref{fig:dotmbproxy}. The lower panels demonstrate that magnetic disturbances lead to (at least) three distinct spectrally-dependent variability profiles, depending on the magnetic regime. 
  }
  \label{fig:XrayUVvstime}
\end{figure*}

\addcontentsline{toc}{subsubsection}{
  \note{Figure~\ref{fig:XrayUVvstime}: X-ray and UV variability during
  inspiral and post-merger evolution}
}

In Figure~\ref{fig:XrayUVvstime} we plot the band-integrated global X-ray and UV flux as a function of time for the three runs. We also plot the fractional RMS variability amplitude over a running window of $\Delta t=5T_{\rm bin,0}\approx2810$M, where $T_{\rm bin,0}$ is the initial orbital period.

The fully MAD run \PROD{18} exhibits increased X-ray and decreased UV luminosity compared to the SANE and mini-MAD cases. Overall the UV and X-ray flux for the mini-MAD runs are similar except at merger. In X-rays, \PROD{20} shows no distinct dip at merger while \PROD{16} drops by an additional factor of 10 below the level reached during decoupling. After merger, the UV luminosity in \PROD{16} increases faster and reaches a higher magnitude than \PROD{20}. Overall, the X-ray and UV light-curves for the two mini-MAD runs dominate the behavior identified above in the bolometric luminosity, and follow the trends with magnetization seen in the accretion rate evolution in Figure \ref{fig:dotmbproxy}. The MAD run shows very little change in either light-curve. 

The RMS variability also shows interesting features. The MAD variability is essentially constant in the X-rays but shows a sudden doubling in amplitude near the time of merger in the UV. Both mini-MAD runs show an approximate halving of their X-ray variability amplitude at merger, though \PROD{20} starts with nearly twice the amplitude seen in \PROD{16} and \PROD{18} during inspiral. 

Lower X-ray variability is consistent with both MAD and SANE regimes, but for different reasons. In the MAD case, magnetic disturbances drive material towards the BH and reduce the high-amplitude modulation from the binary orbit seen in mini-MAD runs. In the SANE case, sustained minidisks contribute a large and steady fraction of the total X-ray emission. On the other hand, the mini-MAD state, where significant X-ray emission comes from the minidisks, but is frequently magnetically disturbed, can be highly variable.

\section{Discussion}
\label{sec:discussion}

3D-GRMHD simulations of binary accretion can capture the behavior of MRI-driven turbulence, jets, relativistic space-time evolution, and in some cases radiation transport. Robust predictions for EM signatures likely require these physical effects. However, these simulations come with high computational cost and code complexity, making efficient exploration of the binary-disk parameter space impossible. 2D simulations provide a complementary tool whereby exploration is efficient, but usually at the expense of key physical processes. We have made a significant step in closing this gap.

In this work we have identified the impact of several simplifications traditionally made in 2D viscous hydrodynamic simulations which have prevented replication of recent 3D studies. To overcome these limitations we introduce physically motivated prescriptions for stochastic fluctuations in viscosity, a floor on the viscosity to account for Maxwell stress, a gravitational potential that approximates relativistic apsidal precession and ISCO, and the dynamical influence of large-scale magnetic fields. 

We have found that the most important of these effects are apsidal precession and magnetic field evolution. Accounting for these largely resolves the apparent difference between predicted EM signatures from LK23 and LE25, two studies chosen here to represent the 2D hydrodynamic and 3D-GRMHD regimes. Simulations performed with our new 2D framework have revealed new ways to use multi-messenger sources to constrain the fundamental physics of accretion disks.

\textit{\bf Post-merger signatures of binaries with wider initial separation.}
Most of our comparison suite of simulations are chosen to start at $a_0=20M$ for a direct comparison to existing 3D-MHD studies which are limited to study shortened inspirals only. We confirm that the impact of a strong large-scale magnetic field is similar in both initially close and wider separation regimes. 

However, a distinct post-merger secondary dimming is seen prominently in initially wider binaries for the unmagnetized run \PROD{7} and SANE magnetization of \PROD{25} in Fig.~\ref{fig:mdotlong}. This is an intriguing feature reported in this work for the first time. We expect that this effect was not seen in $a_0=20$M runs because the cavity was artificially small, preventing the separation of scales of the fast-precession and slow-precession regions of the cavity. Because this dimming is intrinsic to the evolution of the self-intersection shock, and only present in low-magnetization disks, it could potentially be used to reveal spin magnitude of the post-merger BH, as that parameter is directly tied to the apsidal precession rate.

\textit{\bf Entropy versus black-body ($T^4$) cooling.}
It is common practice to use black-body cooling in 2D simulations where the local cooling source term scales as $\mathcal{L} \propto T_{\mathrm{eff}}^4$. The effective surface temperature is defined as,
\begin{equation}
T_{\rm eff}^{4}
\equiv
\frac{4}{3}\,
\frac{T^{4}}{\kappa_\mathrm{es} \Sig} ,
\end{equation}
for evolved gas temperature $T$, and electron scattering opacity $\kappa_\mathrm{es}$. This inherently assumes instantaneous vertical thermodynamic and radiative equilibrium and is the cooling prescription chosen in LK23. 

Instead, we choose to evolve with the target-entropy cooling prescription introduced in \cite{2009ApJ...692..411N}. Our study is the first of its kind using target-entropy cooling in 2D hydrodynamic simulations of accretion disks of a single or a binary BH. When we choose to relax the thermal content to the entropy floor on the orbital timescale $t_\mathrm{cool}=2\pi/\Omega_{\rm K}$, we are able to reproduce much of the behavior seen in 3D-GRMHD simulations.

In \PROD{1} we choose the same laminar viscosity and gravitational potential as LK23, but cool to the target entropy at the orbital period. On the other hand, in \PROD{4} we choose the extreme fast-cooling limit, $t_\mathrm{cool}\sim\Delta t$, where $\Delta t$ is the simulation timestep, and the evolution is approximately isentropic. \PROD{1} produces much more similar behavior to LE25 than \PROD{4}. In the former, we see less of a drop in \dotm\ and a faster post-merger recovery. \PROD{4}, on the other hand, very closely matches the behavior seen in LK23. 

This sensitivity to the cooling rate can be understood in the context of balance between viscous heating and radiative cooling in the SS disk model. For the radiation-pressure dominated inner zone of a moderately sub-Eddington disk, the dynamical, thermal, and viscous timescales are ordered $t_{\rm dyn} \ll t_{\rm th} \ll t_{\rm visc}$. The cooling timescale, over which a small perturbation away from the equilibrium temperature, $\delta T/T_\mathrm{eq}\ll1$, is damped, occurs on nearly the thermal timescale, $t_\mathrm{cool}\approx (3\alpha\Omega_\mathrm{K})^{-1}\sim t_\mathrm{th}$. This is also, by nature of the cooling source term, nearly equal to the radiative diffusion timescale, $t_\mathrm{diff}\approx \tau H/c \approx 4/(9\alpha\Omega_\mathrm{K})$, the limiting rate at which photons can transport heat from inside the disk to the radiative surface. For the gas-pressure dominated SS zone, $t_\mathrm{diff}$ picks up a factor of the ratio of radiation and gas pressures, $P_\mathrm{rad}/P_\mathrm{gas}$, but for the accretion rates we are considering, this zone starts at a much larger radius than our simulations evolve.

However, we find that the highly nonlinear interaction of the binary with the CBD results in localized regions of significant compressional and shock heating that can raise the temperature in small regions by factors of 10 to even 50 times the equilibrium value, on timescales very short compared to the orbital period. This timescale is estimated in Appendix~\ref{app:entvst4} to be $t_\mathrm{impact}\sim 0.3 a^{3/2}\sim 0.3t_\mathrm{dyn}$. For such large ratios of $\delta T/T_\mathrm{eq}$ the response of the disk changes and the cooling timescale picks up a steep scaling with the change in temperature, $t_\mathrm{cool}\sim t_\mathrm{th}(\delta T/T_\mathrm{eq})^{-3} \sim t_\mathrm{diff}(\delta T/T_\mathrm{eq})^{-3}$. Therefore, for fluctuations increasing the temperature by a factor of 10 or more, the cooling response of the normal black-body cooling prescription occurs on a tiny fraction of the radiative diffusion timescale. In this case the assumption of vertical radiative equilibrium made by the black-body prescription is inconsistent with the physics of photon transport. However, a time-scale ordering with $t_\mathrm{cool}$ setting the fastest disk response \textit{is} shared by isothermal evolution.

This analysis may thus explain why studies evolving disks in 2D with black-body cooling typically produce cavity morphologies and narrow streams more similar to isothermal runs than our fiducial target-entropy runs or those in 3D-GRMHD studies, including LE25. In our suite of simulations, the largest cavities are found for our approximately isentropic runs (\PROD{4} and \PROD{5}), consistent with more angular momentum being deposited immediately at the inner edge of the CBD through rapidly-cooled impacts. This behavior is also apparent when comparing recent isothermal \citep{TiedeZrake2025} to adiabatic/black-body \citep{TiedeONeillDOrazio2026} runs those authors used to study the sensitivity of binary accretion to Mach number.

Our target-entropy cooling occurs on the orbital timescale independent of the magnitude and timescale of temperature fluctuations, and preserves the timescale ordering appropriate for the photon diffusion-limited response of the inner radiation-pressure dominated disk around late-stage merging binaries. We therefore reason that the merger response of the disk is more physically self-consistent when evolved with target-entropy cooling compared to the traditional black-body cooling prescription.

\textit{\bf Entropy cooling vs. $\beta$ cooling.}
Along with black-body cooling and a locally isothermal EOS, `$\beta$' cooling, where $\beta\equiv\Omega t_\mathrm{cool}$, is a common choice in 2D simulations. The usual implementation is to remove heat at a rate proportional to $\mathcal{L}\propto (T-T_\mathrm{target})\Omega/\beta$. High $\beta$ results in inefficient cooling and can result in advection dominated accretion in the extreme case, whereas low $\beta$ approaches isothermality \citep{WangBaiLai2023}. Our choice to cool on the orbital timescale would be equivalent to $\beta=2\pi$ if a target-temperature were used instead of target-entropy. The advantage and major difference is that target-entropy cooling allows for adiabatic contraction and expansion (occurring on the $t_\mathrm{impact}$ timescale in the prior Discussion point, for instance) in the quasi-turbulent gas without significantly changing the cooling rate. In other words, cooling requires dissipative heating rather than adiabatic compression in order to radiate energy away. For large fluctuations in temperature, $\beta$ cooling responds linearly,  $t_\mathrm{cool}\propto T^{-1}$. For very large fluctuations this can, again, lead to cooling faster than vertical photon diffusion would allow, but is not nearly as significant a change as that of the $T^{-3}$ scaling of the black-body prescription.

\textit{\bf How sensitive are our results to calibration of the wind magnitude?} In our simulations the Bproxy $r-\phi$ stress applied to regions of high next vertical magnetic flux, is an approximation to the effects of winds seen in 3D-MHD and 3D-GRMHD simulations. This includes a constant of proportionality which we have calibrated to existing 3D-GRMHD simulations of sBH MAD disks. However, this calibration could be a source of systematic error. 

In order to quantify this error, we performed a restart of \PROD{19} with this constant set to 0, so that the proxy magnetic field only affects the flow through pressure gradients leading to enhanced Reynolds stress. This resulted in qualitative differences in the evolution, namely, the flux eruptions had visibly different structure inconsistent with the winding seen in sBH simulations of MAD disks. However, the disturbance of the flow still acted to decrease the overall drop in \dotm, and led to a faster post-merger recovery than the case without magnetization. The qualitative response of the disk to decoupling and merger that is of primary concern to our study was not significantly altered. However, the flow of angular momentum is quantitatively different in this test. 

\textit{\bf Does Maxwell stress provide a floor to local viscosity in the cavity for SANE disks?}
Figure 12 of \cite{NobleKrolik2021} shows the radial and time dependence of the shell-averaged Maxwell stress for disks under conditions similar to both MJA24 and LE25. The Maxwell stress does reach high values inside $r=2a$, comparable to the magnitude further out in the disk at early times before the MRI decays. This justifies our choice for using a target pressure in calculating local kinematic viscosity, preventing artificially low viscosity caused by a thermal pressure drop in the cavity.

\textit{\bf Caveats of our 2D prescription.}
Though our framework offers significant cost savings compared to 3D-GRMHD simulations, and does account for much of the 3D-GRMHD behavior, 2D simulations have inherent limitations. Turbulence has fundamentally different properties in 3D compared to 2D, and the source powering turbulence in 3D-MHD simulations of disks, the MRI, does not occur. We are only able to approximate the correlated Reynolds and Maxwell stress resulting from the MRI. 

Other magnetic behavior we do not track is the full 3D magneto-centrifugal wind, which we approximate, and the jets powered by BH spin through the Blandford-Znajek process \citep{1977MNRAS.179..433B}. Though our framework roughly captures the balance of ram pressure of infalling gas with the magnetic pressure contained on the BH, in 3D, magnetic reconnection leads to the eruption of magnetic flux off the BH, a process we cannot capture in 2D. Reconnection at current sheets is only approximated by 3D-MHD and ultimately requires a fully kinetic implementation to evolve \citep{RipperdaBacchiniPhilippov2020}.

Though we have developed a new post-Newtonian prescription that includes an ISCO and apsidal precession, orbits in the Kerr metric do differ from our prescription in possibly important ways. However, other effects are expected to dominate over the space-time approximations. MJA24 found that even binaries with coplanar disks can develop complex out-of-plane structures in the gas flow. 2D evolution vertically integrates these out. However, they found that this is not a dominant effect during decoupling and inspiral.

Finally, we do not include self-consistent radiation transport. This is currently a limitation we share with nearly all prior studies of binary accretion, though CBD- and mini-disk-only simulations with radiation have recently been performed \citep{TiwariChanBogdanovic2025,TiwariChanBogdanovic2025b,ChanTiwariBogdanovic2025}. We have provided reasons why entropy cooling may be a reasonable approximation, but there is no good substitute for self-consistent radiative feedback on the disk. Radiation feedback from the mini-disks onto the CBD in complex geometries of binary-CBD misalignment may become especially important and motivates future work accounting for both magnetic and radiative effects.

Simulating in full 3D~GRMHD has several advantages, including more self-consistent magnetic evolution, energy and angular momentum transport by turbulence rather than a purely local shear viscosity, and disk thermodynamics can be evolved with full radiation transport. Additionally, in 3D one can account for the conversion of disk and BH angular momentum into winds and jets that feed back on the circum-binary environment. Systems that are not coplanar have no 2D analogue, and MJA24 found that even the coplanar case develops important behavior washed out by vertical integration.

\textit{\bf What are the numerical cost savings of our 2D runs?}
Although more accurate, any 3D-MHD approach comes with significant computational cost. Only very recently has hardware become available that can support, with generous computing allocations, 3D-MHD runs that evolve the equilibrated CBD and binary region simultaneously for long enough to study the inspiral and merger. These simulations regularly cost $10{,}000$ to $100{,}000$ GPU-hrs or CPU-node-hours. 
Important outstanding problems include the behavior in systems with more realistic disk-binary configurations and thermodynamics. The parameter space of binary orbits, mass ratio, and spin is very wide, thus requiring parameter surveys to explore. These are both frontiers requiring $\sim 100-1000$ times the expense of the recent heroic single and few-run studies if 3D-GRMHD methods are used. 

The framework we present offers a way to approach many of these important parameter surveys in 2D, for a small fraction of the cost, and without sacrificing important 3D-GRMHD effects. The reconciliation of the different predictions between 3D-GRMHD and 2D hydro simulations of merger was a necessary first step. We have replicated the accretion response to merger at approximately 100-1000 times lower computational cost, as summarized in Table 2, where we list cost estimates for our simulation and other recent studies for which sufficient information is publically available. 

\begin{table*}
\centering
\caption{Estimated computational expense of accreting binary simulations.}
\label{tab:binary_costs}

\setlength{\tabcolsep}{3pt}
\renewcommand{\arraystretch}{1.1}

\begin{tabularx}{\textwidth}{
  @{}
  >{\raggedright\arraybackslash}p{0.18\textwidth}
  >{\raggedright\arraybackslash}X
  >{\centering\arraybackslash}p{0.08\textwidth}
  >{\raggedright\arraybackslash}p{0.17\textwidth}
  >{\raggedright\arraybackslash}p{0.20\textwidth}
  @{}
}
\toprule
Study
& Physical setup
& $N_{\mathrm{orb}}$
& Code
& CPU node-h or GPU-h (hardware, cores per node)
\\
\midrule
\addlinespace[3pt]

\textbf{This work}
& \textbf{2D hydro+Bproxy; PNP; $20M\rightarrow\mathrm{merger}$}
& \textbf{90--250}
& \textbf{\texttt{Athena++}}
& \textbf{250--500 (CPU, 48)}
\\
\addlinespace[3pt]

\rowcolor{black!6}
MJA24
& 3D GRMHD; matched-PN metric; 3.5PN,
  $a=20M\rightarrow9M$
& 36
& PatchworkMHD + HARM3D
& $\sim10{,}000$ (CPU, 56)
\\
\addlinespace[3pt]

\cite{BowenMewes2019}
& 3D GRMHD; matched-PN metric; 3.5PN,
  $a=20M\rightarrow19M$
& 3
& HARM3D
& $\sim10{,}000$--$100{,}000$ (CPU, 56)
\\
\addlinespace[3pt]

\rowcolor{black!6}
\cite{CombiLopezArmengol2022}
& 3D GRMHD; superposed KS metric; 3.5PN,
  $a=20M\rightarrow17M$
& 12.5-15 
& HARM3D
& $\sim10{,}000$--$100{,}000$ (CPU, 56)
\\
\addlinespace[3pt]

\cite{MostWang2024}
& 3D MHD; Newtonian; fixed $a$; MAD
& 230
& AthenaK
& $135{,}000$ (GPU, V100)
\\
\addlinespace[3pt]

\rowcolor{black!6}
\cite{MostWang2025}
& 3D MHD; Newtonian; Peters $a(t)$; MAD
& $\sim480$
& AthenaK
& $60{,}000$ (GPU, V100)
\\
\addlinespace[3pt]

LE25
& 3D GRMHD; superposed KS metric transitioning to
  numerical relativity; 3.5PN $\rightarrow$ NR; $a(t)$
& $\sim200$
& Cactus/IllinoisGRMHD
& $\sim10{,}000$--$100{,}000$ (CPU, 56)
\\
\addlinespace[3pt]

\rowcolor{black!6}
\cite{TiwariChanBogdanovic2025}
& 3D radiation MHD; Newtonian; fixed $a$; CBD only
& 73
& \texttt{Athena++} (CPU)
& $\sim10{,}000$ (CPU, 56)
\\

\bottomrule
\end{tabularx}
\end{table*}

\textit{\bf Observational implications.}
LK23 predicts nearly a complete drop-out in the X-ray emission by the time of merger, which persists for days for a binary mass of $>10^7\rm M_\odot$, and be potentially detectable with X-ray monitoring. On the other hand, we find consistency with 3D-GRMHD simulations, and predict that the X-ray dimming will not occur if the gas is sufficiently magnetized. Large-scale fields or precession may shorten the dim period, depending on field strength. This short transient accompanying merger might only be captured with very rapid observing cadence, and require an accurate estimate of the time of merger, possibly provided by a chirping signal in LISA or prior information from LSST \citep{XinIsi2026}. However, we still find that the X-ray light-curve offers the most pronounced signatures of binary inspiral and merger.

Having shown that large-scale magnetization plays an essential role in determining the X-ray and UV profiles, our study calls for special attention to the X-ray spectral evolution. X-ray spectral shape, luminosity, and variability are expected to be affected by large-scale magnetic reconnection \citep{ScepiBegelmanDexter2024,HakobyanRipperdaPhilippov2023}. \cite{KrauthDevelaar2026} also discusses the potential for reconnection-driven loss of horizon-threading field to contribute to X-ray emission. Our simulations showcase the complex magnetic state changes that might produce unique X-ray signatures in this way. 

In addition to reconnection-driven X-ray flux contributions, we find a pronounced X-ray feature from the self-intersection shock in the spectra plotted in Fig.~\ref{fig:spectraprod16}. Because this feature is unique to the immediate post-merger disk, it may offer a unique observable transient with well-defined spectral evolution, though further study is necessary to determine its practical observability and diagnostic potential.

Finally, our unification of the magnetic influence on binary accretion during inspiral and merger reveals the unique variability evolution tied to large-scale magnetization (see Fig.~\ref{fig:XrayUVvstime}). Our well-controlled variation of the global field magnitude across runs reveals distinct paths for variability evolution through the merger. The ratio of fractional RMS variability measured during inspiral and about seven hours after merger (for our $10^6\rm M_\cdot$ binary) varies strongly with magnetization. Two windows of high-cadence monitoring could capture this behavior, while allowing for significant uncertainty of the time of merger. 

\textit{A priori}, the magnetization intrinsic to a merging binary candidate would be unknown, but the combination of observed band-specific variability matching the predicted simultaneous spectral evolution for a specific intrinsic magnetization, holds great promise in positively identifying a SMBHB merger. The potential, in our unified model, to match well-defined evolutionary tracks, provides a new way to determine the intrinsic magnetization of AGN disks hosting binaries.

\section{Conclusions}
\label{sec:conclusions}

We have presented a suite of 2D simulations of accretion onto equal-mass, coplanar, prograde SMBHBs inspiraling and merging via the emission of GWs. We have used these simulations to explore the  physical evolution of these systems, and to predict the EM counterparts associated with their merger.  

With relatively simple, but physically motivated prescriptions for viscosity and cooling, a potential that mimics relativistic apsidal precession, and a proxy field accounting for 3D large-scale magnetic field evolution, 
we successfully reproduce much of the behavior described in existing 3D-MHD and even 3D-GRMHD studies.

Comparing simulations run with our novel prescriptions to traditional approaches, we find qualitatively different outcomes, revealing specific sensitivities to both numerical and physical parameters. We have identified the reasons for the apparent tension between predictions of prior 2D-hydro and 3D-GRMHD studies. Numerical choices for disk thermodynamics and viscosity contribute, but the largest differences are due to the influence of relativistic precession and large-scale magnetic fields.

\textbf{Our conclusions related to the influence of large-scale magnetic fields can be summarized as follows:}
\begin{itemize}
    \item Most importantly, we find that varying the large-scale magnetic flux bridges the prior 2D and 3D outcomes into a unified framework, resolving their apparent differences, and leading to new predictions of observational consequence. 
\begin{itemize}
    \item Weakly organized large-scale fields result in a SANE accretion flow and match purely hydrodynamic simulations.
    \item Strong large-scale magnetic fields result in both mini-disks and even the CBD becoming MAD and matches 3D-GRMHD behavior. 
    \item  Intermediate large-scale fields result in a `mini-MAD' regime where only the minidisks and cavity are significantly disturbed by the large-scale magnetic flux. After merger, the total magnetic flux determines if, and how long, the remnant's disk remains MAD.
\end{itemize}
    \item This unification suggests an intriguing possibility -- the identification of an SMBHB with a well-sampled light curve, as well as spectra in X-ray and far UV bands, could be used to constrain the intrinsic magnetization of its accretion flow. 
\end{itemize}

\textbf{A summary of our additional conclusions related to precession effects and our other prescriptions are as follows:}  
\begin{itemize}
    \item A gravitational potential that includes apsidal precession results in a less pronounced drop in accretion rate during decoupling. 
    \item In SANE or mini-MAD binaries, differential precession in the eccentric post-merger inner disk results in a self-intersection spiral shock that can cause a second dip in the accretion rate. It contributes to rapid recovery of the accretion rate and luminosity by directly dissipating energy and helping to refill the cavity. This newly discovered behavior is predicted to have a unique spectral feature that might be observable. 
    \item Prior 2D studies may underestimate viscosity in the cavity. 3D-MHD simulations suggest Maxwell stress there remains high. A 2D viscosity accounting for the Maxwell stress component results in less significant decoupling and faster post-merger recovery, more consistent with 3D-GRMHD. 
    \item The different pre- and post-merger fractional variability in UV and X-rays could provide an additional signature of merging binaries and help reveal the magnetization intrinsic to an AGN disk. 
    \item Slower (target-entropy, Noble-type) cooling leads to a less distinct cavity edge than faster (isothermal or black-body) cooling in 2D simulations. The shortened local effective cooling timescale in compressed regions in the latter case alters the interaction of the binary and CBD, resulting in a larger, cooler cavity at merger, and slower post-merger recovery. 
    \item The merger response is sensitive to the nature of the over-dense lump and effective stress transporting angular momentum in the inner edge of the CBD. Suppressed viscosity in the inner CBD edge results in faster decoupling and the flat accretion rate profile through merger seen in the 3D-MHD simulation of LE25. 
    \item The conflicting predictions for merger response in LK23 and LE25 is fully explained, quantifiably, in this study, by their differing numerical prescriptions for viscosity and thermodynamics, and the absence of precession and large-scale magnetic field effects in LK23.
\end{itemize}

Because we are able to account for behavior previously unique to 3D-GRMHD simulations, but at a fraction of the cost, our framework will enable coplanar binary parameter surveys of mass ratio, eccentricity, disk thermodynamics, and magnetization, that would otherwise be prohibitively expensive. AGN in the sub-Eddington radiatively efficient regime are expected to have much thinner disks (i.e. have Mach numbers greater than 50), and have much longer viscous timescales than current 3D-GRMHD studies can probe. Our framework makes the magnetized version of this regime accessible for the first time. 

For many outstanding problems in the astrophysics of accreting binaries, 3D-GRMHD is necessary, and our framework can aid in selecting the most important regions of parameter space, worth dedicating valuable computational resources.

\section*{Acknowledgments}

The authors wish to thank Dan D'Orazio, Jonathan Zrake, Andrew MacFadyen, Alex Dittmann, and especially Luke Krauth for enlightening discussions. This research was supported in part by grant NSF PHY-2309135 to the Kavli Institute for Theoretical Physics (KITP), through the 2022 ``Bridging the Gap: Accretion and Orbital Evolution in Stellar and Black Hole Binaries" workshop. This work is supported by ERC grant (Bright-BHBs, grant agreement no.101200084). Views and opinions expressed are however those of the author(s) only and do not necessarily reflect those of the European Union or the European Research Council Executive Agency. Neither the European Union nor the granting authority can be held responsible for them.

This work made use of the following software packages:  \texttt{Athena++} \citep{Stone2020}, \texttt{Jupyter} \citep{2007CSE.....9c..21P,kluyver2016jupyter}, \texttt{matplotlib} \citep{Hunter:2007}, \texttt{numpy} \citep{numpy}, \texttt{pandas} \citep{mckinney-proc-scipy-2010}
, \texttt{python} \citep{python}, \texttt{scipy} \citep{2020SciPy-NMeth}
, \texttt{Cython} \citep{cython:2011}, and \texttt{h5py} \citep{collette_python_hdf5_2014}.

Software citation information aggregated using \texttt{\href{https://www.tomwagg.com/software-citation-station/}{The Software Citation Station}} \citep{software-citation-station-paper,software-citation-station-zenodo}.

\bibliographystyle{mnras}
\bibliography{full_clean,mad}

\begin{thebibliography}{}
\makeatletter
\relax
\def\mn@urlcharsother{\let\do\@makeother \do\$\do\&\do\#\do\^\do\_\do\%\do\~}
\def\mn@doi{\begingroup\mn@urlcharsother \@ifnextchar [ {\mn@doi@} {\mn@doi@[]}}
\def\mn@doi@[#1]#2{\def\@tempa{#1}\ifx\@tempa\@empty \href {http://dx.doi.org/#2} {doi:#2}\else \href {http://dx.doi.org/#2} {#1}\fi \endgroup}
\def\mn@eprint#1#2{\mn@eprint@#1:#2::\@nil}
\def\mn@eprint@arXiv#1{\href {http://arxiv.org/abs/#1} {{\tt arXiv:#1}}}
\def\mn@eprint@dblp#1{\href {http://dblp.uni-trier.de/rec/bibtex/#1.xml} {dblp:#1}}
\def\mn@eprint@#1:#2:#3:#4\@nil{\def\@tempa {#1}\def\@tempb {#2}\def\@tempc {#3}\ifx \@tempc \@empty \let \@tempc \@tempb \let \@tempb \@tempa \fi \ifx \@tempb \@empty \def\@tempb {arXiv}\fi \@ifundefined {mn@eprint@\@tempb}{\@tempb:\@tempc}{\expandafter \expandafter \csname mn@eprint@\@tempb\endcsname \expandafter{\@tempc}}}

\bibitem[\protect\citeauthoryear{Abramowicz}{Abramowicz}{2009}]{Abramowicz2009}
Abramowicz M.~A.,  2009, Astronomy \& Astrophysics, 500, 213

\bibitem[\protect\citeauthoryear{{Artymowicz} \& {Lubow}}{{Artymowicz} \& {Lubow}}{1994}]{ArtymowiczLubow1994}
{Artymowicz} P.,  {Lubow} S.~H.,  1994, \mn@doi [\apj] {10.1086/173679}, \href {https://ui.adsabs.harvard.edu/abs/1994ApJ...421..651A} {421, 651}

\bibitem[\protect\citeauthoryear{{Artymowicz} \& {Lubow}}{{Artymowicz} \& {Lubow}}{1996}]{ArtymowiczLubow1996}
{Artymowicz} P.,  {Lubow} S.~H.,  1996, \mn@doi [\apjl] {10.1086/310200}, \href {https://ui.adsabs.harvard.edu/abs/1996ApJ...467L..77A} {467, L77}

\bibitem[\protect\citeauthoryear{{Avara}, {McKinney}  \& {Reynolds}}{{Avara} et~al.}{2016}]{AvaraMcKinney2016}
{Avara} M.~J.,  {McKinney} J.~C.,   {Reynolds} C.~S.,  2016, \mn@doi [\mnras] {10.1093/mnras/stw1643}, \href {https://ui.adsabs.harvard.edu/abs/2016MNRAS.462..636A} {462, 636}

\bibitem[\protect\citeauthoryear{{Avara}, {Krolik}, {Campanelli}, {Noble}, {Bowen}  \& {Ryu}}{{Avara} et~al.}{2024}]{AvaraKrolik2024}
{Avara} M.~J.,  {Krolik} J.~H.,  {Campanelli} M.,  {Noble} S.~C.,  {Bowen} D.,   {Ryu} T.,  2024, \mn@doi [\apj] {10.3847/1538-4357/ad5bda}, \href {https://ui.adsabs.harvard.edu/abs/2024ApJ...974..242A} {974, 242}

\bibitem[\protect\citeauthoryear{{Baker} et~al.,}{{Baker} et~al.}{2019}]{Baker2019Decadal2019}
{Baker} J.,  et~al., 2019, \mn@doi [\baas] {10.48550/arXiv.1903.04417}, \href {https://ui.adsabs.harvard.edu/abs/2019BAAS...51c.123B} {51, 123}

\bibitem[\protect\citeauthoryear{{Balbus} \& {Hawley}}{{Balbus} \& {Hawley}}{1991}]{1991ApJ...376..214B}
{Balbus} S.~A.,  {Hawley} J.~F.,  1991, \mn@doi [\apj] {10.1086/170270}, \href {https://ui.adsabs.harvard.edu/abs/1991ApJ...376..214B} {376, 214}

\bibitem[\protect\citeauthoryear{Barker \& Ogilvie}{Barker \& Ogilvie}{2016}]{BarkerOgilvie2016}
Barker A.~J.,  Ogilvie G.~I.,  2016, \mn@doi [Monthly Notices of the Royal Astronomical Society] {10.1093/mnras/stw580}, 458, 3739

\bibitem[\protect\citeauthoryear{{Begelman}, {Blandford}  \& {Rees}}{{Begelman} et~al.}{1980}]{BegelmanBlandford1980}
{Begelman} M.~C.,  {Blandford} R.~D.,   {Rees} M.~J.,  1980, \mn@doi [\nat] {10.1038/287307a0}, \href {https://ui.adsabs.harvard.edu/abs/1980Natur.287..307B} {287, 307}

\bibitem[\protect\citeauthoryear{{Begelman}, {Scepi}  \& {Dexter}}{{Begelman} et~al.}{2022}]{BegelmanScepi2022}
{Begelman} M.~C.,  {Scepi} N.,   {Dexter} J.,  2022, \mn@doi [\mnras] {10.1093/mnras/stab3790}, \href {https://ui.adsabs.harvard.edu/abs/2022MNRAS.511.2040B} {511, 2040}

\bibitem[\protect\citeauthoryear{Behnel, Bradshaw, Citro, Dalcin, Seljebotn  \& Smith}{Behnel et~al.}{2011}]{cython:2011}
Behnel S.,  Bradshaw R.,  Citro C.,  Dalcin L.,  Seljebotn D.~S.,   Smith K.,  2011, \mn@doi [Computing in Science Engineering] {10.1109/MCSE.2010.118}, 13, 31

\bibitem[\protect\citeauthoryear{{Ben Arosh Arad} \& {Sari}}{{Ben Arosh Arad} \& {Sari}}{2026}]{AradSari2026}
{Ben Arosh Arad} I.,  {Sari} R.,  2026, \mn@doi [\apj] {10.3847/1538-4357/ae433f}, \href {https://ui.adsabs.harvard.edu/abs/2026ApJ...999..254B} {999, 254}

\bibitem[\protect\citeauthoryear{{Bisnovatyi-Kogan} \& {Ruzmaikin}}{{Bisnovatyi-Kogan} \& {Ruzmaikin}}{1974}]{1974Ap&SS..28...45B}
{Bisnovatyi-Kogan} G.~S.,  {Ruzmaikin} A.~A.,  1974, \mn@doi [\apss] {10.1007/BF00642237}, \href {https://ui.adsabs.harvard.edu/abs/1974Ap&SS..28...45B} {28, 45}

\bibitem[\protect\citeauthoryear{{Blandford} \& {Payne}}{{Blandford} \& {Payne}}{1982}]{1982MNRAS.199..883B}
{Blandford} R.~D.,  {Payne} D.~G.,  1982, \mn@doi [\mnras] {10.1093/mnras/199.4.883}, \href {https://ui.adsabs.harvard.edu/abs/1982MNRAS.199..883B} {199, 883}

\bibitem[\protect\citeauthoryear{{Blandford} \& {Znajek}}{{Blandford} \& {Znajek}}{1977}]{1977MNRAS.179..433B}
{Blandford} R.~D.,  {Znajek} R.~L.,  1977, \mn@doi [\mnras] {10.1093/mnras/179.3.433}, \href {https://ui.adsabs.harvard.edu/abs/1977MNRAS.179..433B} {179, 433}

\bibitem[\protect\citeauthoryear{{Bogdanovi{\'c}}, {Miller}  \& {Blecha}}{{Bogdanovi{\'c}} et~al.}{2022}]{BogdanovicMiller2022}
{Bogdanovi{\'c}} T.,  {Miller} M.~C.,   {Blecha} L.,  2022, \mn@doi [Living Reviews in Relativity] {10.1007/s41114-022-00037-8}, \href {https://ui.adsabs.harvard.edu/abs/2022LRR....25....3B} {25, 3}

\bibitem[\protect\citeauthoryear{{Bowen}, {Campanelli}, {Krolik}, {Mewes}  \& {Noble}}{{Bowen} et~al.}{2017}]{BowenCampanelli2017}
{Bowen} D.~B.,  {Campanelli} M.,  {Krolik} J.~H.,  {Mewes} V.,   {Noble} S.~C.,  2017, \mn@doi [\apj] {10.3847/1538-4357/aa63f3}, \href {https://ui.adsabs.harvard.edu/abs/2017ApJ...838...42B} {838, 42}

\bibitem[\protect\citeauthoryear{{Bowen}, {Mewes}, {Campanelli}, {Noble}, {Krolik}  \& {Zilh{\~a}o}}{{Bowen} et~al.}{2018}]{BowenMewes2018}
{Bowen} D.~B.,  {Mewes} V.,  {Campanelli} M.,  {Noble} S.~C.,  {Krolik} J.~H.,   {Zilh{\~a}o} M.,  2018, \mn@doi [\apjl] {10.3847/2041-8213/aaa756}, \href {https://ui.adsabs.harvard.edu/abs/2018ApJ...853L..17B} {853, L17}

\bibitem[\protect\citeauthoryear{{Bowen}, {Mewes}, {Noble}, {Avara}, {Campanelli}  \& {Krolik}}{{Bowen} et~al.}{2019}]{BowenMewes2019}
{Bowen} D.~B.,  {Mewes} V.,  {Noble} S.~C.,  {Avara} M.,  {Campanelli} M.,   {Krolik} J.~H.,  2019, \mn@doi [\apj] {10.3847/1538-4357/ab2453}, \href {https://ui.adsabs.harvard.edu/abs/2019ApJ...879...76B} {879, 76}

\bibitem[\protect\citeauthoryear{{Cattorini} \& {Giacomazzo}}{{Cattorini} \& {Giacomazzo}}{2024}]{CattoriniGiacomazzo2024}
{Cattorini} F.,  {Giacomazzo} B.,  2024, \mn@doi [Astroparticle Physics] {10.1016/j.astropartphys.2023.102892}, \href {https://ui.adsabs.harvard.edu/abs/2024APh...15402892C} {154, 102892}

\bibitem[\protect\citeauthoryear{{Chan}, {Tiwari}, {Bogdanovi{\'c}}, {Jiang}  \& {Davis}}{{Chan} et~al.}{2025}]{ChanTiwariBogdanovic2025}
{Chan} C.-H.,  {Tiwari} V.,  {Bogdanovi{\'c}} T.,  {Jiang} Y.-F.,   {Davis} S.~W.,  2025, \mn@doi [\apj] {10.3847/1538-4357/adf4c9}, \href {https://ui.adsabs.harvard.edu/abs/2025ApJ...991...71C} {991, 71}

\bibitem[\protect\citeauthoryear{{Clyburn} \& {Zrake}}{{Clyburn} \& {Zrake}}{2025}]{ClyburnZrake2025}
{Clyburn} M.,  {Zrake} J.,  2025, \mn@doi [\mnras] {10.1093/mnras/staf585}, \href {https://ui.adsabs.harvard.edu/abs/2025MNRAS.539.1430C} {539, 1430}

\bibitem[\protect\citeauthoryear{Collette}{Collette}{2013}]{collette_python_hdf5_2014}
Collette A.,  2013, Python and HDF5.
O'Reilly

\bibitem[\protect\citeauthoryear{{Colpi} et~al.,}{{Colpi} et~al.}{2024}]{LISAColpi2024}
{Colpi} M.,  et~al., 2024, \mn@doi [arXiv e-prints] {10.48550/arXiv.2402.07571}, \href {https://ui.adsabs.harvard.edu/abs/2024arXiv240207571C} {p. arXiv:2402.07571}

\bibitem[\protect\citeauthoryear{{Combi}, {Lopez Armengol}, {Campanelli}, {Noble}, {Avara}, {Krolik}  \& {Bowen}}{{Combi} et~al.}{2022}]{CombiLopezArmengol2022}
{Combi} L.,  {Lopez Armengol} F.~G.,  {Campanelli} M.,  {Noble} S.~C.,  {Avara} M.,  {Krolik} J.~H.,   {Bowen} D.,  2022, \mn@doi [\apj] {10.3847/1538-4357/ac532a}, \href {https://ui.adsabs.harvard.edu/abs/2022ApJ...928..187C} {928, 187}

\bibitem[\protect\citeauthoryear{{Corrales}, {Haiman}  \& {MacFadyen}}{{Corrales} et~al.}{2010}]{CorralesHaiman2010}
{Corrales} L.~R.,  {Haiman} Z.,   {MacFadyen} A.,  2010, \mn@doi [\mnras] {10.1111/j.1365-2966.2010.16324.x}, \href {https://ui.adsabs.harvard.edu/abs/2010MNRAS.404..947C} {404, 947}

\bibitem[\protect\citeauthoryear{{D'Orazio} \& {Charisi}}{{D'Orazio} \& {Charisi}}{2023}]{DOrazioCharisi2023}
{D'Orazio} D.~J.,  {Charisi} M.,  2023, \mn@doi [arXiv e-prints] {10.48550/arXiv.2310.16896}, \href {https://ui.adsabs.harvard.edu/abs/2023arXiv231016896D} {p. arXiv:2310.16896}

\bibitem[\protect\citeauthoryear{{D'Orazio}, {Haiman}  \& {MacFadyen}}{{D'Orazio} et~al.}{2013}]{DOrazioHaiman2013}
{D'Orazio} D.~J.,  {Haiman} Z.,   {MacFadyen} A.,  2013, \mn@doi [\mnras] {10.1093/mnras/stt1787}, \href {https://ui.adsabs.harvard.edu/abs/2013MNRAS.436.2997D} {436, 2997}

\bibitem[\protect\citeauthoryear{{Dempsey}, {Mu{\~n}oz}  \& {Lithwick}}{{Dempsey} et~al.}{2020}]{DempseyMunoz2020}
{Dempsey} A.~M.,  {Mu{\~n}oz} D.,   {Lithwick} Y.,  2020, \mn@doi [\apjl] {10.3847/2041-8213/ab800e}, \href {https://ui.adsabs.harvard.edu/abs/2020ApJ...892L..29D} {892, L29}

\bibitem[\protect\citeauthoryear{{Dittmann} \& {Ryan}}{{Dittmann} \& {Ryan}}{2021}]{DittmannRyan2021}
{Dittmann} A.~J.,  {Ryan} G.,  2021, \mn@doi [\apj] {10.3847/1538-4357/ac1bbd}, \href {https://ui.adsabs.harvard.edu/abs/2021ApJ...921...71D} {921, 71}

\bibitem[\protect\citeauthoryear{{Dittmann} \& {Ryan}}{{Dittmann} \& {Ryan}}{2022}]{DittmannRyan2022}
{Dittmann} A.~J.,  {Ryan} G.,  2022, \mn@doi [\mnras] {10.1093/mnras/stac935}, \href {https://ui.adsabs.harvard.edu/abs/2022MNRAS.513.6158D} {513, 6158}

\bibitem[\protect\citeauthoryear{{Dittmann}, {Ryan}  \& {Miller}}{{Dittmann} et~al.}{2023}]{DittmannRyan2023}
{Dittmann} A.~J.,  {Ryan} G.,   {Miller} M.~C.,  2023, \mn@doi [\apjl] {10.3847/2041-8213/acd183}, \href {https://ui.adsabs.harvard.edu/abs/2023ApJ...949L..30D} {949, L30}

\bibitem[\protect\citeauthoryear{{Duffell}, {D'Orazio}, {Derdzinski}, {Haiman}, {MacFadyen}, {Rosen}  \& {Zrake}}{{Duffell} et~al.}{2020}]{DuffellDOrazio2020}
{Duffell} P.~C.,  {D'Orazio} D.,  {Derdzinski} A.,  {Haiman} Z.,  {MacFadyen} A.,  {Rosen} A.~L.,   {Zrake} J.,  2020, \mn@doi [\apj] {10.3847/1538-4357/abab95}, \href {https://ui.adsabs.harvard.edu/abs/2020ApJ...901...25D} {901, 25}

\bibitem[\protect\citeauthoryear{{Duffell} et~al.,}{{Duffell} et~al.}{2024}]{DuffellDittmannetal2024}
{Duffell} P.~C.,  et~al., 2024, \mn@doi [\apj] {10.3847/1538-4357/ad5a7e}, \href {https://ui.adsabs.harvard.edu/abs/2024ApJ...970..156D} {970, 156}

\bibitem[\protect\citeauthoryear{{Ennoggi} et~al.,}{{Ennoggi} et~al.}{2025}]{EnnoggiCampanelli2025}
{Ennoggi} L.,  et~al., 2025, \mn@doi [\prd] {10.1103/yc25-v1q4}, \href {https://ui.adsabs.harvard.edu/abs/2025PhRvD.112f3009E} {112, 063009}

\bibitem[\protect\citeauthoryear{{Farris}, {Gold}, {Paschalidis}, {Etienne}  \& {Shapiro}}{{Farris} et~al.}{2012}]{FarrisGold2012}
{Farris} B.~D.,  {Gold} R.,  {Paschalidis} V.,  {Etienne} Z.~B.,   {Shapiro} S.~L.,  2012, \mn@doi [\prl] {10.1103/PhysRevLett.109.221102}, \href {https://ui.adsabs.harvard.edu/abs/2012PhRvL.109v1102F} {109, 221102}

\bibitem[\protect\citeauthoryear{{Farris}, {Duffell}, {MacFadyen}  \& {Haiman}}{{Farris} et~al.}{2014}]{FarrisDuffell2014}
{Farris} B.~D.,  {Duffell} P.,  {MacFadyen} A.~I.,   {Haiman} Z.,  2014, \mn@doi [\apj] {10.1088/0004-637X/783/2/134}, \href {https://ui.adsabs.harvard.edu/abs/2014ApJ...783..134F} {783, 134}

\bibitem[\protect\citeauthoryear{{Farris}, {Duffell}, {MacFadyen}  \& {Haiman}}{{Farris} et~al.}{2015a}]{FarrisDuffell2015a}
{Farris} B.~D.,  {Duffell} P.,  {MacFadyen} A.~I.,   {Haiman} Z.,  2015a, \mn@doi [\mnras] {10.1093/mnrasl/slu160}, \href {https://ui.adsabs.harvard.edu/abs/2015MNRAS.446L..36F} {446, L36}

\bibitem[\protect\citeauthoryear{{Farris}, {Duffell}, {MacFadyen}  \& {Haiman}}{{Farris} et~al.}{2015b}]{FarrisDuffell2015b}
{Farris} B.~D.,  {Duffell} P.,  {MacFadyen} A.~I.,   {Haiman} Z.,  2015b, \mn@doi [\mnras] {10.1093/mnrasl/slu184}, \href {https://ui.adsabs.harvard.edu/abs/2015MNRAS.447L..80F} {447, L80}

\bibitem[\protect\citeauthoryear{Fishbach, Holz  \& Farr}{Fishbach et~al.}{2017}]{FishbachHolzFarr2017}
Fishbach M.,  Holz D.~E.,   Farr B.,  2017, \mn@doi [The Astrophysical Journal Letters] {10.3847/2041-8213/aa7045}, 840, L24

\bibitem[\protect\citeauthoryear{{Franchini}, {Lupi}  \& {Sesana}}{{Franchini} et~al.}{2022}]{FranchiniLupiSesana2022}
{Franchini} A.,  {Lupi} A.,   {Sesana} A.,  2022, \mn@doi [\apjl] {10.3847/2041-8213/ac63a2}, \href {https://ui.adsabs.harvard.edu/abs/2022ApJ...929L..13F} {929, L13}

\bibitem[\protect\citeauthoryear{{Franchini}, {Lupi}, {Sesana}  \& {Haiman}}{{Franchini} et~al.}{2023}]{FranchiniLupiSesanaHaiman2023}
{Franchini} A.,  {Lupi} A.,  {Sesana} A.,   {Haiman} Z.,  2023, \mn@doi [\mnras] {10.1093/mnras/stad1070}, \href {https://ui.adsabs.harvard.edu/abs/2023MNRAS.522.1569F} {522, 1569}

\bibitem[\protect\citeauthoryear{{Franchini}, {Bonetti}, {Lupi}  \& {Sesana}}{{Franchini} et~al.}{2024a}]{FranchiniBonetti2024}
{Franchini} A.,  {Bonetti} M.,  {Lupi} A.,   {Sesana} A.,  2024a, \mn@doi [\aap] {10.1051/0004-6361/202449206}, \href {https://ui.adsabs.harvard.edu/abs/2024A&A...686A.288F} {686, A288}

\bibitem[\protect\citeauthoryear{{Franchini}, {Prato}, {Longarini}  \& {Sesana}}{{Franchini} et~al.}{2024b}]{FranchiniPrato2024}
{Franchini} A.,  {Prato} A.,  {Longarini} C.,   {Sesana} A.,  2024b, \mn@doi [\aap] {10.1051/0004-6361/202449402}, \href {https://ui.adsabs.harvard.edu/abs/2024A&A...688A.174F} {688, A174}

\bibitem[\protect\citeauthoryear{{Gammie}}{{Gammie}}{1999}]{1999ApJ...522L..57G}
{Gammie} C.~F.,  1999, \mn@doi [\apjl] {10.1086/312207}, \href {https://ui.adsabs.harvard.edu/abs/1999ApJ...522L..57G} {522, L57}

\bibitem[\protect\citeauthoryear{{Giacomazzo}, {Baker}, {Miller}, {Reynolds}  \& {van Meter}}{{Giacomazzo} et~al.}{2012}]{GiacomazzoBaker2012}
{Giacomazzo} B.,  {Baker} J.~G.,  {Miller} M.~C.,  {Reynolds} C.~S.,   {van Meter} J.~R.,  2012, \mn@doi [\apjl] {10.1088/2041-8205/752/1/L15}, \href {https://ui.adsabs.harvard.edu/abs/2012ApJ...752L..15G} {752, L15}

\bibitem[\protect\citeauthoryear{{Gold}}{{Gold}}{2019a}]{Gold2019}
{Gold} R.,  2019a, \mn@doi [Galaxies] {10.3390/galaxies7020063}, \href {https://ui.adsabs.harvard.edu/abs/2019Galax...7...63G} {7, 63}

\bibitem[\protect\citeauthoryear{{Gold}}{{Gold}}{2019b}]{Gold2019review}
{Gold} R.,  2019b, \mn@doi [Galaxies] {10.3390/galaxies7020063}, \href {https://ui.adsabs.harvard.edu/abs/2019Galax...7...63G} {7, 63}

\bibitem[\protect\citeauthoryear{{Haiman}}{{Haiman}}{2017}]{Haiman2017}
{Haiman} Z.,  2017, \mn@doi [\prd] {10.1103/PhysRevD.96.023004}, \href {https://ui.adsabs.harvard.edu/abs/2017PhRvD..96b3004H} {96, 023004}

\bibitem[\protect\citeauthoryear{{Haiman} et~al.,}{{Haiman} et~al.}{2023}]{HaimanXin2023}
{Haiman} Z.,  et~al., 2023, \mn@doi [arXiv e-prints] {10.48550/arXiv.2306.14990}, \href {https://ui.adsabs.harvard.edu/abs/2023arXiv230614990H} {p. arXiv:2306.14990}

\bibitem[\protect\citeauthoryear{{Hakobyan}, {Ripperda}  \& {Philippov}}{{Hakobyan} et~al.}{2023}]{HakobyanRipperdaPhilippov2023}
{Hakobyan} H.,  {Ripperda} B.,   {Philippov} A.~A.,  2023, \mn@doi [\apjl] {10.3847/2041-8213/acb264}, \href {https://ui.adsabs.harvard.edu/abs/2023ApJ...943L..29H} {943, L29}

\bibitem[\protect\citeauthoryear{Harris et~al.,}{Harris et~al.}{2020}]{numpy}
Harris C.~R.,  et~al., 2020, \mn@doi [Nature] {10.1038/s41586-020-2649-2}, 585, 357

\bibitem[\protect\citeauthoryear{{Hayasaki}, {Mineshige}  \& {Sudou}}{{Hayasaki} et~al.}{2007}]{HayasakiMineshige2007}
{Hayasaki} K.,  {Mineshige} S.,   {Sudou} H.,  2007, \mn@doi [\pasj] {10.1093/pasj/59.2.427}, \href {https://ui.adsabs.harvard.edu/abs/2007PASJ...59..427H} {59, 427}

\bibitem[\protect\citeauthoryear{{Hogg} \& {Reynolds}}{{Hogg} \& {Reynolds}}{2016}]{HoggReynolds2016}
{Hogg} J.~D.,  {Reynolds} C.~S.,  2016, \mn@doi [\apj] {10.3847/0004-637X/826/1/40}, \href {https://ui.adsabs.harvard.edu/abs/2016ApJ...826...40H} {826, 40}

\bibitem[\protect\citeauthoryear{Hunter}{Hunter}{2007}]{Hunter:2007}
Hunter J.~D.,  2007, \mn@doi [Computing in Science \& Engineering] {10.1109/MCSE.2007.55}, 9, 90

\bibitem[\protect\citeauthoryear{{Igumenshchev}}{{Igumenshchev}}{2008}]{2008ApJ...677..317I}
{Igumenshchev} I.~V.,  2008, \mn@doi [\apj] {10.1086/529025}, \href {https://ui.adsabs.harvard.edu/abs/2008ApJ...677..317I} {677, 317}

\bibitem[\protect\citeauthoryear{{Kauffmann} \& {Haehnelt}}{{Kauffmann} \& {Haehnelt}}{2000}]{KauffmannHaehnelt2000}
{Kauffmann} G.,  {Haehnelt} M.,  2000, \mn@doi [\mnras] {10.1046/j.1365-8711.2000.03077.x}, \href {https://ui.adsabs.harvard.edu/abs/2000MNRAS.311..576K} {311, 576}

\bibitem[\protect\citeauthoryear{{Kelly}, {Etienne}, {Golomb}, {Schnittman}, {Baker}, {Noble}  \& {Ryan}}{{Kelly} et~al.}{2021}]{KellyEtienne2021}
{Kelly} B.~J.,  {Etienne} Z.~B.,  {Golomb} J.,  {Schnittman} J.~D.,  {Baker} J.~G.,  {Noble} S.~C.,   {Ryan} G.,  2021, \mn@doi [\prd] {10.1103/PhysRevD.103.063039}, \href {https://ui.adsabs.harvard.edu/abs/2021PhRvD.103f3039K} {103, 063039}

\bibitem[\protect\citeauthoryear{{Klein} et~al.,}{{Klein} et~al.}{2016}]{KleinBarausse2016}
{Klein} A.,  et~al., 2016, \mn@doi [\prd] {10.1103/PhysRevD.93.024003}, \href {https://ui.adsabs.harvard.edu/abs/2016PhRvD..93b4003K} {93, 024003}

\bibitem[\protect\citeauthoryear{Kluyver et~al.,}{Kluyver et~al.}{2016}]{kluyver2016jupyter}
Kluyver T.,  et~al., 2016, in Loizides F.,  Schmidt B.,  eds, Positioning and Power in Academic Publishing: Players, Agents and Agendas. pp 87--90

\bibitem[\protect\citeauthoryear{{Kocsis}, {Haiman}  \& {Menou}}{{Kocsis} et~al.}{2008}]{KocsisHaiman2008}
{Kocsis} B.,  {Haiman} Z.,   {Menou} K.,  2008, \mn@doi [\apj] {10.1086/590230}, \href {https://ui.adsabs.harvard.edu/abs/2008ApJ...684..870K} {684, 870}

\bibitem[\protect\citeauthoryear{{Kormendy} \& {Ho}}{{Kormendy} \& {Ho}}{2013}]{KormendyHo2013}
{Kormendy} J.,  {Ho} L.~C.,  2013, \mn@doi [\araa] {10.1146/annurev-astro-082708-101811}, \href {https://ui.adsabs.harvard.edu/abs/2013ARA&A..51..511K} {51, 511}

\bibitem[\protect\citeauthoryear{{Kormendy} \& {Richstone}}{{Kormendy} \& {Richstone}}{1995}]{KormendyRichstone1995}
{Kormendy} J.,  {Richstone} D.,  1995, \mn@doi [\araa] {10.1146/annurev.aa.33.090195.003053}, \href {https://ui.adsabs.harvard.edu/abs/1995ARA&A..33..581K} {33, 581}

\bibitem[\protect\citeauthoryear{{Krauth} \& {Davelaar}}{{Krauth} \& {Davelaar}}{2026}]{KrauthDevelaar2026}
{Krauth} L.,  {Davelaar} J.,  2026, \mn@doi [arXiv e-prints] {10.48550/arXiv.2602.11112}, \href {https://ui.adsabs.harvard.edu/abs/2026arXiv260211112K} {p. arXiv:2602.11112}

\bibitem[\protect\citeauthoryear{{Krauth}, {Davelaar}, {Haiman}, {Westernacher-Schneider}, {Zrake}  \& {MacFadyen}}{{Krauth} et~al.}{2023}]{KrauthDavelaar2023}
{Krauth} L.~M.,  {Davelaar} J.,  {Haiman} Z.,  {Westernacher-Schneider} J.~R.,  {Zrake} J.,   {MacFadyen} A.,  2023, \mn@doi [\mnras] {10.1093/mnras/stad3095}, \href {https://ui.adsabs.harvard.edu/abs/2023MNRAS.526.5441K} {526, 5441}

\bibitem[\protect\citeauthoryear{{Lai} \& {Mu{\~n}oz}}{{Lai} \& {Mu{\~n}oz}}{2023}]{LaiMunoz2023}
{Lai} D.,  {Mu{\~n}oz} D.~J.,  2023, \mn@doi [\araa] {10.1146/annurev-astro-052622-022933}, \href {https://ui.adsabs.harvard.edu/abs/2023ARA&A..61..517L} {61, 517}

\bibitem[\protect\citeauthoryear{{Lin} \& {Papaloizou}}{{Lin} \& {Papaloizou}}{1979}]{LinPapaloizou1979}
{Lin} D.~N.~C.,  {Papaloizou} J.,  1979, \mn@doi [\mnras] {10.1093/mnras/188.2.191}, \href {https://ui.adsabs.harvard.edu/abs/1979MNRAS.188..191L} {188, 191}

\bibitem[\protect\citeauthoryear{{Lopez Armengol} et~al.,}{{Lopez Armengol} et~al.}{2021}]{LopezArmengolCombi2021}
{Lopez Armengol} F.~G.,  et~al., 2021, \mn@doi [\apj] {10.3847/1538-4357/abf0af}, \href {https://ui.adsabs.harvard.edu/abs/2021ApJ...913...16L} {913, 16}

\bibitem[\protect\citeauthoryear{{Lyubarskii}}{{Lyubarskii}}{1997}]{Lyubarskii1997}
{Lyubarskii} Y.~E.,  1997, \mn@doi [\mnras] {10.1093/mnras/292.3.679}, \href {https://ui.adsabs.harvard.edu/abs/1997MNRAS.292..679L} {292, 679}

\bibitem[\protect\citeauthoryear{{MacFadyen} \& {Milosavljevi{\'c}}}{{MacFadyen} \& {Milosavljevi{\'c}}}{2008}]{MacFadyenMilosavljevic2008}
{MacFadyen} A.~I.,  {Milosavljevi{\'c}} M.,  2008, \mn@doi [\apj] {10.1086/523869}, \href {https://ui.adsabs.harvard.edu/abs/2008ApJ...672...83M} {672, 83}

\bibitem[\protect\citeauthoryear{{Mangiagli}, {Caprini}, {Volonteri}, {Marsat}, {Vergani}, {Tamanini}  \& {Inchausp{\'e}}}{{Mangiagli} et~al.}{2022}]{MangiagliCaprini2022}
{Mangiagli} A.,  {Caprini} C.,  {Volonteri} M.,  {Marsat} S.,  {Vergani} S.,  {Tamanini} N.,   {Inchausp{\'e}} H.,  2022, \mn@doi [\prd] {10.1103/PhysRevD.106.103017}, \href {https://ui.adsabs.harvard.edu/abs/2022PhRvD.106j3017M} {106, 103017}

\bibitem[\protect\citeauthoryear{{Manikantan} \& {Paschalidis}}{{Manikantan} \& {Paschalidis}}{2025}]{2025PhRvD.112j3050M}
{Manikantan} V.,  {Paschalidis} V.,  2025, \mn@doi [\prd] {10.1103/g98b-m8m1}, \href {https://ui.adsabs.harvard.edu/abs/2025PhRvD.112j3050M} {112, 103050}

\bibitem[\protect\citeauthoryear{{McKinney}, {Tchekhovskoy}  \& {Blandford}}{{McKinney} et~al.}{2012}]{2012MNRAS.423.3083M}
{McKinney} J.~C.,  {Tchekhovskoy} A.,   {Blandford} R.~D.,  2012, \mn@doi [\mnras] {10.1111/j.1365-2966.2012.21074.x}, \href {https://ui.adsabs.harvard.edu/abs/2012MNRAS.423.3083M} {423, 3083}

\bibitem[\protect\citeauthoryear{{Menou}, {Haiman}  \& {Narayanan}}{{Menou} et~al.}{2001}]{MenouHaiman2001}
{Menou} K.,  {Haiman} Z.,   {Narayanan} V.~K.,  2001, \mn@doi [\apj] {10.1086/322310}, \href {https://ui.adsabs.harvard.edu/abs/2001ApJ...558..535M} {558, 535}

\bibitem[\protect\citeauthoryear{{Milosavljevi{\'c}} \& {Merritt}}{{Milosavljevi{\'c}} \& {Merritt}}{2003}]{MilosavljevicMerritt2003}
{Milosavljevi{\'c}} M.,  {Merritt} D.,  2003, in {Centrella} J.~M.,  ed.,  American Institute of Physics Conference Series Vol. 686, The Astrophysics of Gravitational Wave Sources. AIP, pp 201--210 (\mn@eprint {arXiv} {astro-ph/0212270}), \mn@doi{10.1063/1.1629432}

\bibitem[\protect\citeauthoryear{{Miranda}, {Mu{\~n}oz}  \& {Lai}}{{Miranda} et~al.}{2017}]{MirandaMunoz2017}
{Miranda} R.,  {Mu{\~n}oz} D.~J.,   {Lai} D.,  2017, \mn@doi [\mnras] {10.1093/mnras/stw3189}, \href {https://ui.adsabs.harvard.edu/abs/2017MNRAS.466.1170M} {466, 1170}

\bibitem[\protect\citeauthoryear{{Moody}, {Shi}  \& {Stone}}{{Moody} et~al.}{2019}]{MoodyShi2019}
{Moody} M. S.~L.,  {Shi} J.-M.,   {Stone} J.~M.,  2019, \mn@doi [\apj] {10.3847/1538-4357/ab09ee}, \href {https://ui.adsabs.harvard.edu/abs/2019ApJ...875...66M} {875, 66}

\bibitem[\protect\citeauthoryear{{Most} \& {Wang}}{{Most} \& {Wang}}{2024}]{MostWang2024}
{Most} E.~R.,  {Wang} H.-Y.,  2024, \mn@doi [\apjl] {10.3847/2041-8213/ad7713}, \href {https://ui.adsabs.harvard.edu/abs/2024ApJ...973L..19M} {973, L19}

\bibitem[\protect\citeauthoryear{{Most} \& {Wang}}{{Most} \& {Wang}}{2025}]{MostWang2025}
{Most} E.~R.,  {Wang} H.-Y.,  2025, \mn@doi [\prd] {10.1103/PhysRevD.111.L081304}, \href {https://ui.adsabs.harvard.edu/abs/2025PhRvD.111h1304M} {111, L081304}

\bibitem[\protect\citeauthoryear{{Mu{\~n}oz}, {Miranda}  \& {Lai}}{{Mu{\~n}oz} et~al.}{2019}]{MunozMiranda2019}
{Mu{\~n}oz} D.~J.,  {Miranda} R.,   {Lai} D.,  2019, \mn@doi [\apj] {10.3847/1538-4357/aaf867}, \href {https://ui.adsabs.harvard.edu/abs/2019ApJ...871...84M} {871, 84}

\bibitem[\protect\citeauthoryear{{Narayan}, {Igumenshchev}  \& {Abramowicz}}{{Narayan} et~al.}{2003}]{2003PASJ...55L..69N}
{Narayan} R.,  {Igumenshchev} I.~V.,   {Abramowicz} M.~A.,  2003, \mn@doi [\pasj] {10.1093/pasj/55.6.L69}, \href {https://ui.adsabs.harvard.edu/abs/2003PASJ...55L..69N} {55, L69}

\bibitem[\protect\citeauthoryear{{Noble}, {Krolik}  \& {Hawley}}{{Noble} et~al.}{2009}]{2009ApJ...692..411N}
{Noble} S.~C.,  {Krolik} J.~H.,   {Hawley} J.~F.,  2009, \mn@doi [\apj] {10.1088/0004-637X/692/1/411}, \href {https://ui.adsabs.harvard.edu/abs/2009ApJ...692..411N} {692, 411}

\bibitem[\protect\citeauthoryear{{Noble}, {Mundim}, {Nakano}, {Krolik}, {Campanelli}, {Zlochower}  \& {Yunes}}{{Noble} et~al.}{2012}]{NobleMundim2012}
{Noble} S.~C.,  {Mundim} B.~C.,  {Nakano} H.,  {Krolik} J.~H.,  {Campanelli} M.,  {Zlochower} Y.,   {Yunes} N.,  2012, \mn@doi [\apj] {10.1088/0004-637X/755/1/51}, \href {https://ui.adsabs.harvard.edu/abs/2012ApJ...755...51N} {755, 51}

\bibitem[\protect\citeauthoryear{{Noble}, {Krolik}, {Campanelli}, {Zlochower}, {Mundim}, {Nakano}  \& {Zilh{\~a}o}}{{Noble} et~al.}{2021}]{NobleKrolik2021}
{Noble} S.~C.,  {Krolik} J.~H.,  {Campanelli} M.,  {Zlochower} Y.,  {Mundim} B.~C.,  {Nakano} H.,   {Zilh{\~a}o} M.,  2021, \mn@doi [\apj] {10.3847/1538-4357/ac2229}, \href {https://ui.adsabs.harvard.edu/abs/2021ApJ...922..175N} {922, 175}

\bibitem[\protect\citeauthoryear{Ogilvie}{Ogilvie}{2001}]{Ogilvie2001}
Ogilvie G.~I.,  2001, \mn@doi [Monthly Notices of the Royal Astronomical Society] {10.1046/j.1365-8711.2001.04416.x}, 325, 231

\bibitem[\protect\citeauthoryear{{Paczy{\'n}sky} \& {Wiita}}{{Paczy{\'n}sky} \& {Wiita}}{1980}]{PaczynskyWiita1980}
{Paczy{\'n}sky} B.,  {Wiita} P.~J.,  1980, \aap, \href {https://ui.adsabs.harvard.edu/abs/1980A&A....88...23P} {88, 23}

\bibitem[\protect\citeauthoryear{{Penna}, {McKinney}, {Narayan}, {Tchekhovskoy}, {Shafee}  \& {McClintock}}{{Penna} et~al.}{2010}]{2010MNRAS.408..752P}
{Penna} R.~F.,  {McKinney} J.~C.,  {Narayan} R.,  {Tchekhovskoy} A.,  {Shafee} R.,   {McClintock} J.~E.,  2010, \mn@doi [\mnras] {10.1111/j.1365-2966.2010.17170.x}, \href {https://ui.adsabs.harvard.edu/abs/2010MNRAS.408..752P} {408, 752}

\bibitem[\protect\citeauthoryear{{Perez} \& {Granger}}{{Perez} \& {Granger}}{2007}]{2007CSE.....9c..21P}
{Perez} F.,  {Granger} B.~E.,  2007, \mn@doi [Computing in Science and Engineering] {10.1109/MCSE.2007.53}, \href {https://ui.adsabs.harvard.edu/abs/2007CSE.....9c..21P} {9, 21}

\bibitem[\protect\citeauthoryear{{Peters}}{{Peters}}{1964}]{Peters1964}
{Peters} P.~C.,  1964, \mn@doi [Physical Review] {10.1103/PhysRev.136.B1224}, \href {https://ui.adsabs.harvard.edu/abs/1964PhRv..136.1224P} {136, 1224}

\bibitem[\protect\citeauthoryear{{Plummer}}{{Plummer}}{1911}]{Plummer1911}
{Plummer} H.~C.,  1911, \mn@doi [\mnras] {10.1093/mnras/71.5.460}, \href {https://ui.adsabs.harvard.edu/abs/1911MNRAS..71..460P} {71, 460}

\bibitem[\protect\citeauthoryear{{Ragusa}, {Lodato}  \& {Price}}{{Ragusa} et~al.}{2016}]{RagusaLodato2016}
{Ragusa} E.,  {Lodato} G.,   {Price} D.~J.,  2016, \mn@doi [\mnras] {10.1093/mnras/stw1081}, \href {https://ui.adsabs.harvard.edu/abs/2016MNRAS.460.1243R} {460, 1243}

\bibitem[\protect\citeauthoryear{{Ripperda}, {Bacchini}  \& {Philippov}}{{Ripperda} et~al.}{2020}]{RipperdaBacchiniPhilippov2020}
{Ripperda} B.,  {Bacchini} F.,   {Philippov} A.~A.,  2020, \mn@doi [\apj] {10.3847/1538-4357/ababab}, \href {https://ui.adsabs.harvard.edu/abs/2020ApJ...900..100R} {900, 100}

\bibitem[\protect\citeauthoryear{{Roedig}, {Dotti}, {Sesana}, {Cuadra}  \& {Colpi}}{{Roedig} et~al.}{2011}]{RoedigDotti2011}
{Roedig} C.,  {Dotti} M.,  {Sesana} A.,  {Cuadra} J.,   {Colpi} M.,  2011, \mn@doi [\mnras] {10.1111/j.1365-2966.2011.18927.x}, \href {https://ui.adsabs.harvard.edu/abs/2011MNRAS.415.3033R} {415, 3033}

\bibitem[\protect\citeauthoryear{{Scepi}, {Begelman}  \& {Dexter}}{{Scepi} et~al.}{2024}]{ScepiBegelmanDexter2024}
{Scepi} N.,  {Begelman} M.~C.,   {Dexter} J.,  2024, \mn@doi [\mnras] {10.1093/mnras/stad3299}, \href {https://ui.adsabs.harvard.edu/abs/2024MNRAS.527.1424S} {527, 1424}

\bibitem[\protect\citeauthoryear{{Shakura} \& {Sunyaev}}{{Shakura} \& {Sunyaev}}{1973}]{ShakuraSunyaev1973}
{Shakura} N.~I.,  {Sunyaev} R.~A.,  1973, \aap, \href {https://ui.adsabs.harvard.edu/abs/1973A&A....24..337S} {24, 337}

\bibitem[\protect\citeauthoryear{{Shi}, {Krolik}, {Lubow}  \& {Hawley}}{{Shi} et~al.}{2012}]{ShiKrolik2012}
{Shi} J.-M.,  {Krolik} J.~H.,  {Lubow} S.~H.,   {Hawley} J.~F.,  2012, \mn@doi [\apj] {10.1088/0004-637X/749/2/118}, \href {https://ui.adsabs.harvard.edu/abs/2012ApJ...749..118S} {749, 118}

\bibitem[\protect\citeauthoryear{{Siwek}, {Weinberger}, {Mu{\~n}oz}  \& {Hernquist}}{{Siwek} et~al.}{2023}]{SiwekWeinberger2023}
{Siwek} M.,  {Weinberger} R.,  {Mu{\~n}oz} D.~J.,   {Hernquist} L.,  2023, \mn@doi [\mnras] {10.1093/mnras/stac3263}, \href {https://ui.adsabs.harvard.edu/abs/2023MNRAS.518.5059S} {518, 5059}

\bibitem[\protect\citeauthoryear{{Sorathia}, {Reynolds}, {Stone}  \& {Beckwith}}{{Sorathia} et~al.}{2012}]{2012ApJ...749..189S}
{Sorathia} K.~A.,  {Reynolds} C.~S.,  {Stone} J.~M.,   {Beckwith} K.,  2012, \mn@doi [\apj] {10.1088/0004-637X/749/2/189}, \href {https://ui.adsabs.harvard.edu/abs/2012ApJ...749..189S} {749, 189}

\bibitem[\protect\citeauthoryear{Stone, Tomida, White  \& Felker}{Stone et~al.}{2020}]{Stone2020}
Stone J.~M.,  Tomida K.,  White C.~J.,   Felker K.~G.,  2020, \mn@doi [The Astrophysical Journal Supplement Series] {10.3847/1538-4365/ab929b}, 249, 4

\bibitem[\protect\citeauthoryear{{Tang}, {MacFadyen}  \& {Haiman}}{{Tang} et~al.}{2017}]{TangMacFadyen2017}
{Tang} Y.,  {MacFadyen} A.,   {Haiman} Z.,  2017, \mn@doi [\mnras] {10.1093/mnras/stx1130}, \href {https://ui.adsabs.harvard.edu/abs/2017MNRAS.469.4258T} {469, 4258}

\bibitem[\protect\citeauthoryear{{Tang}, {Haiman}  \& {MacFadyen}}{{Tang} et~al.}{2018}]{TangHaiman2018}
{Tang} Y.,  {Haiman} Z.,   {MacFadyen} A.,  2018, \mn@doi [\mnras] {10.1093/mnras/sty423}, \href {https://ui.adsabs.harvard.edu/abs/2018MNRAS.476.2249T} {476, 2249}

\bibitem[\protect\citeauthoryear{{Tchekhovskoy}, {McKinney}  \& {Narayan}}{{Tchekhovskoy} et~al.}{2009}]{2009ApJ...699.1789T}
{Tchekhovskoy} A.,  {McKinney} J.~C.,   {Narayan} R.,  2009, \mn@doi [\apj] {10.1088/0004-637X/699/2/1789}, \href {https://ui.adsabs.harvard.edu/abs/2009ApJ...699.1789T} {699, 1789}

\bibitem[\protect\citeauthoryear{{Tchekhovskoy}, {McKinney}  \& {Narayan}}{{Tchekhovskoy} et~al.}{2012}]{2012JPhCS.372a2040T}
{Tchekhovskoy} A.,  {McKinney} J.~C.,   {Narayan} R.,  2012, in Journal of Physics Conference Series. IOP, p. 012040 (\mn@eprint {arXiv} {1202.2864}), \mn@doi{10.1088/1742-6596/372/1/012040}

\bibitem[\protect\citeauthoryear{{Tiede}, {Zrake}, {MacFadyen}  \& {Haiman}}{{Tiede} et~al.}{2020}]{TiedeZrake2020}
{Tiede} C.,  {Zrake} J.,  {MacFadyen} A.,   {Haiman} Z.,  2020, \mn@doi [\apj] {10.3847/1538-4357/aba432}, \href {https://ui.adsabs.harvard.edu/abs/2020ApJ...900...43T} {900, 43}

\bibitem[\protect\citeauthoryear{{Tiede}, {Zrake}, {MacFadyen}  \& {Haiman}}{{Tiede} et~al.}{2022}]{TiedeZrake2022}
{Tiede} C.,  {Zrake} J.,  {MacFadyen} A.,   {Haiman} Z.,  2022, \mn@doi [\apj] {10.3847/1538-4357/ac6c2b}, \href {https://ui.adsabs.harvard.edu/abs/2022ApJ...932...24T} {932, 24}

\bibitem[\protect\citeauthoryear{{Tiede}, {Zrake}, {MacFadyen}  \& {Haiman}}{{Tiede} et~al.}{2025}]{TiedeZrake2025}
{Tiede} C.,  {Zrake} J.,  {MacFadyen} A.,   {Haiman} Z.,  2025, \mn@doi [\apj] {10.3847/1538-4357/adc727}, \href {https://ui.adsabs.harvard.edu/abs/2025ApJ...984..144T} {984, 144}

\bibitem[\protect\citeauthoryear{{Tiede}, {O'Neill}  \& {D'Orazio}}{{Tiede} et~al.}{2026}]{TiedeONeillDOrazio2026}
{Tiede} C.,  {O'Neill} D.,   {D'Orazio} D.~J.,  2026, \mn@doi [\apj] {10.3847/1538-4357/ae914e}, \href {https://ui.adsabs.harvard.edu/abs/2026ApJ..1008...26T} {1008, 26}

\bibitem[\protect\citeauthoryear{{Tiwari}, {Chan}, {Bogdanovi{\'c}}, {Jiang}  \& {Davis}}{{Tiwari} et~al.}{2025a}]{TiwariChanBogdanovic2025b}
{Tiwari} V.,  {Chan} C.-H.,  {Bogdanovi{\'c}} T.,  {Jiang} Y.-F.,   {Davis} S.~W.,  2025a, \mn@doi [arXiv e-prints] {10.48550/arXiv.2510.13955}, \href {https://ui.adsabs.harvard.edu/abs/2025arXiv251013955T} {p. arXiv:2510.13955}

\bibitem[\protect\citeauthoryear{{Tiwari}, {Chan}, {Bogdanovi{\'c}}, {Jiang}, {Davis}  \& {Ferrel}}{{Tiwari} et~al.}{2025b}]{TiwariChanBogdanovic2025}
{Tiwari} V.,  {Chan} C.-H.,  {Bogdanovi{\'c}} T.,  {Jiang} Y.-F.,  {Davis} S.~W.,   {Ferrel} S.,  2025b, \mn@doi [\apj] {10.3847/1538-4357/add408}, \href {https://ui.adsabs.harvard.edu/abs/2025ApJ...986..158T} {986, 158}

\bibitem[\protect\citeauthoryear{{Turner} \& {Reynolds}}{{Turner} \& {Reynolds}}{2023}]{TurnerReynolds2023}
{Turner} S. G.~D.,  {Reynolds} C.~S.,  2023, \mn@doi [\mnras] {10.1093/mnras/stad2275}, \href {https://ui.adsabs.harvard.edu/abs/2023MNRAS.525.2287T} {525, 2287}

\bibitem[\protect\citeauthoryear{{Uttley} \& {McHardy}}{{Uttley} \& {McHardy}}{2001}]{UttleyMcHardy2001}
{Uttley} P.,  {McHardy} I.~M.,  2001, \mn@doi [\mnras] {10.1046/j.1365-8711.2001.04496.x}, \href {https://ui.adsabs.harvard.edu/abs/2001MNRAS.323L..26U} {323, L26}

\bibitem[\protect\citeauthoryear{Van~Rossum \& Drake}{Van~Rossum \& Drake}{2009}]{python}
Van~Rossum G.,  Drake F.~L.,  2009, Python 3 Reference Manual.
CreateSpace, Scotts Valley, CA

\bibitem[\protect\citeauthoryear{Virtanen et~al.,}{Virtanen et~al.}{2020}]{2020SciPy-NMeth}
Virtanen P.,  et~al., 2020, \mn@doi [Nature Methods] {10.1038/s41592-019-0686-2}, \href {https://rdcu.be/b08Wh} {17, 261}

\bibitem[\protect\citeauthoryear{{Volonteri}, {Haardt}  \& {Madau}}{{Volonteri} et~al.}{2002}]{VolonteriHaardt2002}
{Volonteri} M.,  {Haardt} F.,   {Madau} P.,  2002, \mn@doi [\apss] {10.1023/A:1019573531536}, \href {https://ui.adsabs.harvard.edu/abs/2002Ap&SS.281..501V} {281, 501}

\bibitem[\protect\citeauthoryear{{Wagg} \& {Broekgaarden}}{{Wagg} \& {Broekgaarden}}{2024}]{software-citation-station-paper}
{Wagg} T.,  {Broekgaarden} F.~S.,  2024, arXiv e-prints, \href {https://ui.adsabs.harvard.edu/abs/2024arXiv240604405W} {p. arXiv:2406.04405}

\bibitem[\protect\citeauthoryear{Wagg, Broekgaarden, Van-Lane, Wu  \& Gültekin}{Wagg et~al.}{2025}]{software-citation-station-zenodo}
Wagg T.,  Broekgaarden F.,  Van-Lane P.,  Wu K.,   Gültekin K.,  2025, TomWagg/software-citation-station: v1.4, \mn@doi{10.5281/zenodo.17654855}, \url {https://doi.org/10.5281/zenodo.17654855}

\bibitem[\protect\citeauthoryear{{Wang}, {Bai}, {Lai}  \& {Lin}}{{Wang} et~al.}{2023}]{WangBaiLai2023}
{Wang} H.-Y.,  {Bai} X.-N.,  {Lai} D.,   {Lin} D. N.~C.,  2023, \mn@doi [\mnras] {10.1093/mnras/stad2884}, \href {https://ui.adsabs.harvard.edu/abs/2023MNRAS.526.3570W} {526, 3570}

\bibitem[\protect\citeauthoryear{{Wang}, {Most}  \& {Hopkins}}{{Wang} et~al.}{2025}]{2025arXiv250816855W}
{Wang} H.-Y.,  {Most} E.~R.,   {Hopkins} P.~F.,  2025, \mn@doi [arXiv e-prints] {10.48550/arXiv.2508.16855}, \href {https://ui.adsabs.harvard.edu/abs/2025arXiv250816855W} {p. arXiv:2508.16855}

\bibitem[\protect\citeauthoryear{{W}es {M}c{K}inney}{{W}es {M}c{K}inney}{2010}]{mckinney-proc-scipy-2010}
{W}es {M}c{K}inney 2010, in {S}t\'efan van~der {W}alt {J}arrod {M}illman eds, {P}roceedings of the 9th {P}ython in {S}cience {C}onference. pp 56--61, \mn@doi{10.25080/Majora-92bf1922-00a}

\bibitem[\protect\citeauthoryear{{Westernacher-Schneider}, {Zrake}, {MacFadyen}  \& {Haiman}}{{Westernacher-Schneider} et~al.}{2022}]{WesternacherSchneiderZrake2022}
{Westernacher-Schneider} J.~R.,  {Zrake} J.,  {MacFadyen} A.,   {Haiman} Z.,  2022, \mn@doi [\prd] {10.1103/PhysRevD.106.103010}, \href {https://ui.adsabs.harvard.edu/abs/2022PhRvD.106j3010W} {106, 103010}

\bibitem[\protect\citeauthoryear{{Xin}, {Isi}, {Farr}  \& {Haiman}}{{Xin} et~al.}{2026}]{XinIsi2026}
{Xin} C.,  {Isi} M.,  {Farr} W.~M.,   {Haiman} Z.,  2026, \mn@doi [\apj] {10.3847/1538-4357/ae40b3}, \href {https://ui.adsabs.harvard.edu/abs/2026ApJ...999..149X} {999, 149}

\bibitem[\protect\citeauthoryear{Zanazzi \& Ogilvie}{Zanazzi \& Ogilvie}{2020}]{ZanazziOgilvie2020}
Zanazzi J.~J.,  Ogilvie G.~I.,  2020, \mn@doi [Monthly Notices of the Royal Astronomical Society] {10.1093/mnras/staa3127}, 499, 5562

\bibitem[\protect\citeauthoryear{{Zrake}, {Tiede}, {MacFadyen}  \& {Haiman}}{{Zrake} et~al.}{2021}]{ZrakeTiede2021}
{Zrake} J.,  {Tiede} C.,  {MacFadyen} A.,   {Haiman} Z.,  2021, \mn@doi [\apjl] {10.3847/2041-8213/abdd1c}, \href {https://ui.adsabs.harvard.edu/abs/2021ApJ...909L..13Z} {909, L13}

\makeatother
\end{thebibliography}

\appendix

\section{Precession-capturing Gravitational potential}
\label{app:gravpot}

We sought a central pseudo-Newtonian force which equatorial apsidal precession more closely following that of Kerr than the Paczy\'nski--Wiita (PW) potential, and while still resulting in an innermost stable circular orbit (ISCO). Let the inward radial force magnitude per unit area and mass be written as
\begin{equation}
F_r(r) = \frac{GM}{r^2}\,g(u),
\qquad
u \equiv \frac{GM}{r},
\end{equation}
so that the circular orbital frequency is
\begin{equation}
\Omega^2(r) = \frac{F_r(r)}{r}
= \frac{GM}{r^3}\,g(u).
\end{equation}
For any central force, the radial epicyclic frequency is
\begin{equation}
\kappa^2(r)
= \frac{1}{r^3}\frac{d}{dr}\!\left[r^3 F_r(r)\right].
\end{equation}
Substituting the above form gives
\begin{equation}
\kappa^2(r)
= \frac{GM}{r^3}\left[g(u)-u g'(u)\right],
\end{equation}
and therefore
\begin{equation}
\frac{\kappa^2}{\Omega^2}
=
\frac{g-u g'}{g}
=
1-u\frac{d\ln g}{du}.
\end{equation}

We then require this ratio to match the equatorial Kerr form, which is exact,
\begin{equation}
\frac{\kappa^2}{\Omega^2}
=
1 - 6u + 8\chi u^{3/2} - 3\chi^2 u^2,
\end{equation}
where
\begin{equation}
\chi \equiv \frac{a}{M}
\end{equation}
is the dimensionless spin parameter. Equating the two expressions yields
\begin{equation}
1-u\frac{d\ln g}{du}
=
1 - 6u + 8\chi u^{3/2} - 3\chi^2 u^2,
\end{equation}
or
\begin{equation}
\frac{d\ln g}{du}
=
6 - 8\chi u^{1/2} + 3\chi^2 u.
\end{equation}
Integrating gives
\begin{equation}
\ln g
=
6u - \frac{16}{3}\chi u^{3/2} + \frac{3}{2}\chi^2 u^2 + C.
\end{equation}
Imposing the Newtonian limit, \(g\to 1\) as \(u\to 0\), specifies \(C=0\) so that,
\begin{equation}
g(u)
=
\exp\!\left[
6u - \frac{16}{3}\chi u^{3/2} + \frac{3}{2}\chi^2 u^2
\right].
\end{equation}
The resulting fitted force, corresponding to our `PNP' potential, is 
\begin{equation}
F_r(r)
=
\frac{GM}{r^2}
\exp\!\left[
\frac{6GM}{r}
-\frac{16}{3}\chi\left(\frac{GM}{r}\right)^{3/2}
+\frac{3}{2}\chi^2\left(\frac{GM}{r}\right)^2
\right].
\end{equation}

The corresponding circular orbital frequency is
\begin{equation}
\Omega^2(r)
=
\frac{GM}{r^3}
\exp\!\left[
\frac{6GM}{r}
-\frac{16}{3}\chi\left(\frac{GM}{r}\right)^{3/2}
+\frac{3}{2}\chi^2\left(\frac{GM}{r}\right)^2
\right],
\end{equation}
while, by construction,
\begin{equation}
\frac{\kappa^2}{\Omega^2}
=
1 - 6u + 8\chi u^{3/2} - 3\chi^2 u^2.
\end{equation}
The marginally stable orbit is defined by \(\kappa^2=0\), i.e.
\begin{equation}
1 - 6u + 8\chi u^{3/2} - 3\chi^2 u^2 = 0.
\end{equation}
Equivalently, with \(x \equiv \sqrt{u}\),
\begin{equation}
1 - 6x^2 + 8\chi x^3 - 3\chi^2 x^4 = 0.
\end{equation}

For nearly circular orbits, the apsidal advance per radial period is
\begin{equation}
\Delta\varpi
=
2\pi\left(\frac{\Omega}{\kappa}-1\right),
\end{equation}
while a useful measure of differential precession is
\begin{equation}
\frac{d\Delta\varpi}{dr}.
\end{equation}

In practice, to avoid excessively steep inner forces in numerical integration, we employ a three-zone hybrid version of the fitted force. Let \(r_t\) denote an outer transition radius and \(r_d<r_t\) an inner damping radius. For \(r\ge r_t\), the force is the exact fitted force,
\begin{equation}
F_r(r)=F_{\rm fit}(r).
\end{equation}
At \(r_t\), define
\begin{equation}
F_t \equiv F_{\rm fit}(r_t),
\qquad
F_t' \equiv \left.\frac{dF_{\rm fit}}{dr}\right|_{r_t},
\end{equation}
and continue inward with the tangent line
\begin{equation}
F_{\rm lin}(r)=F_t + F_t'(r-r_t),
\qquad
r_d \le r < r_t.
\end{equation}
For \(r<r_d\), we smoothly roll over the linear branch using
\begin{equation}
F_r(r)
=
F_{\rm lin}(r)\,
\frac{r^2/(r^2+\epsilon^2)}{r_d^2/(r_d^2+\epsilon^2)},
\qquad
r<r_d,
\end{equation}
where \(\epsilon\) sets the inner smoothing scale. Thus the full hybrid force is
\begin{equation}
F_r(r)=
\begin{cases}
F_{\rm fit}(r), & r \ge r_t, \\[1ex]
F_t + F_t'(r-r_t), & r_d \le r < r_t, \\[1ex]
\displaystyle
\left[F_t + F_t'(r-r_t)\right]
\frac{r^2/(r^2+\epsilon^2)}{r_d^2/(r_d^2+\epsilon^2)},
& r < r_d.
\end{cases}
\end{equation}
The Cartesian acceleration is then obtained from the radial force magnitude as
\begin{equation}
a_x = -\frac{F_r(r)}{r}x,
\qquad
a_y = -\frac{F_r(r)}{r}y,
\qquad
a_z = -\frac{F_r(r)}{r}z,
\end{equation}
for coordinates {x,y,z} centered on one of the BHs.

For simulations in this work performed with the PNP prescription, for $BH_i$ with mass $M_i$, we choose transition values of $r_t=4M_i$, $r_d=M_i$, $\epsilon=0.1M_i$. 

\begin{figure*}
\label{fig:ecctrajcomp}
\centering
\includegraphics[width=\textwidth]{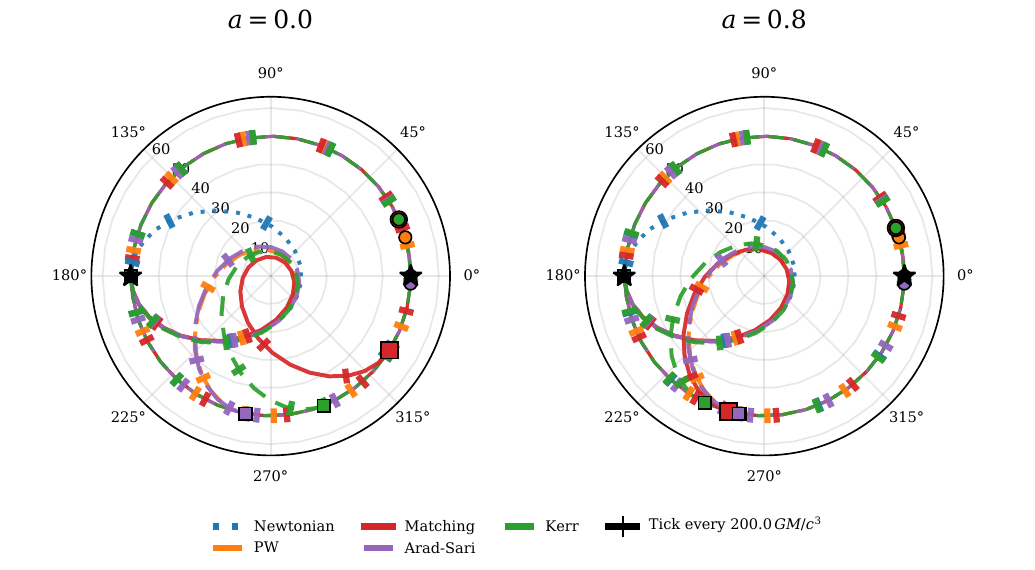}
\caption{For spins 0.0 and 0.8, the trajectories of a near-circular and highly eccentric infalling particle are shown for Newtonian (blue), Paczy\'{n}ski-Wiita (orange), our precession-matching potential (red), a PNP potential of the type exampled in \protect\cite{AradSari2026} (purple), and a geodesic for Kerr. Ticks mark equal coordinate time spacing of 150M along each trajectory.}
\end{figure*}

We test the robustness of our hybrid potential in capturing true Kerr precession by integrating eccentric particle trajectories around a single stationary point mass. Trajectories for the PNP potential, Kerr, Plummer, PW, and one solved for using the method introduced in \cite{AradSari2026} are shown in Figure~\ref{fig:ecctrajcomp}. The trajectories are integrated from the same injection point at r=50M, approximating an origin from the CBD inner edge for the separations considered in our suite of simulations, and have specific angular momentum $l=f_l l_c$, where $l_c$ is the specific angular momentum of a circular orbit at that radius in each potential. $f_l$ values of 0.999 and 0.6 are chosen. We choose 0.999 rather than circular so that precession over the integration of one epicycle 
is well defined. 

For zero spin, and the highly eccentric trajectory, the single-orbit precession angle of PW and \cite{AradSari2026} are identical, about 35$^{\circ}$ behind Kerr. Our potential is about 35$^{\circ}$ ahead of Kerr. On the other hand, for perturbative eccentricity, we match Kerr exactly. 

For spin of $a=\chi=0.8$ chosen to represent an equal-mass MBHB merger remnant, while PW and \cite{AradSari2026} match Kerr better only for this spin, our potential even more closely matches Kerr, with a difference in precession angle of less than 10$^{\circ}$. The coordinate time of all orbits for the 0.8 spin case are within about 10\% of Kerr, though generally traverse faster.

\section{Entropy versus black-body ($T^4$) cooling}
\label{app:entvst4}

We have demonstrated that target-entropy cooling on a very rapid timescale mimics the behavior of isothermal evolution, and have reasoned that black-body cooling also leads to this behavior when large temperature fluctuations are present, as they are in the cavity and inner CBD of an accreting binary. Here we will derive the relevant timescales referenced in the Discussion. 

For black-body cooling, let us consider the equation for energy evolution of an orbiting parcel of gas, representing the balance of local viscous heating $Q^+$ and radiative cooling $Q^-$,

\begin{equation}
\label{eqn:balance}
\begin{aligned}
\frac{{\rm d}U_{\rm g}}{{\rm d}t}
&= Q^{+}-Q^{-}
\\[3pt]
&=
\frac{9}{4}\alpha P\Omega_{\rm K}
-
\frac{8}{3}
\frac{\sigma_{\rm SB}T^{4}}{\kappa\Sigma}.
\end{aligned}
\end{equation}
For a gas-pressure dominated disk, $U_g=P_g/(\gamma-1)=\Sig T/(\gamma-1)$. While the physical scaling in LK23 would formally result in a radiation pressure dominated disk over the radial ranges evolved, as it would in our study, it is evolved as a gas pressure dominated disk, with $\gamma=5/3$, and all dissipated thermal energy goes into $U_g$. 

Eqn.~\ref{eqn:balance} is at the heart of the SS disk solution, which has a timescale ordering of the dynamical time ($t_\mathrm{dyn}$), thermal time ($t_\mathrm{th}$), and viscous time ($t_\mathrm{visc}$), of $t_{\rm dyn} \ll t_{\rm th} \ll t_{\rm visc}$, where
\begin{align}
t_{\rm dyn}
&\equiv \Omega_{\rm K}^{-1},
\\[3pt]
t_{\rm th}
&\equiv \frac{U_{\rm g}}{Q^{+}}
\sim \frac{1}{\alpha\Omega_{\rm K}}
=\alpha^{-1}t_{\rm dyn},
\\[3pt]
t_{\rm visc}
&\equiv \frac{r^{2}}{\nu}
\sim
\frac{1}{\alpha}
\left(\frac{r}{H}\right)^{2}
\Omega_{\rm K}^{-1}
=
\left(\frac{r}{H}\right)^{2}t_{\rm th},
\end{align}
and $\alpha,H/r\ll1$. Our fiducial cooling thus occurs slightly faster than the thermal timescale, and slightly slower than the dynamical timescale.

The region where most shock heating and significant compression occurs is where gas propelled outwards by the binary’s non-axisymmetric gravitational torque strikes the cavity wall. We find that the highly nonlinear interaction of the binary with the CBD results in localized regions of significant compressional and shock heating that can raise the temperature in small regions by factors of 10 to even 50 times the equilibrium value, $T_\mathrm{eq}$, on timescales very short compared to the orbital period. An estimate of this timescale, $t_\mathrm{impact}$, can be made by calculating the speed at which the material flung outwards by the binary traverses a distance equal to the scale height of the disk. First, to estimate the stream speed at impact, we take the ratio of the stream traversal distance to the time it takes for a full stream to pass a specific azimuthal angle. This gives estimates the stream velocity to be $v_s\sim a/(\pi/2\Omega_b)$, since the stream occupies about $\pi/2$ in azimuth. Then, to traverse the height of the disk at the inner edge of the CBD, $2(H/r)a$, takes time $t_\mathrm{impact}\sim 0.3 a^{3/2}\sim 0.3t_\mathrm{dyn}$. Therefore, our fiducial entropy cooling occurs with $t_\mathrm{impact}\ll t_\mathrm{cool}$, and impact response is adiabatic.

How does blackbody cooling respond to these impacts? Under the change of variables $x\equiv{\delta T}/{T_{\rm eq}}$, and for a fluctuation decaying from peak temperature $x_0=\delta T_0/T_\mathrm{eq}$, Eqn.~\ref{eqn:balance} can be written,

\begin{equation}
\frac{{\rm d}x}{{\rm d}t}
=
\frac{1}{t_{\rm th}}
\left[(1+x)-(1+x)^{4}\right].
\end{equation}
This has solutions for small and large temperature fluctuations,
\begin{equation}
x(t)\simeq
\begin{cases}
x_0\exp\!\left(-3t/t_{\rm th}\right)
&\ x\ll1
\\[6pt]
\left(x_0^{-3}+3t/t_{\rm th}\right)^{-1/3}
& \ x\gg1.
\end{cases}
\end{equation}
By inspection, in the $x\ll1$ limit, cooling occurs on the timescale $t_{\rm cool}\sim t_{\rm th}/3 \sim (\alpha \Omega_\mathrm{K})^{-1}$, and for $x\gg1$ it becomes $t_{\rm cool}(x)\sim t_{\rm th}/(3x^3)\sim (\delta T/T)^{-3}$. Therefore, our measured perturbations of $x\sim10-50$, occurring on the $t_\mathrm{impact}$ timescale, can result in cooling excess heat away $10^3 - 50^3$ times faster. This results in a new effective ordering of $t_{\rm cool} \ll t_\mathrm{impact} < t_{\rm dyn} \ll t_{\rm visc}$. If the thermal timescale is interpreted as the fastest timescale of temperature change, then $t_\mathrm{th}\sim t_\mathrm{cool}$, and the timescale ordering becomes $t_{\rm th} \ll t_{\rm dyn} \ll t_{\rm visc}$, the same as isothermal evolution.

\bsp
\label{lastpage}
\end{document}